\documentclass[lineno]{jfm}

\usepackage{graphicx}
\usepackage{newtxtext}
\usepackage{newtxmath}
\usepackage{natbib}
\usepackage{xcolor}
\usepackage{hyperref}
\usepackage[percent]{overpic}
\hypersetup{
    colorlinks = true,
    urlcolor   = blue,
    citecolor  = black,
}

\newcommand{\RomanNumeralCaps}[1]
\linenumbers

\newcommand{\pd}[2]{\frac{\partial#1}{\partial#2}}

\newcommand{\Pdc}[3]{\frac{\partial^{2} #1 }{\partial #2 \partial #3}}
\newcommand{\pdline}[2]{\partial #1 / \partial #2}

\newcommand{\avg}[1]{\langle #1 \rangle}

\newcommand{\beq}{\begin{equation}}
\newcommand{\eeq}{\end{equation}}

\newcommand{\st}[1]{{\color{black} #1}}
\newcommand{\ie}{{i.e.}}

\title{The impact of finite span and wing-tip vortices on a turbulent NACA0012 wing}

\author{Siavash Toosi\aff{1,2,3}
  \corresp{\email{siavash.toosi@fau.de}},
  A. Peplinski\aff{1,2}, P. Schlatter\aff{3,1,2}
 \and R. Vinuesa\aff{1,2} \corresp{\email{rvinuesa@mech.kth.se}}}

\affiliation{
\aff{1} FLOW, Engineering Mechanics, KTH Royal Institute of Technology, SE-100 44 Stockholm, Sweden
\aff{2} Swedish e-Science Research Centre (SeRC), Stockholm, Sweden
\aff{3} Institute of Fluid Mechanics (LSTM), Friedrich-Alexander-Universität (FAU) Erlangen-Nürnberg, 91058 Erlangen, Germany
}

\begin{document}
\maketitle

\begin{abstract}

High-fidelity simulations are conducted to investigate the turbulent boundary layers around a finite-span NACA0012 wing
with a rounded wing-tip geometry
at a chord-based Reynolds number of $Re_c=200\,000$
and at various angles of attack up to $10^\circ$. 
The study aims to discern the differences between the boundary layers on the finite-span wing 
and those on infinite-span wings at equivalent angles of attack. 
The finite-span boundary layers exhibit:
(i) an altered streamwise and a non-zero spanwise pressure gradient as a result of the variable downwash induced by the wing-tip vortices 
(an inviscid effect typical of finite-span wings);
(ii) differences in the flow history at different wall-normal distances, caused by the variable flow angle in the wall-normal direction
(due to constant pressure gradients and variable momentum normal to the wall);
(iii) laminar flow entrainment into the turbulent boundary layers near the wing tip
(due to a laminar/turbulent interface); and
(iv) variations in boundary layer thickness across the span, attributed to the variable wall-normal velocity in that direction
(a primarily inviscid effect).
These physical effects are then used to explain the differences 
in the Reynolds stress profiles and other boundary layer quantities,
including the reduced near-wall peak of the streamwise Reynolds stress
and the elevated Reynolds stress levels near the boundary layer edge, 
both observed in the finite-span wings.
Other aspects of the flow, such as the downstream development of wing-tip vortices and their interactions with the surrounding flow,
are reserved for future investigations.

\end{abstract}


{\bf MSC Codes }  %{\it(Optional)} Please enter your MSC Codes here
%\st{I am thinking of these: }
76D10, %\st{(Boundary-layer theory, separation and reattachment, higher-order effects)},
76F40, %\st{(Turbulent boundary layers)},
76F65, %\st{(Direct numerical and large eddy simulation of turbulence)},
76G25. %\st{(General aerodynamics and subsonic flows)}.

%\clearpage

\section{Introduction \label{sec:intro}}

A direct result of the pressure difference between the suction and pressure sides of a finite-span wing
is the formation of wing-tip vortices and 
their induced three-dimensionality.
Despite the formation of these vortices near the tip, 
they exert
a global influence, 
affecting the entire span of the wing and giving rise to a complex, 
three-dimensional (3D) flow.

\begin{figure}
  \centering
    \includegraphics[width=50mm,clip=true,trim=25mm 1mm 40mm 1mm]{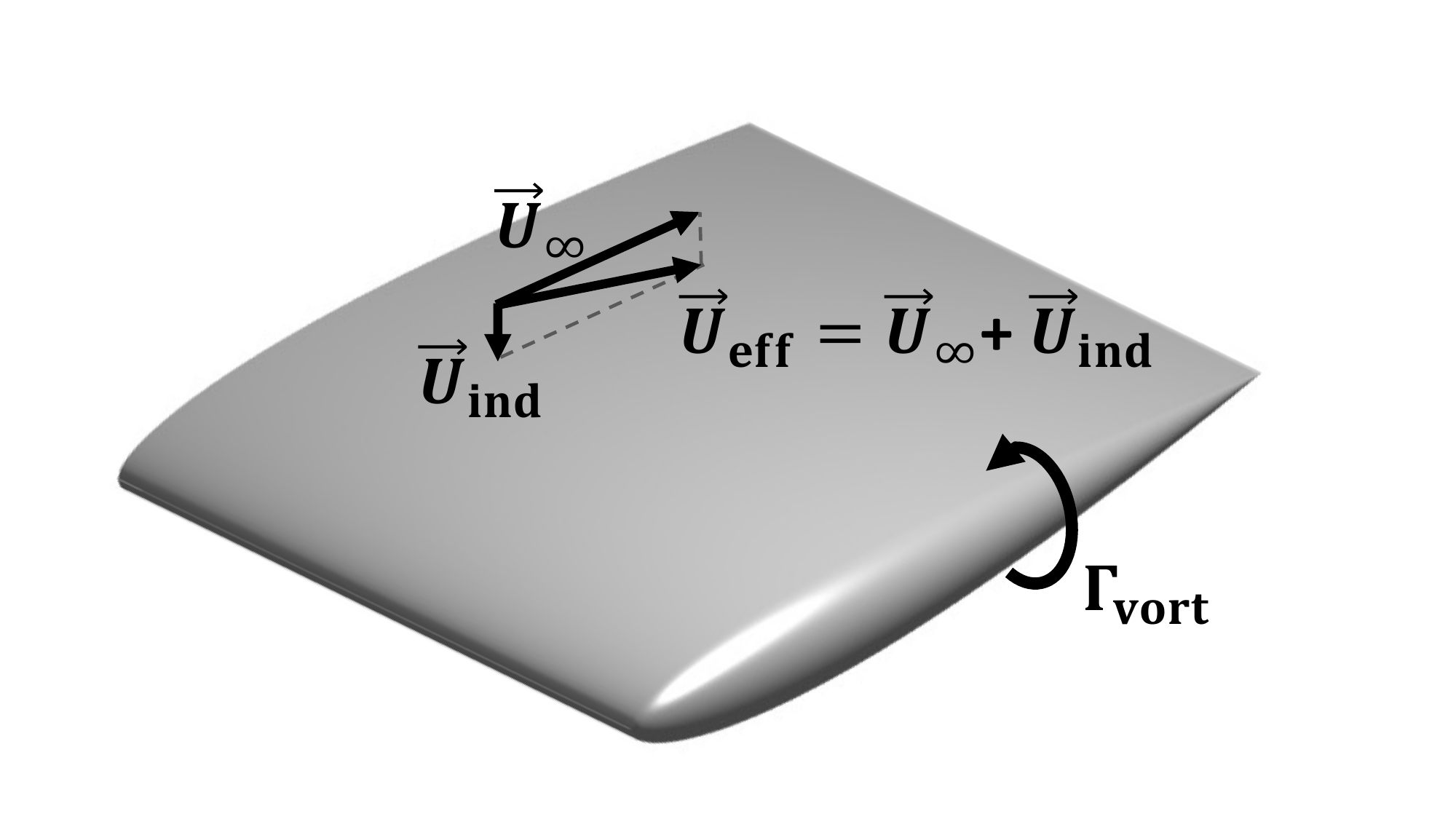}\llap{\parbox[b]{48mm}{(a)\\\rule{0ex}{30mm}}}
    \includegraphics[width=80mm,clip=true,trim=1mm 30mm 1mm 16mm]{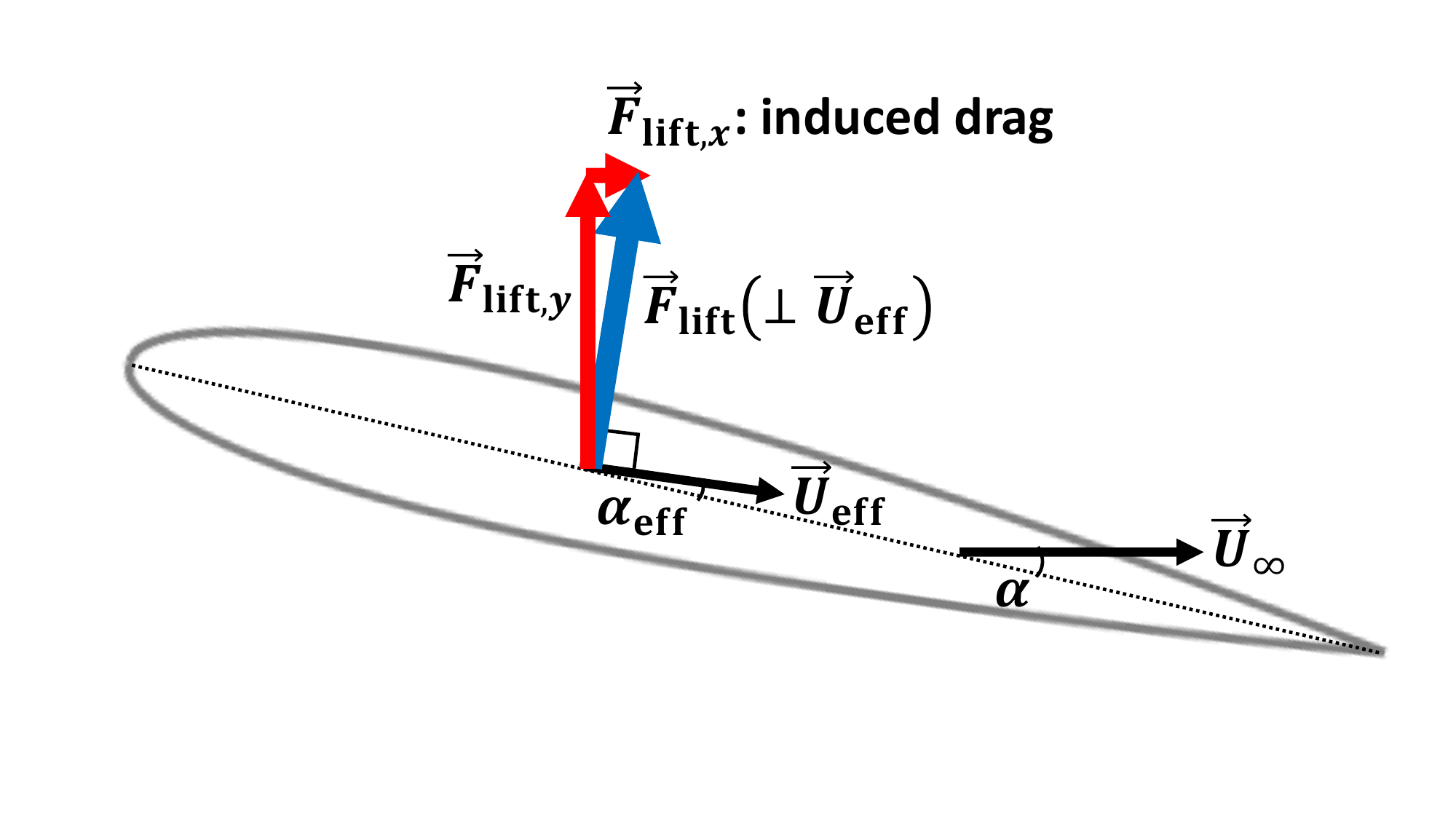}\llap{\parbox[b]{73mm}{(b)\\\rule{0ex}{30mm}}}%\\
  \caption{
  \label{fig:intro-wingtip}
  The impact of wing-tip vortices on the effective angle of attack and induced drag: 
  (a) vortex with strength (circulation) $\Gamma_{\rm vort}$ 
  induces a downward velocity (downwash) of 
  $\vec{U}_{\rm ind}\approx \Gamma_{\rm vort}/4\pi d_{\rm vort}$ 
  (where $d_{\rm vort}$ is the distance along the span to the vortex core)
  on the wing~\citep[cf.][]{Houghton:book}; 
  (b) change in the effective angle of attack
  and generation of a new component of drag, $\vec{F}_{{\rm lift},x}$, 
  as a result of deflection of the effective free-stream direction, $\vec{U}_{\rm eff}$, 
  compared to the geometric free-stream %(\ie, flight) 
  direction, $\vec{U}_\infty$.
  }
\end{figure}

The most significant impact of the wing-tip vortices on the flow is their induced downwash on the 
wing,~\ie, an induced inviscid velocity normal to the free-stream in the downward direction~\citep[cf.][]{Houghton:book}.
As illustrated in figure~\ref{fig:intro-wingtip},
this downwash changes the free-stream direction, 
reducing the effective angle of attack and consequently the lift, 
while altering the pressure distribution in the process.
The change in the pressure distribution 
(and thus
the streamwise pressure gradient)
significantly impacts the development of the boundary layers. 
In addition, as the induced downwash is inversely proportional to the distance from the vortex core
(refer to the caption of figure~\ref{fig:intro-wingtip}), 
different spanwise locations on the wing encounter varying free-stream directions 
and a non-uniform pressure distribution along the span. 
Note that this is a simplified explanation valid for the geometry of figure~\ref{fig:intro-wingtip}, 
but the conclusion is generally valid for other wing %planforms and 
configurations~\citep[cf.][]{Houghton:book}.
This leads to 
a spanwise pressure gradient which varies across the chord and span, 
a non-zero spanwise acceleration, 
and the formation of %three-dimensional 
boundary layers 
that exhibit three-dimensional behavior,
such as skewed velocity profiles,
across a large portion of the wing's span. 
This non-zero and variable spanwise velocity introduces additional complexities to the boundary layers,
which are
added to those already present due to adverse or favorable streamwise 
pressure gradients~\citep[cf.][]{spalart:93,perry:02,aubertine:05,monty:11,harun:13,bobke:17,bross:19,pozuelo:22,devenport:22}, 
as well as spanwise pressure gradients 
and other three-dimensional effects~\citep[cf.][]{johnston:60,perry:65,VDB:75,rotta:79,pirece:83,bradshaw:85,spalart:89,moin:90,degani:93,johnston:96,olcmen:95,coleman:00,kannepalli:00,schlatter-brandt:10,kevin:19,suardi:20,devenport:22}.

%Our 
The goal here is to better understand the flow in the vicinity of the wing,
and in particular,
how the turbulent boundary layers are influenced by %3D effects. 
the induced three-dimensionality of the 
finite-span geometry and the resulting
wing-tip vortices.
This is done by identifying the behaviors present in the finite-span wings
which are absent in the limit of infinite span. 
To do this, a set of 
high-fidelity simulations
is carried out
for wings with a symmetric NACA0012 profile and rounded wing-tip geometry
at a chord-based Reynolds number of 
$Re_c=U_\infty c/\nu =200\,000$ 
(where $U_\infty$ is the free-streem velocity, $c$ is the chord,
and $\nu$ is the kinematic viscosity).
%We consider free-flight conditions at multiple angles of attack of $\alpha=0^\circ$, $5^\circ$, and $10^\circ$.
Free-flight conditions are considered at angles of attack of $\alpha=0^\circ$, $5^\circ$, and $10^\circ$.
Additionally, %we perform 
another set of simulations 
is performed
for infinite-span (i.e., periodic) wings 
with the same profile, Reynolds number, and angle of attack, 
including an additional configuration of $\alpha=2^\circ$, which matches 
the near-root effective angle of attack of the finite-span wing at $\alpha=5^\circ$.
%We then compare the boundary layers of the finite-span and infinite-span wings, 
%identify the differences, 
%and propose explanations for these differences and the underlying mechanisms that lead to them.
The boundary layers of the finite-span and infinite-span wings are then compared,
the discrepancies are identified, 
and explanations are proposed for the observed differences 
as well as the underlying mechanisms responsible for them.

%Our 
The primary focus is on the differences between the finite- and infinite-span wings;
thus, common behaviors such as the response to favorable and adverse pressure gradients are excluded from this study.
Furthermore, we
limit our scope to the regions of the wing 
which are
not dominated by wing-tip vortices or trailing edge effects.
For instance, the flow on the suction and pressure sides of the wing possess opposing spanwise components 
(directed towards the root on the suction side and towards the tip on the pressure side). 
At the trailing edge, these flows (with different spanwise, streamwise, and wall-normal velocity components) intersect, 
leading to additional shear and vorticity which could influence 
both the upstream boundary layers and the downstream wake.
Similarly, closer to the tip, especially on the suction side,
there are stronger three-dimensional effects and secondary flows present at a more local scale
in close vicinity of the wing-tip vortex.
%Here, we focus 
The focus here is
on regions of the wing that are not dominated by these effects.
%and will reserve further investigation of the trailing edge and near-tip effects for future work.

Although not directly related to this work, it is worth noting that wing-tip vortices contribute to an additional drag component, 
known as the induced drag~\citep[also called lift-induced drag, or vortex drag; cf.][]{Houghton:book,phak}. 
A brief explanation of this is provided in the caption of figure~\ref{fig:intro-wingtip}. 
More detailed discussions
about this phenomenon, including potential methods to mitigate its effects, 
can be found in~\citet{Houghton:book},~\citet{phak},~\citet{Kroo:01},~\citet{Ceron-Munoz:13} and~\citet{Phillips:19}.

Moreover, %we do 
this work does
not present an in-depth study of the structure of wing-tip vortices, their formation, development, and downstream influence. 
A wealth of research has been dedicated to these topics and is readily available in literature. 
A few significant studies worth highlighting here include the work of~\citet{Spalart:98,Spalart:08},
focusing on the development of these vortices in the far wake of an aircraft,
and studies by~\citet{Devenport:96},~\citet{Chow:97:aiaa,Chow:97:nasa}, and~\citet{Giuni:2013} 
mainly exploring the near-wake behavior of these vortices in more canonical settings. 
Other interesting works include studies into the meandering motion of these vortices for rigid and stationary wings~\citep[cf.][]{Dghim:21},
the impact of the heaving motion of the wing on the structure and development of the wing-tip vortices~\citep[cf.][]{Garmann:17,Fishman:17},
the interactions of wing-tip vortices with other co-rotating or counter-rotating vortices~\citep{Devenport:97,Devenport:99},
and the interaction of streamwise-oriented vortices (e.g., wing-tip vortices) 
upon their incidence on other downstream aerodynamic surfaces~\citep[cf.][]{Rockwell:98,Garmann:15,McKenna:17}.
We are not aware of any work specifically investigating the influence of the wing-tip vortices on the turbulent boundary layers formed on the wings, 
which motivated the present study.

The remainder of the paper is structured as follows:
Section~\ref{sec:numerical-setup} details the numerical setup, 
including the solver, computational domain, boundary conditions, grid, and other parameters. 
%Our 
The approach for %remove initial transients and 
obtaining accurate statistics 
is also outlined in this section, with more details provided in appendix~\ref{app:averaging}. 
Section~\ref{sec:flowfield} describes the general flow field and its most important features,
and section~\ref{sec:BL},
which is the main focus of this work,
takes a closer look at the impact of finite span and three-dimensionality 
on the turbulent boundary layers.
Finally, section~\ref{sec:conclusions} summarizes the findings, 
acknowledges the limitations, and discusses potential future directions.

\section{Numerical setup \label{sec:numerical-setup}}

\subsection{Numerical solver}

The numerical solutions presented in this paper are obtained by solving the incompressible Navier--Stokes equations,
\beq 
\pd{u_i}{t}+ u_j \pd{u_i}{x_j}
=
-\frac{1}{\rho}\pd{p}{x_i} + \nu \Pdc{u_i}{x_j}{x_j}
\, ,
\eeq
under the divergence free constraint, $\pdline{u_i}{x_i}=0$,
where $u_i$ and $p$ are the instantaneous velocity and pressure fields, 
$x_i$ and $t$ are the spatial coordinates and time,
and $\rho$ and $\nu$ are the fluid density and kinematic viscosity. 

These equations are discretized in space and integrated in time using the high-order solver Nek5000,
developed by~\citet{nek5000}, with adaptive mesh refinement (AMR) capabilities developed at KTH~\citep{peplinski:18,offermans:thesis,offerman:amr,tanarro:amr}.
Nek5000 is based on the spectral-element method~\citep{patera:84}, essentially a high-order finite-element method, which combines the flexibility of the finite-element formulation in meshing complex geometries with the numerical accuracy of spectral methods~\citep{deville:book}. 
Inside each element, the velocity field is represented by a polynomial of order $p$ (here $p=7$ in all cases) using Lagrange interpolants on the Gauss--Lobatto--Legendre (GLL) points ($N=p+1$ GLL points in each direction), while the pressure is represented on $p-1$ Gauss--Legendre (GL) points 
%(resulting in the same polynomial expansion order for the pressure and velocity fields) 
following the $P_N-P_{N-2}$ formulation~\citep{maday:87}.
The nonlinear convective term is overintegrated on a grid with $3N/2$ GL points in each direction, to avoid (or reduce) aliasing errors. 
Time stepping is performed by an implicit third-order backward-differentiation scheme for the viscous terms 
and an explicit third-order extrapolation for the nonlinear terms~\citep{karniadakis:91}.
A high-pass filter relaxation term~\citep{schlatter:hpfrt} is added to the right-hand side of the Navier--Stokes equations. 
This term provides numerical stability and acts as a subgrid-scale (SGS) dissipation.
This specific setup has been used and verified in several previous studies~\citep{schlatter:10,eitel-amor:14}, 
including wing simulations~\citep{negi:wing,vinuesa:wing:18}
and flow around obstacles~\citep{lazpita:22,atzori:23}.

The standard version of Nek5000 is based on hexahedral elements of conforming topology (\ie, no ``hanging nodes'' are allowed). The AMR version adds the capability of handling non-conforming hexahedral elements with hanging nodes, and so, adds an $h$-refinement capability where each element can be refined individually. 
Solution continuity at non-conforming interfaces is ensured by interpolating from the ``coarse side'' onto the ``fine side''. 
The AMR version includes some modifications to the pressure solver and preconditioner~\citep{peplinski:18}, 
as well as the stiffness matrix and its direct summation operations~\citep{offermans:thesis,offerman:amr}; 
however, the majority of the code is identical to the standard version.
The AMR version of the code has gone through an extensive verification and validation (V\&V) process, including wing simulations~\citep[cf.][]{tanarro:amr} and other flows~\citep[cf.][]{offerman:amr}.

\subsection{Computational domain and setup}

\begin{figure}
  \centering
    \includegraphics[width=100mm,clip=true,trim=25mm 20mm 10mm 20mm]{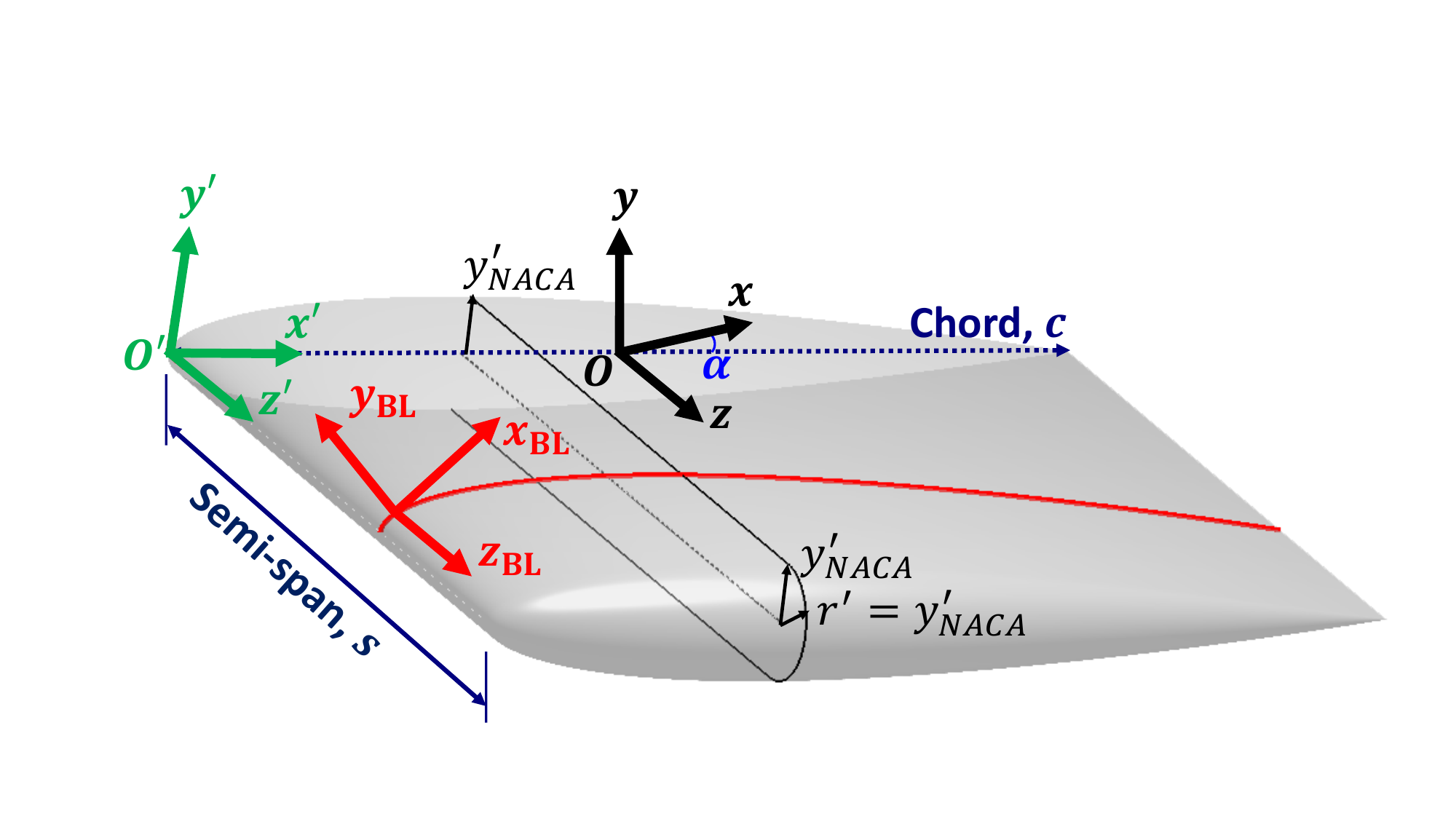}
  \caption{
  \label{fig:wing-geometry}
  A schematic of the wing-surface definition and its
  placement in the computational domain with the global Cartesian coordinate system $(x,y,z)$ and its origin $O$ represented in black, the rotated wing coordinate system $(x',y',z')$ and its origin $O'$ represented in green, 
  and the boundary layer coordinate system $(x_{\rm BL},y_{\rm BL},z_{\rm BL})$ represented in red. 
  Note that $(x_{\rm BL},y_{\rm BL},z_{\rm BL})$ is shown at a random location on the wing
  for visual clarity, while its origin $O_{\rm BL}$ is in fact on $O'$.
  The figure also shows the rounded (semi-circle) wing-tip geometry and its definition. 
  The wing has a chord length of $c$ and a semi-span of $s=b/2=0.75c$ (dark blue), 
  with a geometric angle of attack of $\alpha$ (blue).
  }
\end{figure}

There are a total of seven different configurations studied in this work. This includes four periodic (infinite-span) wing sections and three finite-span wings.
All cases are based on a symmetric NACA0012 profile. 
Both finite- and infinite-span wings
have a non-tapered and non-swept planform with zero dihedral angle and no twist. 
This means that the airfoil chord and angle of attack are constant in the spanwise direction, 
and the line that connects the leading-edge of the different spanwise sections of the wing
is normal to both the free-stream direction and the lift direction (figure~\ref{fig:wing-geometry}).
The finite-span wings have an aspect ratio (equal to the full-span-to-chord ratio for rectangular planforms) of 
$b/c=2s/c=1.5$ 
and a rounded wing-tip geometry described by a semi-circle centered at $(y',z')/c=(0,0.75)$ 
with a radius equal to the profile thickness at that $x'$ location (see figure~\ref{fig:wing-geometry}).

The wings are located such that their mid-chord, $(x',y',z')/c=(0.5,0,0)$, 
coincides with the origin of the global Cartesian coordinate system, $(x,y,z)_O=(0,0,0)$, 
and have a no-slip, no-penetration boundary condition. 
The computational domain has a rectangular cross-section in the $z$-normal plane that extends $20c$ upstream, $30c$ downstream, and $20c$ in positive and negative $y$ directions.
Different angles of attack are achieved by rotating the wing along the $z$ axis around its center $(x',y')/c=(0.5,0)$, located at $(x,y)=(0,0)$, 
without changing the inflow boundary condition. 
This specific design is to allow for the use of the ``outflow-normal'' boundary condition~\citep[cf.][]{deville:book} on $y$-normal boundaries, which allows for a non-zero $y$ component of velocity (inward at $y=20c$ and outward at $y=-20c$ in lift-generating configurations). The inflow boundary at $x=-20c$ has a Dirichlet boundary condition with 
$(u_1,u_2,u_3)=(U_\infty,0,0)=\vec{\boldsymbol{U}}_\infty$, 
while the outlet boundary at $x=30c$ has the outflow boundary condition.
In order to avoid backflow at the outlet there is a sponge region for all $x \geq 10c$ with a gradually increasing forcing term that acts to bring the velocity to its free-stream condition. The effect of the sponge region on the flow has been tested in a number of preliminary runs 
and deemed negligible %for $x/c \leq 9$.
for all quantities of interest here.

The finite-span wing domain extends $20c$ in the spanwise direction $z$ from the wing root (located at $z=0$) 
and has an ``outflow-normal'' boundary condition at $z=20c$. 
A symmetry boundary condition is used at the $z=0$ plane (\ie, the wing root).
While the symmetry condition forces the $z$ component of the velocity to be zero at the wing root (and thus, for instance, the corresponding Reynolds stress components), 
the impact of this boundary condition on turbulence quantities appears to be negligible for $z/c \geq 0.05$; 
therefore, simulation of full-span wings was not deemed necessary.
The infinite-span (periodic) wings share the same domain design and boundary conditions in the $xy$-plane but have a spanwise width of $L_z=0.6c$ with periodic boundary conditions in $z$.
The wider $L_z$ in these simulations 
\citep[compared to previous works, e.g.,][]{hosseini:wing,negi:wing,vinuesa:wing:18}
is to allow for a more accurate simulation of the wake region 
(not discussed here).

The boundary layers are tripped on both the suction and pressure sides in all seven cases. 
The implemented trip is a randomized time-dependent wall-normal force 
proposed by~\cite{schlatter:trip}~\citep[also see][]{hosseini:wing}
aimed at minimizing the required development length and history effects from the trip.
The finite-span wings are only tripped on the main section of the wing ($z<0.75c$), and not on the wing tip region. 
On the suction side, the tripping is always located at $x'=0.1c$ regardless of the angle of attack, 
while on the pressure side it is at $x'=0.1c$ for the $0^\circ$, $2^\circ$ and $5^\circ$ angles of attack and at $x'=0.25c$ for $\alpha=10^\circ$.
This change in the tripping location on the pressure side of the periodic wing at $\alpha=10^\circ$ was necessary
since the acceleration parameter $K=\left(\nu/U_e^2\right){\rm d}U_e/{\rm d}x$ (where $U_e$ is the velocity at the boundary layer edge) far exceeded the rule-of-thumb value of $2.5-3\times10^{-6}$ for relaminarization~\citep{spalart:86,yuan:15} and the added energy in the tripping region was dissipated before generating turbulence. 
While this relaminarization was only present in the periodic case, 
an early decision was made to match the location of the trip for finite- and infinite-span wings
at the same geometric angle of attack $\alpha$.
%to make them more directly comparable. 
As will be seen in section~\ref{sec:BL}, this was not an optimal choice;
however, since it did not have an impact on the quantities studied here and
due to the excessive cost of the simulations, this was not modified later.

\subsection{Computational grids}

The production grids used in this study are generated by iteratively adapting (\ie, refining) an initial grid using the solution-based spectral error indicator introduced by~\cite{mavriplis:90} for turbulent flows. 
From a mathematical point of view, this error indicator 
is an approximation of the interpolation error
in the numerically-obtained velocity field
compared to its estimated exact counterpart 
(in an $L^2$-norm sense),
computed by estimating the truncation and quadrature errors~\citep{offerman:amr, tanarro:amr}.
From a physical perspective, this error indicator estimates the sum of the small-scale and unresolved turbulent kinetic energy~\citep{toosi:17}, 
and corresponds to the numerical, modeling (from the LES model), 
and projection errors~\citep{toosi:thesis}.

The initial grid for the periodic cases is conformal and originally two-dimensional, which is then 
extruded (\ie, copied to an appropriate number of spanwise locations) 
to make sure that the spanwise homogeneity of the mesh is maintained. Different angles of attack share the same near-wing mesh (up to a few boundary layer thicknesses), which is rotated with the wing. 
The root of the finite-span wing shares a nearly-identical mesh with their periodic counterpart.

The production grids are generated by iterative refinement of the initial grids, where at each iteration elements with the highest contribution to solution error,
identified by 
the volume-weighted error indicator~\citep[cf.][]{lapenta:03,park:03,toosi:caf:20},
are selected for refinement. 
The convergence process is accelerated by some manual input from the user; for instance, by manually marking the wall elements for refinement.
The initial grids are designed to reach their desired wall resolution after four refinements of the near-wall elements. 
The automatic refinement is continued for a few iterations (where in the first four the wall elements are manually marked for refinement), 
and the refinement regions are then manually extended for three more iterations (\ie, each element is refined if any of its neighboring elements is refined). 
This last step helps to avoid repeating the refinement process further (a process which becomes expensive for these progressively finer grids) and leads to a smoother and more uniform mesh. 
The adaptation process is terminated after reaching the resolution criteria from 
literature~\citep[such as those used in][]{vinuesa:wing:17,vinuesa:wing:18},
expressed in terms of the viscous length scale,
$\delta_\nu=\nu\sqrt{\rho/\tau_w}$
(where $\tau_w$ is the wall shear stress)
for the boundary layer mesh,
and the Kolmogorov length scale
$\eta=(\nu^3/\epsilon)^{1/4}$ 
(where $\epsilon$ is the local isotropic dissipation rate)
for the wake region~\citep[cf.][]{pope:00}.

Table~\ref{table:grids} summarizes some of the characteristics of the production grids used in this work. 
%As can be seen from the table,
%all grids have resolutions that are only slightly coarser than 
%a full direct numerical simulation (DNS).
Note that the finite-span wings require significantly larger numbers of grid points 
compared to the periodic cases.
This is partly to resolve the tip region and the larger span of these wings,
and partly because of the decision to perform these simulations at slightly higher resolutions
due to the potential insufficiency of the resolution criteria originally verified for canonical flows.
For similar reasons, wings at higher angles of attack require higher numbers of grid points to accurately 
resolve all the important features of their more complex flow field
with stronger vortices and secondary flows.

\begin{table}
  \begin{center}
\def~{\hphantom{0}}
\begin{tabular}{lccccccc}
%  \begin{tabular}{lcccccccc}
%      Case  & $\alpha$  & $N_{\rm GLL}$ & 
%      $\Delta t U_\infty/c$ &
%      $(\Delta x^+_{\rm BL},\delta_1 y^+_{\rm BL},\Delta z^+_{\rm BL})$ & 
%      $(\Delta x_{\rm tip},\Delta{y_{\rm tip}},\Delta z_{\rm tip})/c$ & 
%      $(\Delta x_{\rm wake},\Delta y_{\rm wake},\Delta z_{\rm wake})/\eta$ \\[3pt]
%      P-0    & $0^\circ$ 	& $376\times10^6$ 	& $15 \times 10^{-6}$ & (10.3,0.72,8.7) & $-$ & (5.8,3.6,3.7)  \\
%      P-2    & $2^\circ$ 	& $383\times10^6$ 	& $18 \times 10^{-6}$ & (10.3,0.71,8.7) & $-$ & (5.8,3.5,3.8)  \\
%      P-5    & $5^\circ$ 	& $376\times10^6$ 	& $3.0 \times 10^{-6}$ & (10.5,0.73,8.5) & $-$ & (5.2,3.2,3.3)  \\
%      P-10    & $10^\circ$ 	& $438 \times10^6$	& $7.9 \times 10^{-6}$ & (12.0,0.80,9.0) & $-$ & (6.1,3.8,4.0)  \\
%      RWT-0  & $0^\circ$  	& $952\times10^6$ 	& $16 \times 10^{-6}$ & (11.4,0.70,6.9)  & $(11,1.7,2.2)\times10^{-4}$ & (5.5,2.6,1.7)  \\
%      RWT-5  & $5^\circ$ 	& $1.56\times10^9$ 	& $8.5 \times 10^{-6}$ & (10.6,0.75,6.0)  & $(11,1.7,2.2)\times10^{-4}$ & (5.7,2.8,1.7)  \\
%      RWT-10 & $10^\circ$ 	& $2.16\times10^9$ 	& $6.3 \times 10^{-6}$ & (10.6,0.80,5.4)  & $(11,1.7,2.2)\times10^{-4}$ & (5.6,3.0,1.8)        
       Case  & $N_{\rm GLL}$ & 
      $\Delta t U_\infty/c$ &
      $(\Delta x^+_{\rm BL},\delta_1 y^+_{\rm BL},\Delta z^+_{\rm BL})$ & 
      $(\Delta x_{\rm tip},\Delta{y_{\rm tip}},\Delta z_{\rm tip})/c$ & 
      $(\Delta x_{\rm wake},\Delta y_{\rm wake},\Delta z_{\rm wake})/\eta$ \\[3pt]
      P-0    		& $376\times10^6$ 	& $15 \times 10^{-6}$ & (10.3, 0.72, 8.7) & $-$ & (5.8, 3.6, 3.7)  \\
      P-2    		& $383\times10^6$ 	& $18 \times 10^{-6}$ & (10.3, 0.71, 8.7) & $-$ & (5.8, 3.5, 3.8)  \\
      P-5    		& $376\times10^6$ 	& $3.0 \times 10^{-6}$ & (10.5, 0.73, 8.5) & $-$ & (5.2, 3.2, 3.3)  \\
      P-10    	& $438 \times10^6$	& $7.9 \times 10^{-6}$ & (12.0, 0.80, 9.0) & $-$ & (6.1, 3.8, 4.0)  \\
      RWT-0  	& $952\times10^6$ 	& $16 \times 10^{-6}$ & \st{(11.4, 0.83, 6.4)}  & $(11,1.7,2.2)\times10^{-4}$ & (5.5, 2.6, 1.7)  \\
      RWT-5  	& $1.56\times10^9$ 	& $8.5 \times 10^{-6}$ & \st{(10.6, 0.77, 5.9)}  & $(11,1.7,2.2)\times10^{-4}$ & \st{(5.6, 2.8, 1.8)}  \\
      RWT-10 	& $2.16\times10^9$ 	& $6.3 \times 10^{-6}$ & \st{(9.9, 0.72, 5.5)}  & $(11,1.7,2.2)\times10^{-4}$ & \st{(5.6, 3.0, 1.9)}
  \end{tabular}
  \caption{\label{table:grids}
  Description of the production grids used in this study. 
  The naming convention distinguishes the different setups
  by P-$\alpha$ for periodic wings at an angle of attack of $\alpha$ 
  and RWT-$\alpha$ for finite-span wings at $\alpha^\circ$ angle of attach.
  $N_{\rm GLL}$ is the total number of GLL points
  (the number of independent grid points is around $0.67 N_{\rm GLL}$).
  All $\Delta$'s are based on the mean resolution computed as the element size divided by the polynomial order, 
  whereas $\delta_1 y_{\rm BL}$ is the distance from the wall of the first GLL point off the wall.
  Wall resolutions, including both $\Delta \ast_{\rm BL}$ and $\Delta \ast_{\rm tip}$,
  are reported along the wall coordinates $(x_{\rm BL},y_{\rm BL},z_{\rm BL})$ (figure~\ref{fig:wing-geometry}).
  The boundary layer resolutions  $\Delta \ast_{\rm BL}$
  are normalized by the viscous length
  $\delta_\nu$ and reported at $(x',z')/c=(0.7,0.3)$ for the element on the wall. 
  Tip resolutions $\Delta \ast_{\rm tip}$ are normalized by the chord $c$ and reported at $x'/c=0.7$ and $z'=z'_{\rm max}$.
  Wake resolutions $\Delta \ast_{\rm wake}$ are 
  reported at $(x,z)/c=(2,0.3)$
  at the location of minimum mean velocity.
  }
  \end{center}
\end{table}

\begin{figure}
  \centering
    \includegraphics[width=100mm,clip=true,trim=1mm 1mm 1mm 1mm]{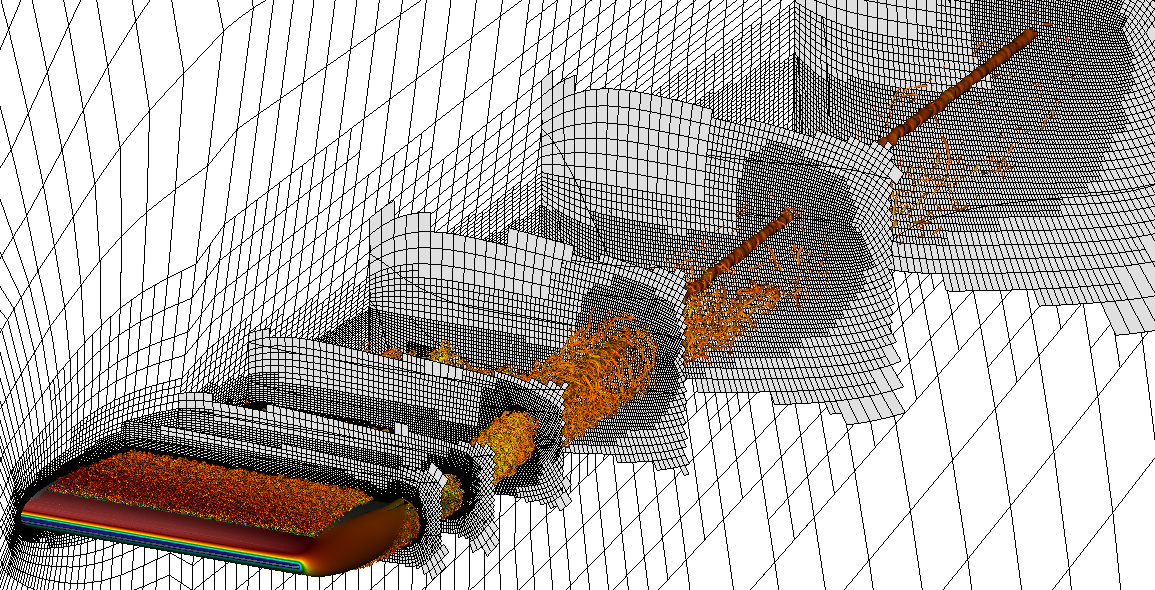}
  \caption{
  \label{fig:a10rwt_elements}
  An example of the grids used in this study.
  Edges of the spectral elements of RWT-10 
  (with 4.23 million spectral elements; see table~\ref{table:grids}) 
  are shown by black lines on a number of planar slices. 
  The grid is generated using the $h$-adaptation capabilities of the AMR version of Nek5000.
  The figure also shows instantaneous vortical structures represented by isosurfaces of $\lambda_2 c^2/U_\infty^2=-100$ %~\citep[see][]{jeong:95} 
  colored by streamwise velocity ranging from low (blue) to high (red)
  for visual reference. 
  }
\end{figure}

Figure~\ref{fig:a10rwt_elements} 
shows the spectral elements of the RWT-10 grid,
as a representative example of the grids used in this study,
with instantaneous vortical structures of the flow visualized using the 
$\lambda_2$ vortex identification method~\citep{jeong:95} added for visual reference.

\subsection{Flow transients and statistical averaging}

Flow transients are removed both during the grid-adaptation stage and after reaching the production grid. 
In total, %we discard 
a minimum of approximately $80c/U_\infty$ (equivalent to 80 convective time units, $\mathcal{T}_{\rm conv}=c/U_\infty$) 
of integration time is discarded
as transients. 
This comprises $50\mathcal{T}_{\rm conv}$ on the coarse initial grid, 
roughly $25\mathcal{T}_{\rm conv}$ on various adapted grids before reaching the production grid, 
and an additional $4\mathcal{T}_{\rm conv}$ on the production grid. 
%Due to cost constraints, RWT-10 has a shorter transient period on the production grid, 
%which is partially compensated for by a longer integration time
%on finer adapted grids which are only slightly coarser than the final grid.

After the transient period, 
turbulence statistics are collected over a period of
$5\mathcal{T}_{\rm conv}$ or above in the periodic cases (P-0, P-2, P-5, P-10),
around $8 \mathcal{T}_{\rm conv}$ for RWT-0, 
\st{around $14 \mathcal{T}_{\rm conv}$ for RWT-10,
and for a longer period of around $23 \mathcal{T}_{\rm conv}$ in RWT-5.}
Table~\ref{table:trans} summarizes the averaging time $t_{\rm avg}$ used in each simulation
in terms of the convective time unit. 
A conversion ratio is provided to relate $\mathcal{T}_{\rm conv}$ to other time scales of the flow, including:
the boundary layer eddy-turnover time $\mathcal{T}_{\rm ETT}=\delta_{99}/u_\tau$, 
where $\delta_{99}$ is the $99\%$ boundary layer thickness and $u_\tau=\sqrt{\tau_w/\rho}$ is the friction velocity;
the shear-layer (wake) time scale $\mathcal{T}_{\rm shear}=\delta_{\rm 0.5, shear}/U_\infty$,
where $\delta_{\rm 0.5, shear}$ is a measure of the wake thickness~\citep[see][]{pope:00};
the vortex-rotation time scale  $\mathcal{T}_{\rm vort}=\pi d_{\rm vort}/u_{\theta,{\rm max}}$,
where $u_{\theta,{\rm max}}$ is the maximum azimuthal velocity around the vortex core
and $d_{\rm vort}$ is the vortex diameter
measured as the distance between the two peaks in the azimuthal velocity around the core.
The RWT-0 (which is symmetric \st{around the $y=0$ plane} and can be averaged in that direction) 
and RWT-5 have similar effective values of $t_{\rm avg}/\mathcal{T}_{\rm ETT,ss}$
(which is the most relevant time scale for the quantities studied here), 
while RWT-10 has a shorter integration time due to computational-cost constraints.

\begin{table}
  \begin{center}
\def~{\hphantom{0}}
  \begin{tabular}{lccccccc}
%      Case  & 
%      $t_{\rm start}/\mathcal{T}_{\rm conv}$  & 
%      $t_{\rm tran}/\mathcal{T}_{\rm conv}$  & 
%      $t_{\rm avg}/\mathcal{T}_{\rm conv}$  & 
%      $\mathcal{T}_{\rm conv}/\mathcal{T}_{\rm ETT,ss}$ & 
%      $\mathcal{T}_{\rm conv}/\mathcal{T}_{\rm ETT,ps}$ & 
%      $\mathcal{T}_{\rm conv}/\mathcal{T}_{\rm shear}$ & 
%      $\mathcal{T}_{\rm conv}/\mathcal{T}_{\rm vort}$ 
%       \\[3pt]
%      P-0    		& 78.5 & 6.3 & 6.1 & 2.4 & 2.4 & 16.4 & $-$  \\
%      P-2    		& 110 & 4.0 & 4.9 & 2.1 & 2.7 & 16.4 & $-$  \\
%      P-5    		& 74.8  & 4.3 & 5.8 & 1.6 & 2.9 & 13.0 & $-$  \\
%      P-10   		& 86.8 & 4.4 & 5.7 & 0.54 & 3.7 & 9.0 & $-$  \\
%      RWT-0 	& 73.8 & 4.4 & 8.2 & 2.5 & 2.5 & 14.1 & $-$  \\
%      RWT-5  	& 71.4 & 6.2 & 7.6 & 1.9 & 3.0 & 14.1 & 2.3 \\
%      RWT-10 	& 84.7 & 2.5 & 3.1 & 1.4 & 3.2 & 13.7 & 2.6 \\
      Case  & 
      $t_{\rm avg}/\mathcal{T}_{\rm conv}$  & 
      $\mathcal{T}_{\rm conv}/\mathcal{T}_{\rm ETT,ss}$ & 
      $\mathcal{T}_{\rm conv}/\mathcal{T}_{\rm ETT,ps}$ & 
      $\mathcal{T}_{\rm conv}/\mathcal{T}_{\rm shear}$ & 
      $\mathcal{T}_{\rm conv}/\mathcal{T}_{\rm vort}$ 
       \\[3pt]
      P-0    		& 6.1 & 2.4 & 2.4 & 16.4 & $-$  \\
      P-2    		& 4.9 & 2.1 & 2.7 & 16.4 & $-$  \\
      P-5    		& 5.8 & 1.6 & 2.9 & 13.0 & $-$  \\
      P-10   		& 5.7 & 0.54 & 3.7 & 9.0 & $-$  \\
      RWT-0 	& 8.2 & 2.5 & 2.5 & \st{14.0} & $-$  \\
      RWT-5  	& 23.2 & 1.9 & 3.0 & \st{14.3} & 2.3 \\
      RWT-10 	& \st{14.1} & 1.4 & 3.2 & \st{14.3} & 2.6 \\

  \end{tabular}
  \caption{
  \label{table:trans}
  A summary of the averaging time $t_{\rm avg}$ used in each case compared to different time scales in the flow. 
  $\mathcal{T}_{\rm conv}$ is the convective (flow-over) time scale,
  $\mathcal{T}_{\rm ETT}$ is the boundary layer eddy turnover time reported at $(x',z')/c=(0.7,0.3)$,
  $\mathcal{T}_{\rm shear}$ is the shear time scale in the wake reported at $(x,z)/c=(2.0,0.3)$,
  and
  $\mathcal{T}_{\rm vort}$ is the time scale of vortex rotation reported at $x/c=2.0$.
  Subscripts $\cdot_{\rm ss}$ and $\cdot_{\rm ps}$ stand for the suction side and pressure side of the wing.
  Definition of time scales is given in the text.
  }
  \end{center}
\end{table}

The statistics are collected on the fly 
(see appendix~\ref{app:averaging} for more details) 
at a sampling rate that is around an order of magnitude higher than the 
highest frequency of the flow, 
here dictated by the viscous time scale $\nu/u_\tau^2$
($\geq 2 \times10^{-3} \mathcal{T}_{\rm conv}$ for cases studied here).
The periodic cases are homogeneous in the spanwise direction
and therefore an ensemble average in that direction is also performed
when computing the statistics. 
While RWT-0, RWT-5, and RWT-10 are fully three-dimensional flows
and exhibit a variation of solution statistics along the span,
these variations are 
smooth over the turbulent boundary layer section of the wings. 
This allows for a spanwise filtering of the statistics 
using a (wide) Gaussian filter with a variable filter width that is adjusted based on flow physics
and resembles an averaging process.
This procedure is explained in more detail in appendix~\ref{app:averaging}.

The temporally and spatially averaged (or filtered) fields are denoted by $\avg{\cdot}$;
e.g., $\avg{u_i}$ and $\avg{p}$ for the mean velocity and mean pressure fields.
The fluctuating part is then defined as the (pointwise) difference between the
instantaneous and mean values; 
e.g., $u_i'=u_i-\avg{u_i}$ and $p'=p-\avg{p}$ 
for the fluctuating velocity and pressure fields.

In each case, the full statistical record is divided into four \st{or more} batches of equal size,
which are used to estimate the uncertainty in solution statistics by computing
the confidence intervals of each quantity of interest using the 
non-overlapping batch method~\citep[cf.][]{conway:63}.
Given the higher sensitivity of the Reynolds stresses, particularly for finite-span wings, 
their approximate error bars are included
in the comparisons in section~\ref{sec:3D-vs-2D}
and appendix~\ref{app:sec-BL-params}.
With the use of ensemble averaging along the span in periodic cases
and the equivalent filtering in the wing-tip cases,
the averaging times used in this work are deemed sufficient 
for the discussions here.

\section{Flow around the finite-span wings \label{sec:flowfield}}

This section describes the important features of
the flow around the finite-span wings of this study
%%with a focus 
%focusing
%on features that are %most relevant 
%important
relevant 
to the discussions in section~\ref{sec:BL}
and, to a certain degree, serves as an introduction to that section.

\subsection{The instantaneous flow field}

\begin{figure}
  \centering
    \includegraphics[width=130mm,clip=true,trim=1mm 200mm 1mm 0mm]{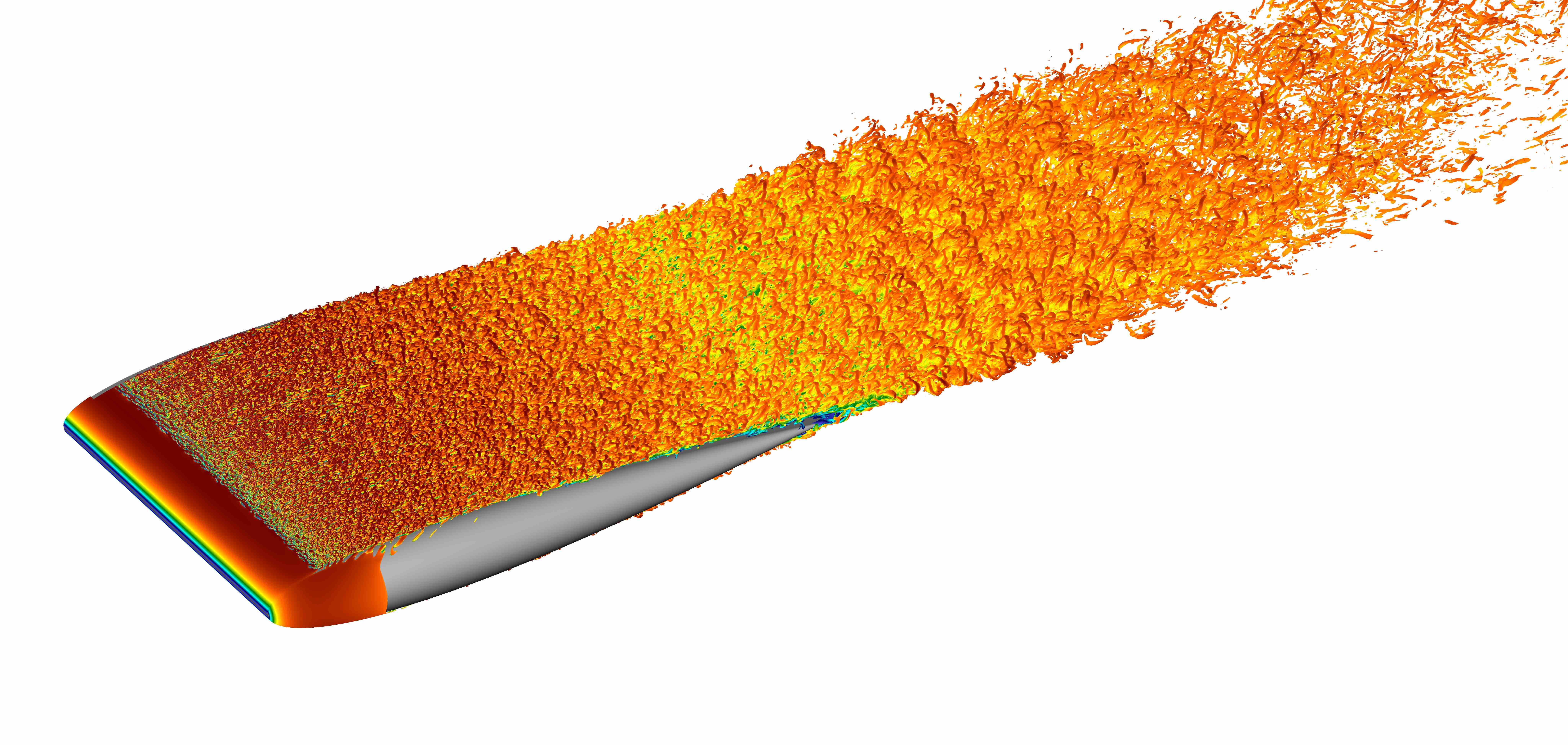}\llap{\parbox[b]{125mm}{(a)\\\rule{0ex}{10mm}}}
    \includegraphics[width=130mm,clip=true,trim=1mm 250mm 1mm 100mm]{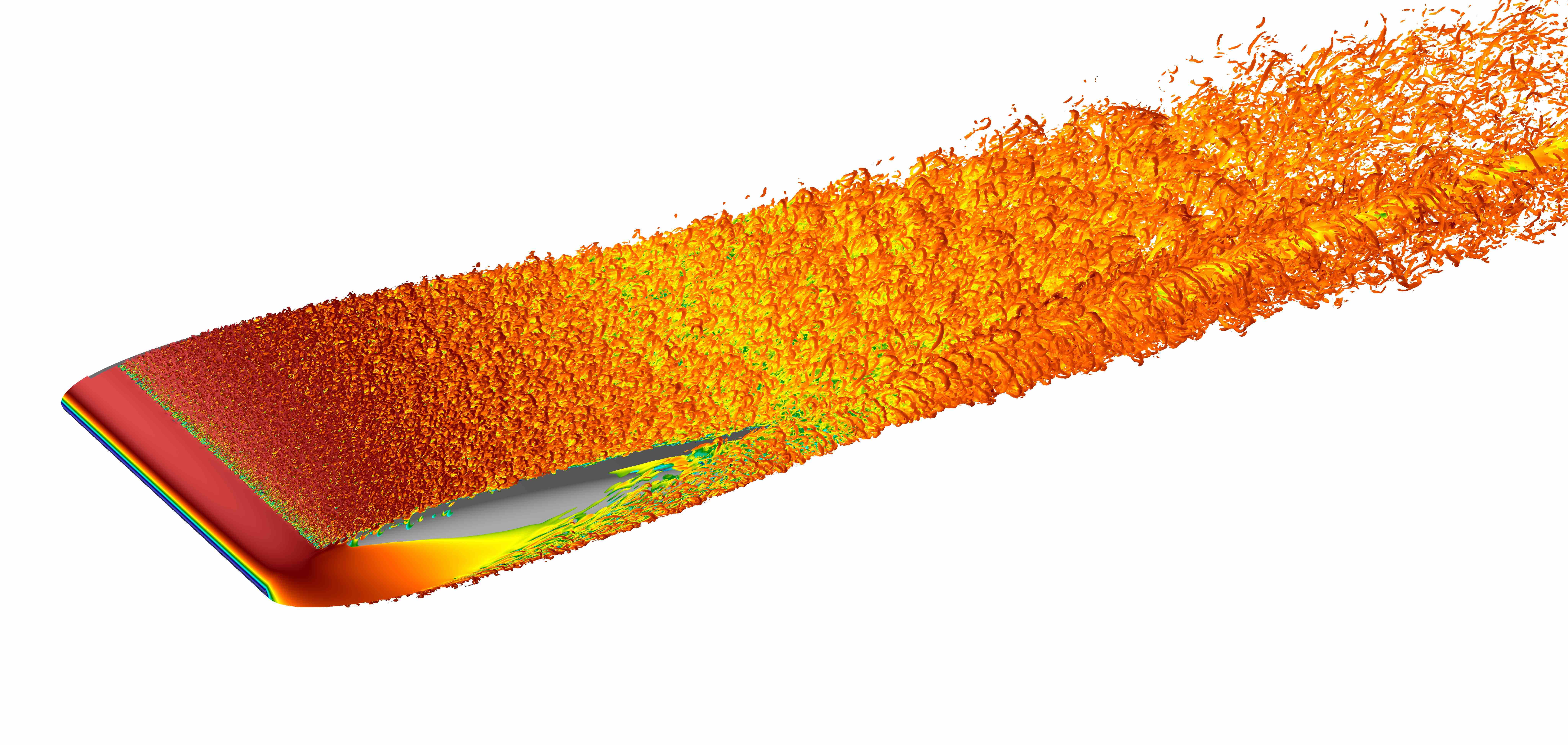}\llap{\parbox[b]{125mm}{(b)\\\rule{0ex}{10mm}}}
    \includegraphics[width=130mm,clip=true,trim=1mm 280mm 1mm 150mm]{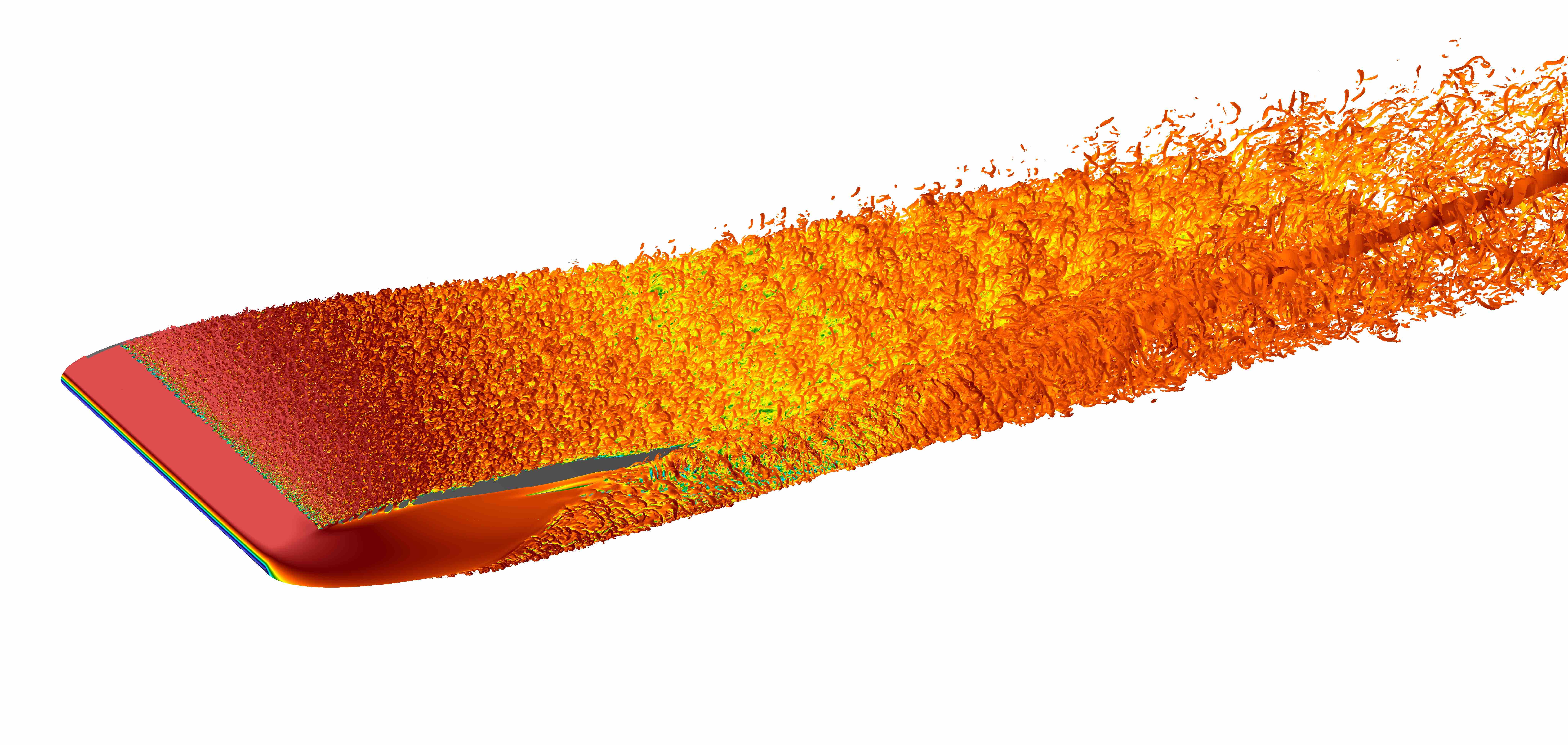}\llap{\parbox[b]{125mm}{(c)\\\rule{0ex}{10mm}}}
  \caption{
  \label{fig:GFF-viz}
  An overview of the flow around the finite-span wings: (a) RWT-0, (b) RWT-5, and (c) RWT-10. 
  Plots show the instantaneous vortical structures, visualized by iso-surfaces of 
  $\lambda_2 c^2/U_\infty^2=-100$ colored by $u_1/U_\infty$ from -0.3 (blue) to 1.3 (red). 
  The wing surface is shown in light grey. 
  }
\end{figure}

The flow around RWT-0, RWT-5 and RWT-10 is illustrated 
in figure~\ref{fig:GFF-viz} 
by the instantaneous vortical structures of the flow
using the $\lambda_2$ visualization method 
of~\cite{jeong:95}.
The figure visualizes the turbulent boundary layers formed on the wings,
the turbulent wake,
and the wing-tip vortex 
identified as a cylindrical structure 
(surrounded by turbulent structures)
that separates from the wing
somewhere close to the tip 
and remains coherent for a long distance downstream of the wing
(the entire field of view in figure~\ref{fig:GFF-viz}).
Note that the wing-tip vortex is only present in lift-generating configurations
and absent in RWT-0.
As expected, RWT-10 has a stronger wing-tip vortex 
(leading to a larger vortex diameter for a fixed value of $\lambda_2$)
which impacts a larger portion of the wing, and more strongly, 
compared to RWT-5. 
%It is also interesting 
While not directly relevant to the discussions of section~\ref{sec:BL}, it is interesting
to note that the vortex core in RWT-10 has a higher streamwise velocity compared to RWT-5,
%This has been 
%associated in the literature to the stronger wing-tip vortex, 
%%which leads 
%leading to lower core pressures and thus flow entrainment into the core region~\citep[cf.][]{Lee:10}.
which is associated in the literature to the lower core pressure of the stronger wing-tip vortex
and thus flow entrainment into the core region~\citep[cf.][]{Lee:10}.

% these figures were plotted at 8192 horizontal pixels and fixed aspect ratio on the 27-monitor in the office

The location of tripping and its effectiveness is visually clear in figure~\ref{fig:GFF-viz}
by the absence of turbulent structures upstream of the trip 
and the presence of hairpin vortices 
(which the trip introduces through wall-normal forcing)
downstream of the tripping line. 
The relatively low friction Reynolds number of the flow 
($200 \lesssim Re_\tau \lesssim 300$, 
see tables~\ref{tab:app-rwt0-vs-p0},~\ref{tab:app-rwt5-vs-p2}, and~\ref{tab:app-rwt10-vs-p5} in appendix~\ref{app:sec-BL-params}) 
is apparent from the hairpin-dominated structure of the turbulent boundary layers~\citep[cf.][]{eitel-amor:15}.
Note that
on both the suction and pressure sides 
the tripping line extends throughout the span of the wing 
from the root to the tip, 
but does no include the wing tip region (the semi-circle shown in figure~\ref{fig:wing-geometry}). 
This results in a laminar flow around the wing tip and
a laminar/turbulent interface at the spanwise end
of the tripping line.

The figure shows a
strong flow convection from
the pressure side to the suction side
(visualized by the convected turbulent structures
from the pressure side), 
starting from the leading edge of the wing.
This underscores the
global impact of the finite span of the wing and the wing-tip vortices 
on the entire flow field. 
RWT-10 generates a higher lift and thus
has stronger pressure gradients, 
leading to a stronger flow convection,
stronger wing-tip vortices, and stronger three-dimensionality in the flow field.

It is also important to note the appearance of a non-turbulent area on the suction side 
of RWT-5 and RWT-10 in a region close to the wing-tip vortex.
This might initially suggest a relaminarization of the turbulent boundary layer
due to the increased 
pressure gradient, rotation and flow acceleration
in the vicinity of the wing-tip vortex.
However, upon further investigation this was proved not to be the case,
since the turbulent region of the boundary layer was almost exactly following the
streamlines of the flow released at the edge of the boundary layer 
near the spanwise end of the tripping line.
In other words, the appearance of this laminar region should be primarily attributed to
the non-zero spanwise velocity towards the root
that convects the spanwise laminar/turbulent interface of the boundary layer %towards the root
in that direction
and causes a laminar region to appear near the wing tip region.
%We should emphasize that 
Nevertheless,
the present behavior 
does not mean that extending the tripping line
to include the tip region will necessarily lead to a fully turbulent suction side, 
since the strong flow acceleration, %around the tip
rotation and pressure gradients might still be sufficiently high to result in relaminarization.

\subsection{Mean-flow streamlines}

\begin{figure}
  \centering
 \begin{overpic}[angle=0,origin=c,width=44mm,clip=true,trim=40mm 30mm 50mm 30mm]{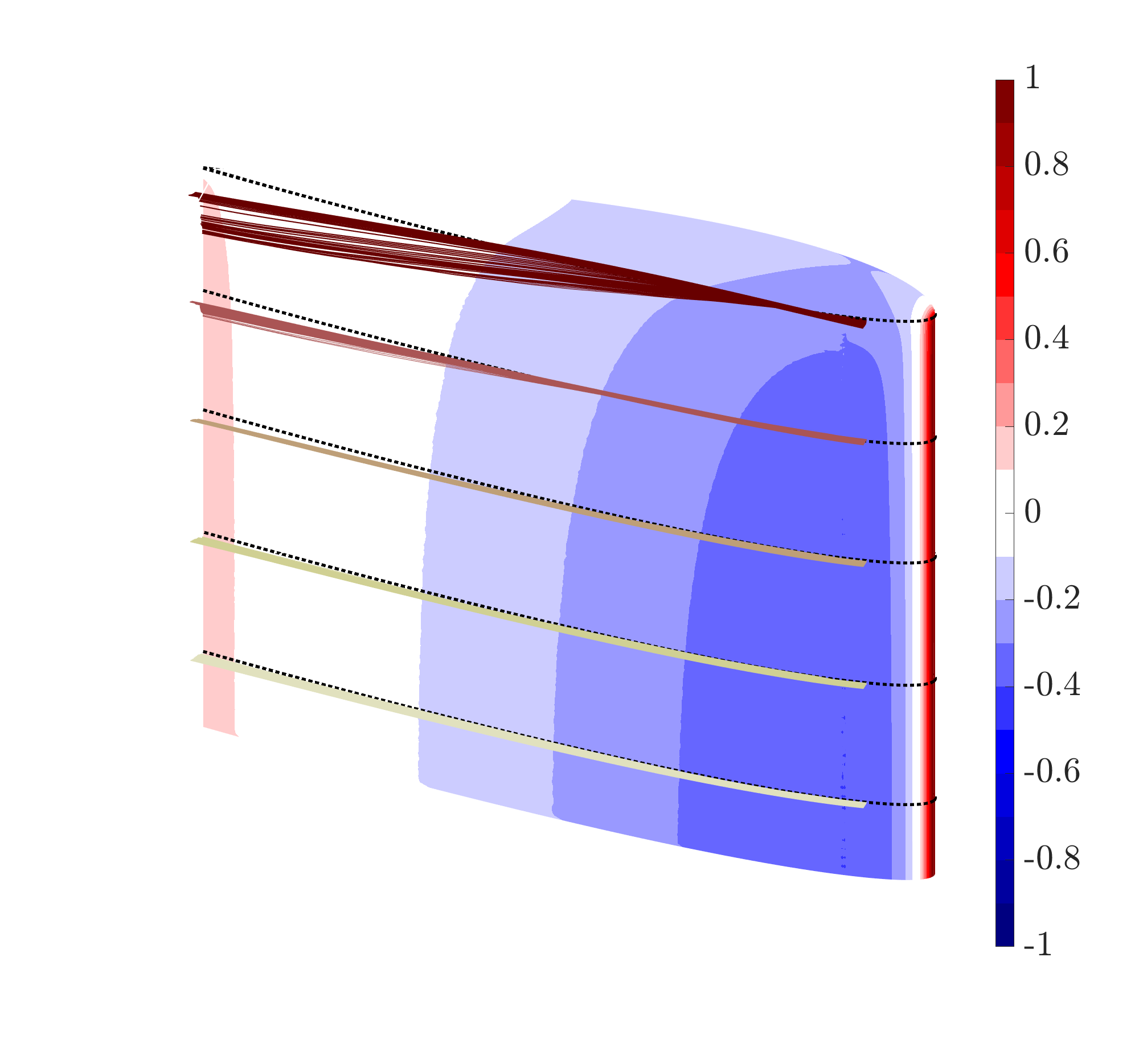}
  \put(60,7){\vector(-4,1){30}} 
  \put(40,3){$\vec{\boldsymbol{U}}_\infty$}
  \put(10,17.5){(a)}
  \end{overpic}
  \begin{overpic}[angle=0,origin=c,width=44mm,clip=true,trim=40mm 30mm 50mm 30mm]{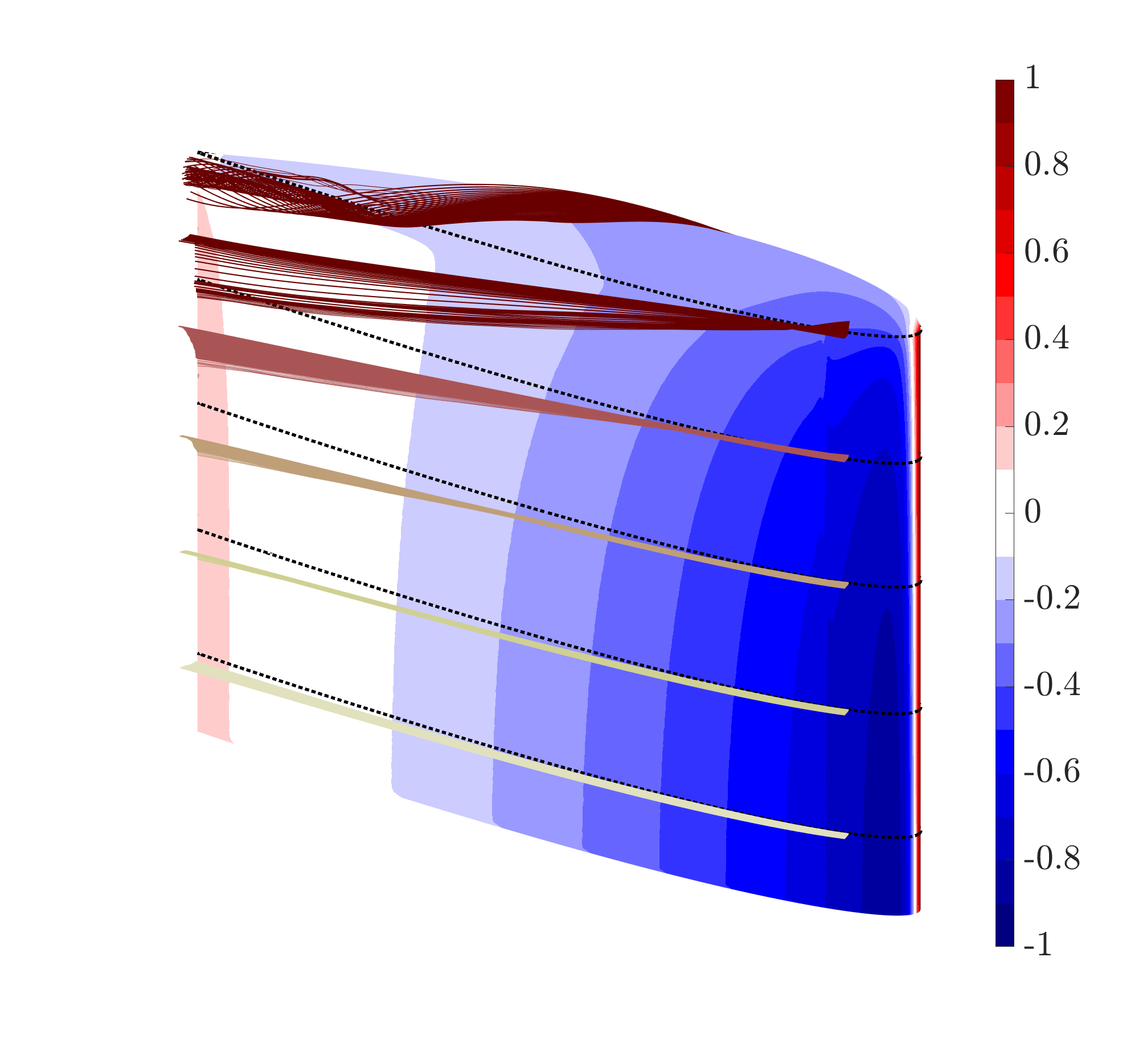}
  \put(60,7){\vector(-4,1){30}} 
  \put(40,3){$\vec{\boldsymbol{U}}_\infty$}
  \put(10,17.5){(b)}
  \end{overpic}
  \begin{overpic}[angle=0,origin=c,width=44mm,clip=true,trim=40mm 30mm 50mm 30mm]{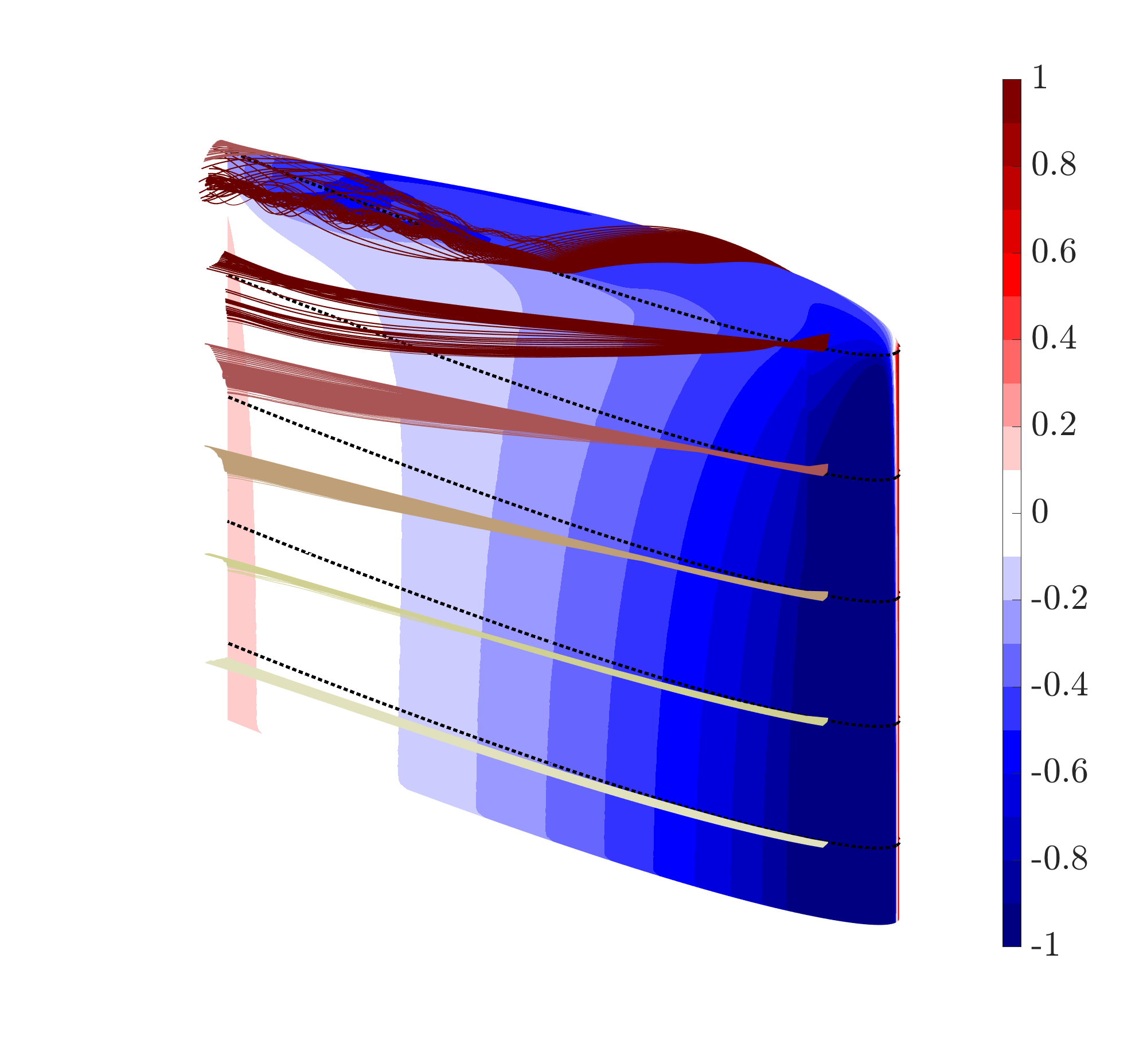}
  \put(60,7){\vector(-4,1){30}} 
  \put(40,3){$\vec{\boldsymbol{U}}_\infty$}
  \put(10,17.5){(c)}
  \end{overpic}\\
\begin{overpic}[angle=0,origin=c,width=44mm,clip=true,trim=40mm 1mm 50mm 30mm]{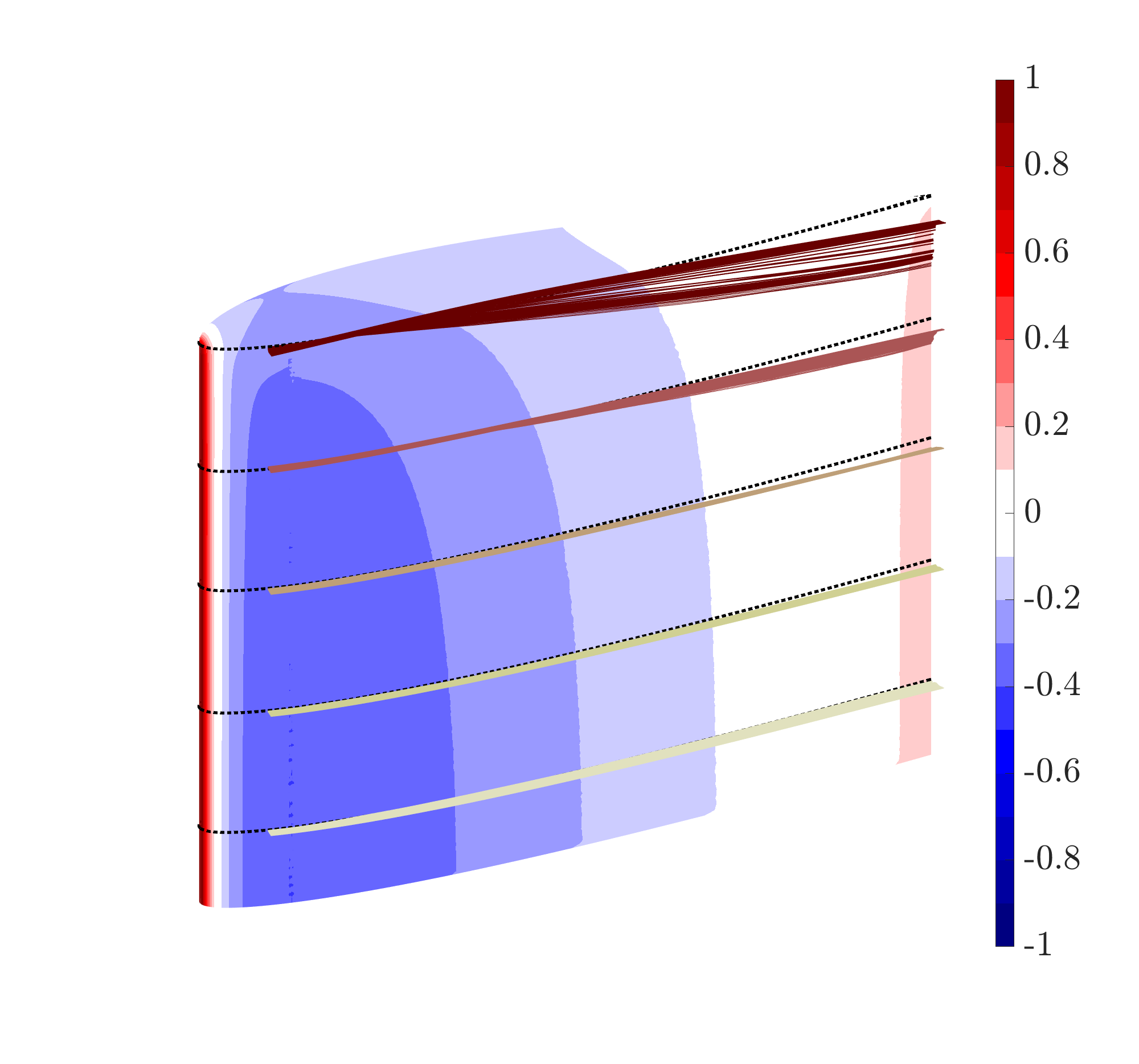}
  \put(30,10){\vector(4,1){30}} 
  \put(43,6.5){$\vec{\boldsymbol{U}}_\infty$}
  \put(5,10){(d)}
  \end{overpic}
  \begin{overpic}[angle=0,origin=c,width=44mm,clip=true,trim=40mm 1mm 50mm 30mm]{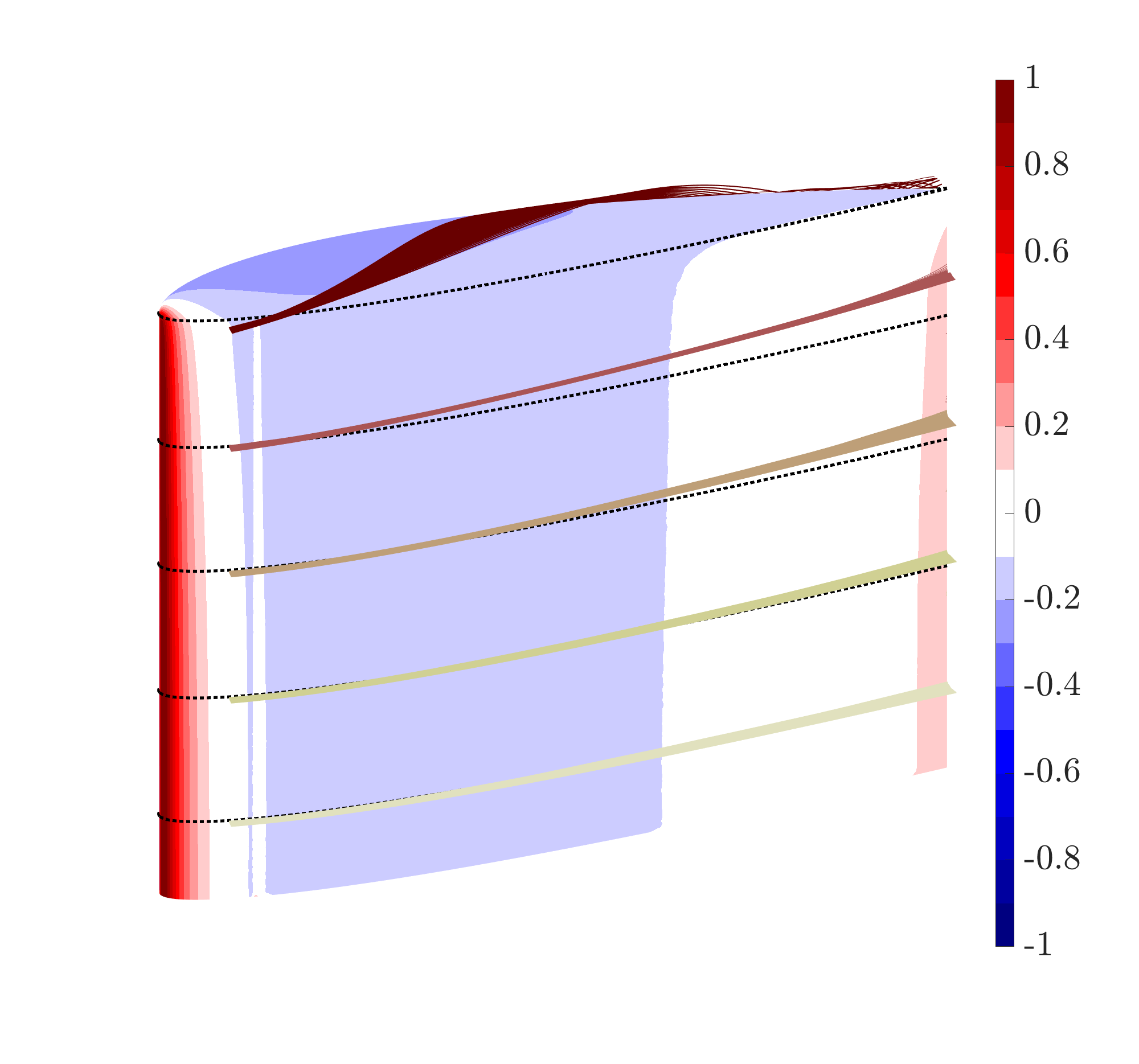}
  \put(30,10){\vector(4,1){30}} 
  \put(43,6.5){$\vec{\boldsymbol{U}}_\infty$}
  \put(5,10){(e)}
  \end{overpic}
  \begin{overpic}[angle=0,origin=c,width=44mm,clip=true,trim=40mm 1mm 50mm 30mm]{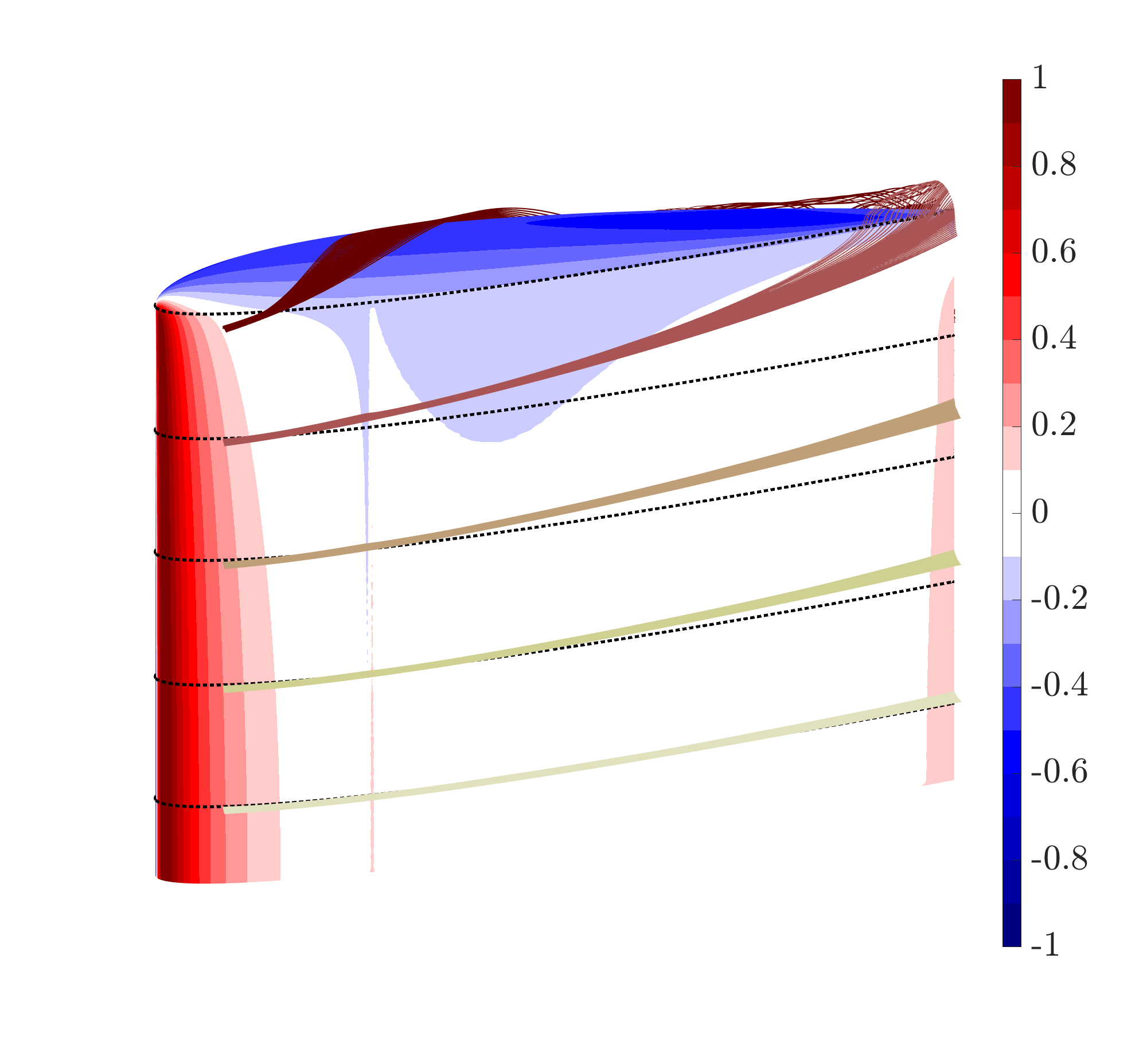}
  \put(30,10){\vector(4,1){30}} 
  \put(43,6.5){$\vec{\boldsymbol{U}}_\infty$}
  \put(5,10){(f)}
  \end{overpic}
  \caption{
  \label{fig:BL_streamline}
  Contour plots of the pressure coefficient $c_p$ on the wing surface for (a, d) RWT-0, (b, e) RWT-5 and (c, f) RWT-10 
  on the suction side (a, b, c) and pressure side (d, e, f) of the wings.
  Contour plots are overlaid with streamlines released at $x'/c=0.12$ at different $z'/c$ locations 
  (0.1, 0.26, 0.43, 0.59, 0.75).
  The contour levels are from -1 (dark blue) to 1 (dark red) in increments of 0.1. 
  The dotted black lines on the wing surface show the chord-wise direction 
  (\ie, constant $z'$, equivalent to the approaching free-stream direction) 
  at the same spanwise locations as the streamlines are released
  as a visual reference for their deflection. 
  Arrows indicate the approximate direction of the free-stream velocity $U_\infty$.
  }
\end{figure}
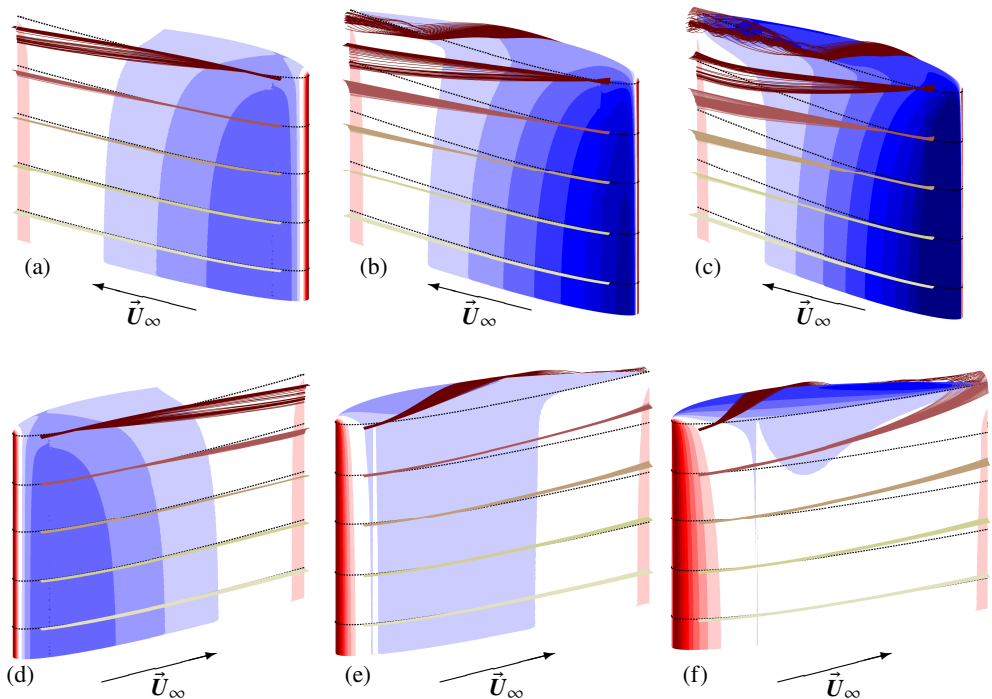

The extent of three-dimensionality of the flow
and its impact on the boundary layers
are shown in figure~\ref{fig:BL_streamline}
using the streamlines obtained from
the mean velocity field $\avg{u_i}$.
The large deflection angle of the streamlines %is quite clear 
can be easily observed
by comparison to the 
approaching free-stream direction (chord-wise lines).
As expected, this deflection increases for all cases closer to the tip. 
%We note that %, as discussed,
Note that
the deflection from the free-stream direction starts at
the beginning of the boundary layer development and
spans across a large region of the wing
(almost the entire semi-span for these low aspect ratio wings).
An interesting observation is that 
the streamlines of RWT-0 
(which does not have a wing-tip vortex) 
are still %noticeably 
impacted by the finite-span of the wing and
deflected towards the root,
albeit with a lower deflection angle compared to RWT-5 and RWT-10.
This is the reason for the appearance of a small laminar region 
close to the trailing edge of RWT-0 in figure~\ref{fig:GFF-viz} (a).
The other interesting observation is 
the large spanwise extent of the impacted boundary layer streamlines on the pressure side of RWT-5 and RWT-10
which is %quite 
comparable to their suction side.
On both the suction and pressure sides, the deflection angle approaches zero at the root of the wing as a result of the symmetry.

% Office monitor

The more %striking 
interesting
observation from figure~\ref{fig:BL_streamline} 
is that the deflection angle of streamlines varies %quite significantly
in the wall-normal direction
across the boundary layer thickness $\delta_{99}$. 
The streamlines closer to the wall 
have a larger deflection angle 
compared to those farther from the wall,
and this is
true on both the suction and pressure sides 
and at all spanwise locations. 
This is a common feature observed in many three-dimensional boundary 
layers~\citep[cf.][]{johnston:60,perry:65,pirece:83,olcmen:95,devenport:22}
which happens due to the lateral pressure gradient encountered by the boundary layer and
the variable balance between the different terms of the momentum equation,
such that
the fluid closer to the wall (which has a lower momentum) 
responds faster to the pressure gradient~\citep[cf.][and references therein]{devenport:22}.
The variable deflection angle of the streamlines
in the streamwise, wall-normal, and spanwise directions
has a number of consequences on the boundary layers which
will be discussed further in section~\ref{sec:BL}.

The wing-tip vortices are visible in RWT-5 and RWT-10 
in figures~\ref{fig:BL_streamline} (b) and (c)
as streamlines that are clustered together and
formed into spirals on the suction side of the wings.
Additional details about the flow in the vicinity of the tip
can be found in appendix~\ref{app:general-vortex}. 
The generated wing-tip vortex impacts the surrounding flow 
differently depending on the distance. 
This is primarily related to the velocity induced by the vortex 
(the Biot--Savart law),
which is proportional to the inverse of the distance from the vortex core 
(refer to the caption of figure~\ref{fig:intro-wingtip})
and approximately in the azimuthal direction of a cylindrical
coordinate system with the vortex core at its origin.
Due to the large variations in both magnitude and direction of the induced velocity, 
the flow in the vicinity of the vortex core is highly non-homogeneous.
The flow acceleration due to pressure gradient is also larger near the tip.
At larger spanwise distances from the vortex%
---and farther downstream from the location of its initial formation---%
the induced velocity on the wing surface
becomes a nearly uni-directional downwash 
with a variable magnitude
(and thus, a variable effective angle of attack along the span).
We will take advantage of this behavior of finite-span wings
to facilitate the analyses in section~\ref{sec:BL}.

Figure~\ref{fig:BL_streamline} also shows the contour lines of the pressure coefficient 
$c_p=2(p-p_\infty)/\rho U_\infty^2$ 
on the surface of the wings.  
Note that the wall-parallel component of
pressure-gradient 
(at the wall and across the boundary layer thickness)
is orthogonal to the constant $c_p$ 
lines represented in the figure. 
%%Given the nearly constant pressure across the boundary layer,
%%these contour lines are also representative of the wall-parallel pressure gradient that the boundary layers experience.
%Looking at RWT-0, 
%we can see an increasing trend in the pressure coefficient (and therefore pressure)
%by approaching the tip region.
An increasing trend in the pressure coefficient (and therefore pressure) 
can be observed approaching the tip region.
This pressure gradient 
is the main
cause for the observed streamline deflections in RWT-0. 
The boundary layer streamlines on the suction side of
RWT-5 and RWT-10 are deflected towards the root
as a result of these pressure gradients 
and the high momentum of the flow approaching the suction side from the pressure side.
There is also a pressure gradient on the pressure sides of
RWT-5 and RWT-10, albeit weaker,
which causes the flow to accelerate towards the tip of the wing. 
On both %the suction and pressure 
sides,
the pressure gradient decreases in magnitude
and becomes more aligned with the streamlines closer to the wing root
(characterized by contour level lines that are more aligned with the spanwise direction
and farther apart from each other).
This implies that the fluid particles following the streamlines are subjected to reduced accelerations due to pressure gradient. 
Consequently, the streamlines approximate straighter lines, leading to less variation in the deflection along a given streamline. 
This in turn mitigates some of the three-dimensional 
effects that result from the skewed velocity profile.

\section{Turbulent boundary layers \label{sec:BL}}

The finite span of the wings and the induced three-dimensionality %of the flow field 
by the wing-tip vortices have
a number of important impacts
on the boundary layers developing on RWT-0, RWT-5, and RWT-10.
The goal of this section is to separate and
simplify these effects as much as possible in sections~\ref{sec:BL-alpha} and~\ref{sec:BL-2D3D}, 
before %we analyze 
analyzing the remaining finite-span effects 
more closely in section~\ref{sec:3D-vs-2D}.

\subsection{The impact of the effective angle of attack \label{sec:BL-alpha}}

\begin{figure}
  \centering
 \includegraphics[angle=0,origin=c,width=65mm,clip=true,trim=0mm 0mm 10mm 1mm]{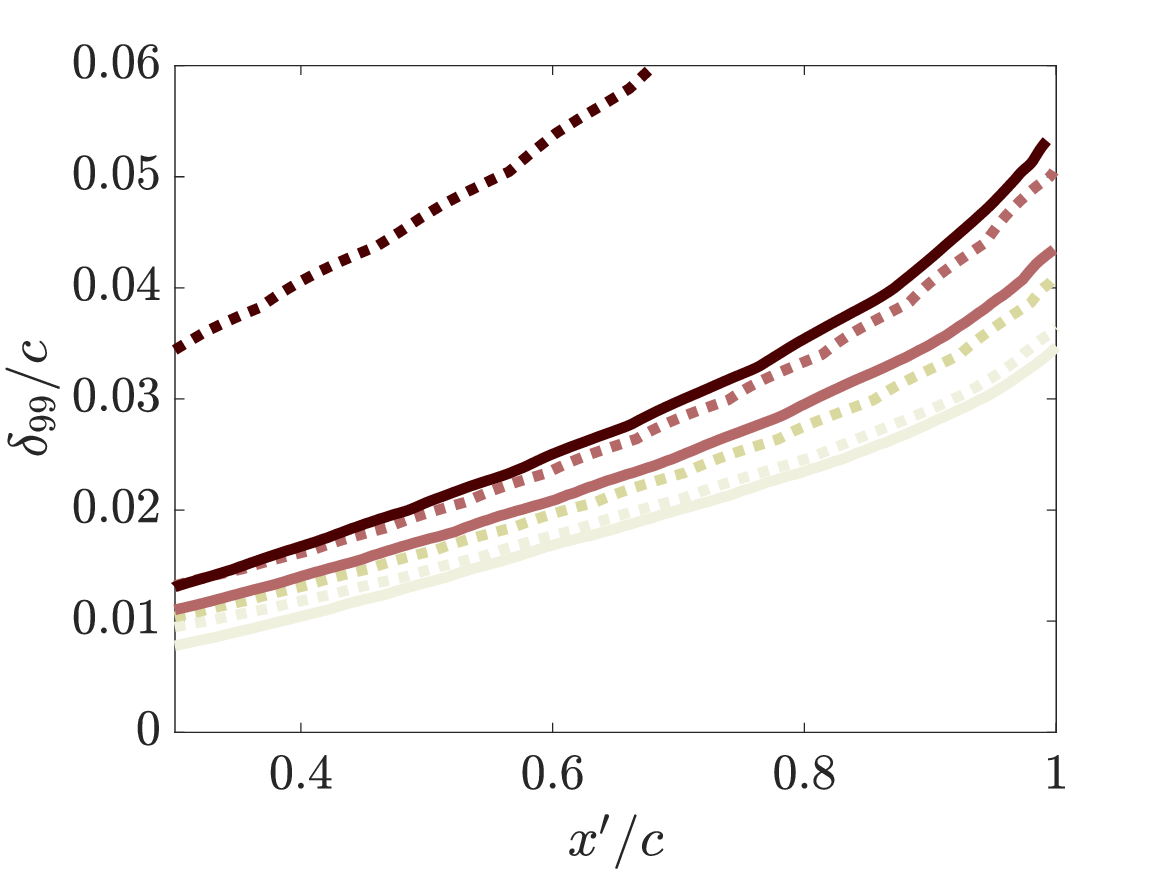}\llap{\parbox[b]{53mm}{(a)\\\rule{0ex}{43mm}}}
 \includegraphics[angle=0,origin=c,width=65mm,clip=true,trim=0mm 0mm 10mm 1mm]{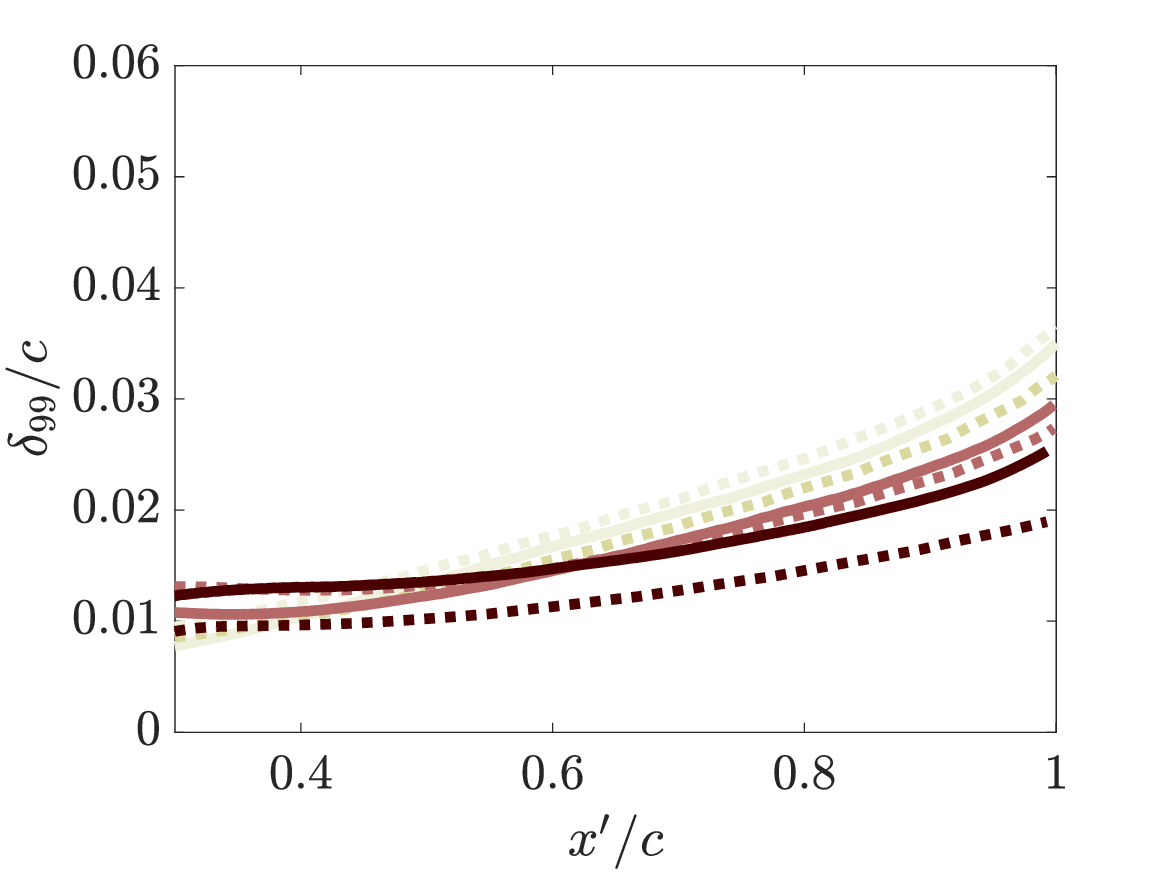}\llap{\parbox[b]{53mm}{(b)\\\rule{0ex}{43mm}}}
 \\
  \includegraphics[angle=0,origin=c,width=65mm,clip=true,trim=0mm 0mm 10mm 1mm]{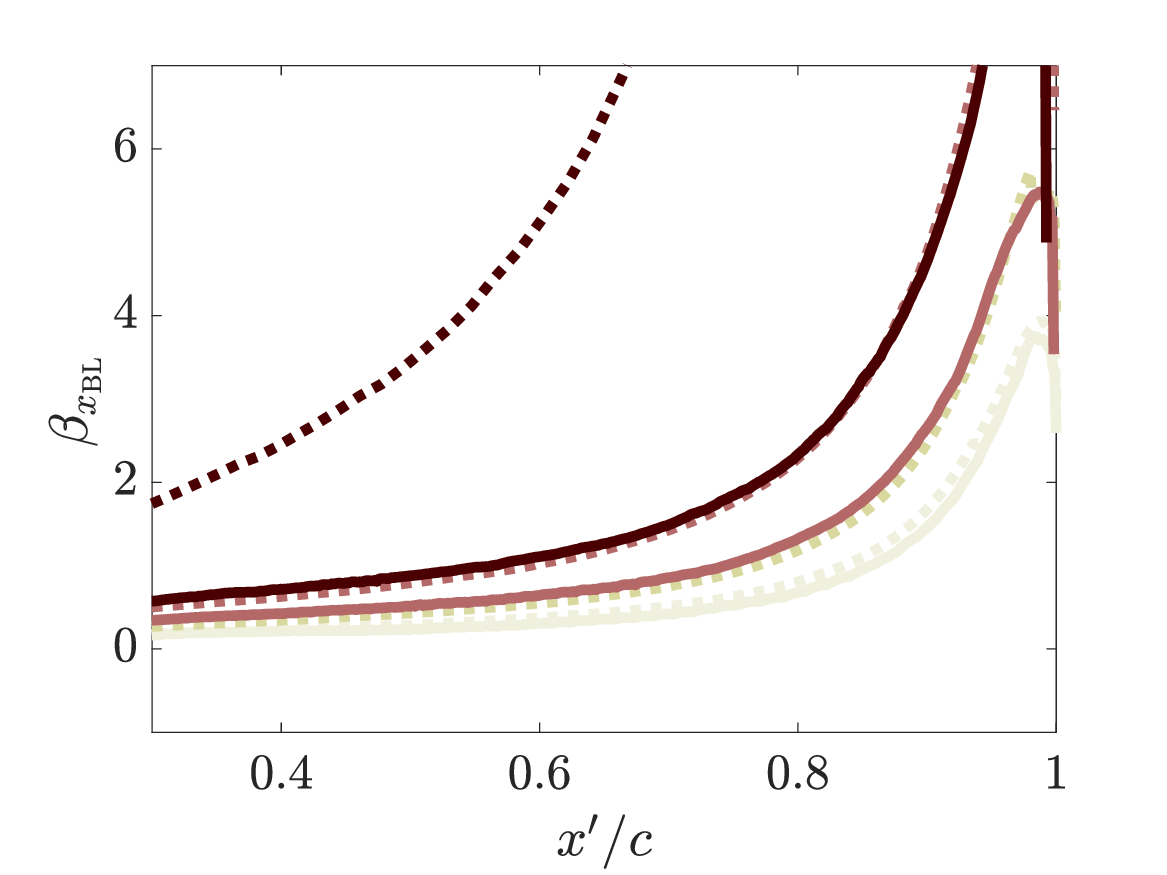}\llap{\parbox[b]{54mm}{(c)\\\rule{0ex}{43mm}}}
  \includegraphics[angle=0,origin=c,width=65mm,clip=true,trim=0mm 0mm 10mm 1mm]{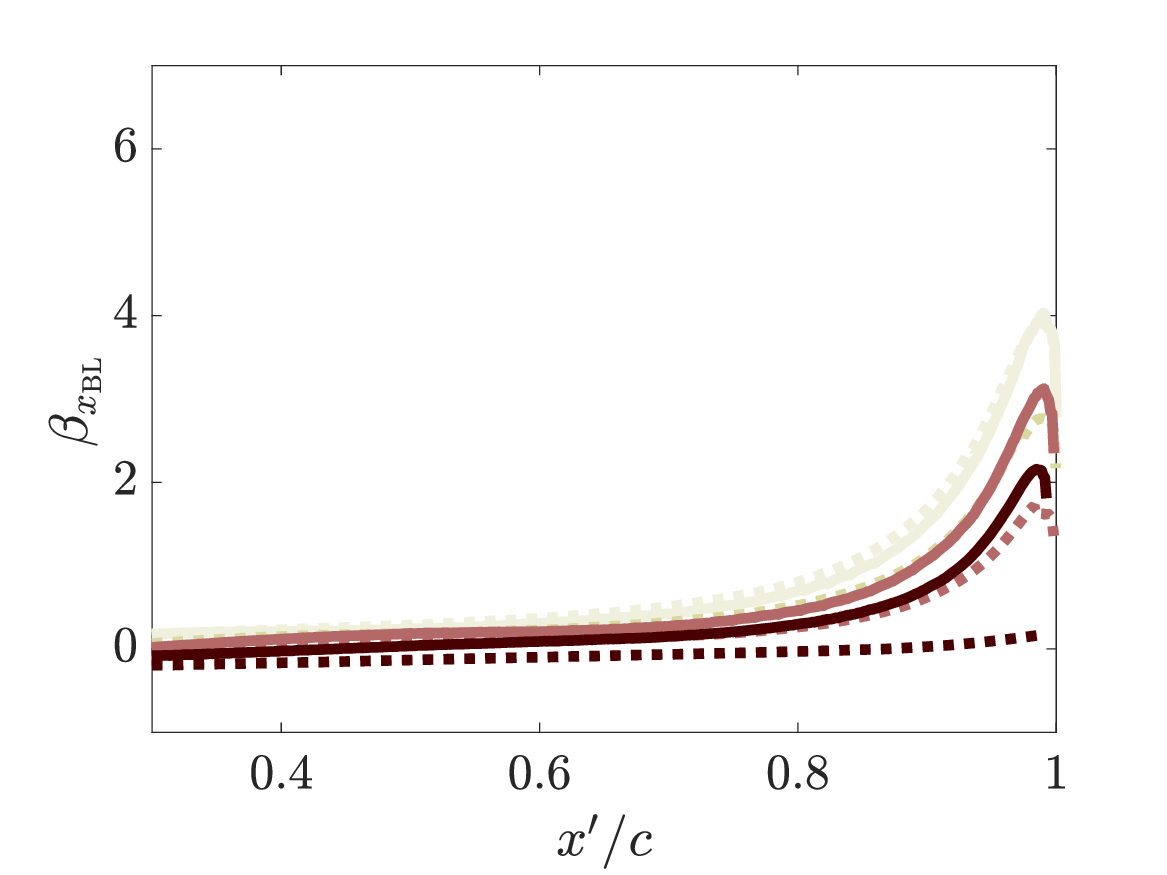}\llap{\parbox[b]{54mm}{(d)\\\rule{0ex}{43mm}}}
  \caption{
  \label{fig:BL_betaX-delta99}
  The $99\%$ boundary layer thickness $\delta_{99}$ (a, b) 
  and the Clauser pressure-gradient parameter $\beta_{x_{\rm BL}}$ (c, d) 
  plotted at $z'/c=0.1$ (close to the root)
  on the suction side (a, c) and pressure side (b, d) of
  the wings of this study. 
  \st{The finite-span wings are plotted in solid lines, while periodic airfoils are shown by the dotted lines.
  Colors, from light to dark, correspond to angles of attack of $\alpha=0^\circ$, $2^\circ$, $5^\circ$, and $10^\circ$, 
  respectively, for infinite-span wings.
  The three finite-span wings are shown with matched colors to the infinite-span ones at the same $\alpha$.}
%  and $\alpha=0^\circ$, $5^\circ$, and $10^\circ$ for finite-span wings with
%  matched colors for matched $\alpha$.}
%  The plots mainly demonstrate 
  The observed differences are mainly attributed to
  the impact of the reduced effective angle of attack $\alpha_{\rm eff}$ 
  \st{(an inviscid effect caused by the induced downwash of the wing-tip vortex; see figure~\ref{fig:intro-wingtip})}
  on the boundary layers.
  }
\end{figure}

Figure~\ref{fig:BL_betaX-delta99} plots the $99\%$ boundary layer thickness $\delta_{99}$
defined based on the diagnostic scaling~\citep{vinuesa:16}
and the Clauser pressure-gradient parameter $\beta_{x_{\rm BL}}$~\citep{clauser:54,clauser:56},
at a spanwise location near the root.
The Clauser parameter is computed 
along $x_{\rm BL}$ (defined in figure~\ref{fig:wing-geometry})
as,
\beq \label{eq:clauser-x}
\beta_{x_{\rm BL}} = 
\frac{\delta^*}{\left\vert \tau_{w} \right\vert} \pd{P_e}{x_{\rm BL}}
\, ,
\eeq
where $\delta^*$ is the boundary layer displacement thickness,
$\left\vert \tau_{w} \right\vert$ is the shear-stress magnitude at the wall,
and $P_e=\avg{p(x_{\rm BL},\delta_{99},z_{\rm BL})}$ is the pressure at the edge of the boundary layer.

\begin{figure}
  \centering
    \includegraphics[width=65mm,clip=true,trim=0mm 0mm 10mm 1mm]{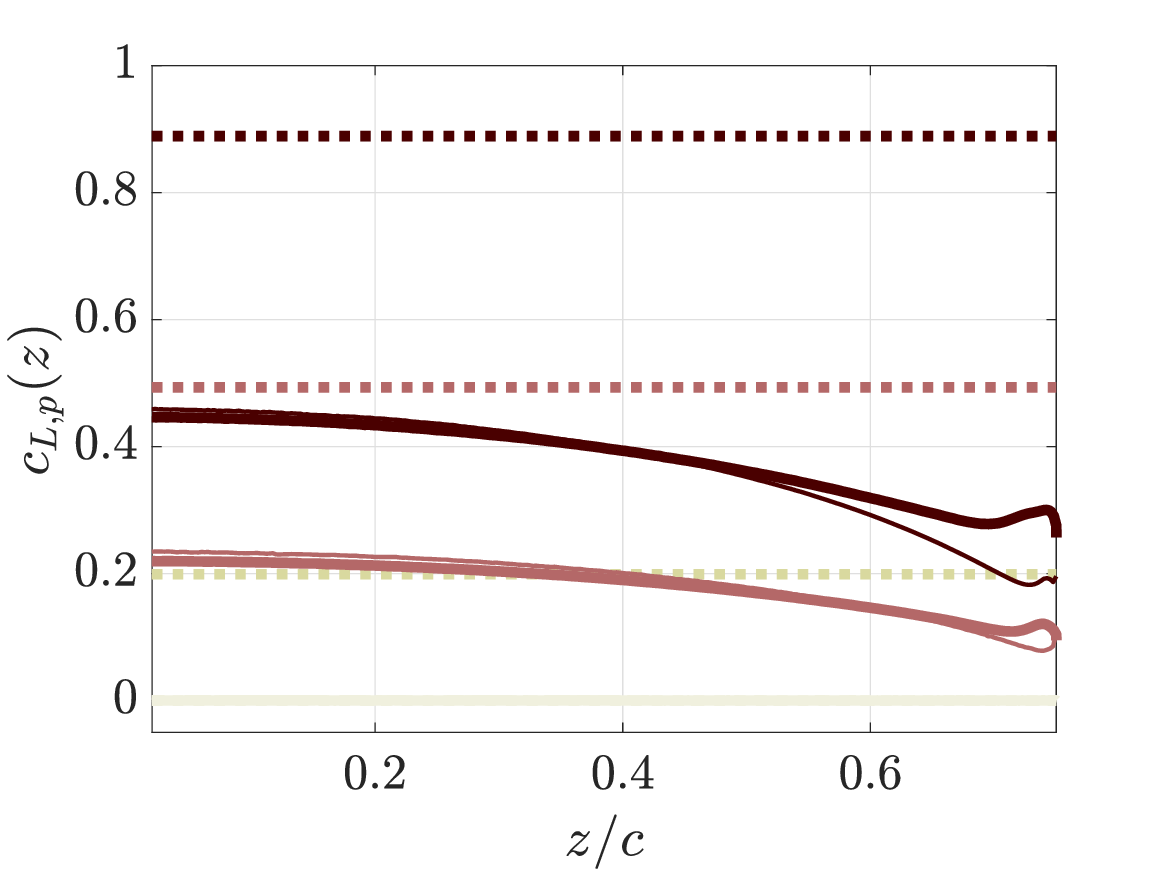}\llap{\parbox[b]{55mm}{(a)\\\rule{0ex}{44mm}}}
    \includegraphics[width=65mm,clip=true,trim=0mm 0mm 10mm 1mm]{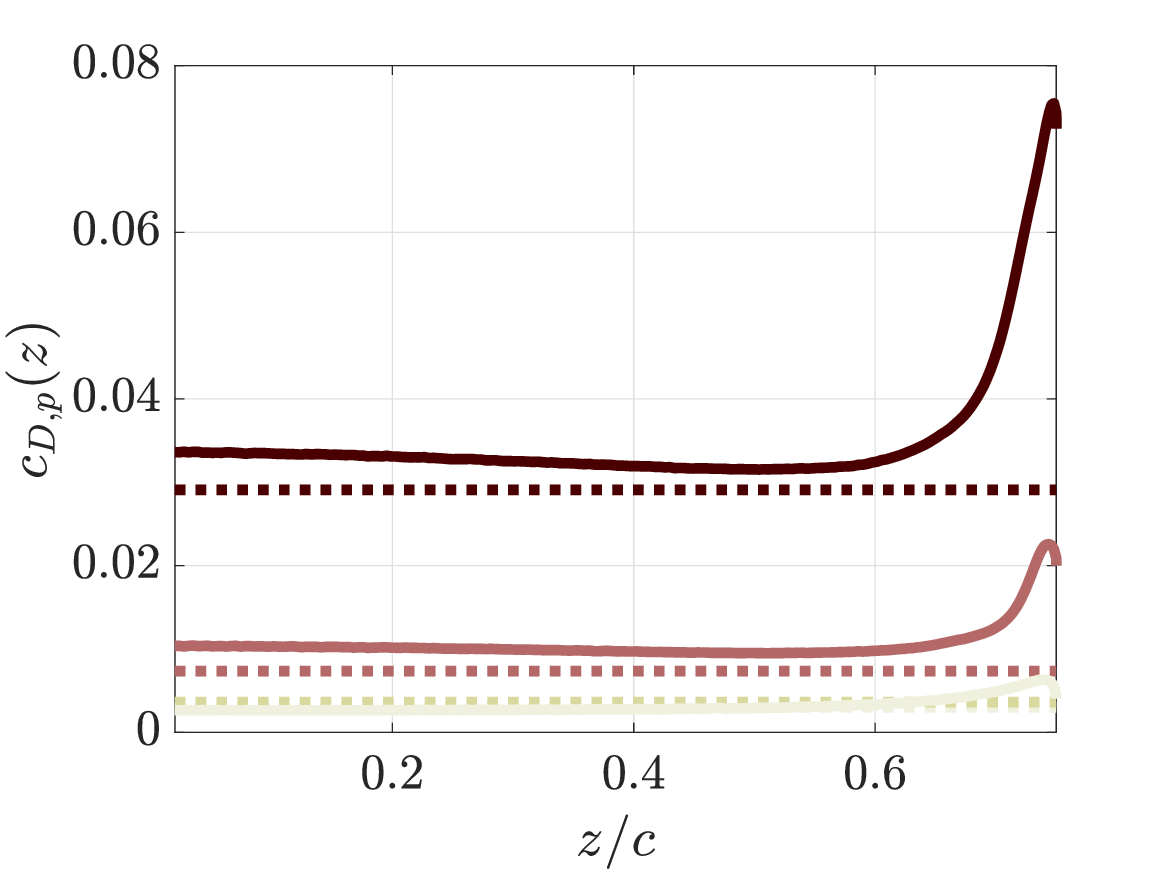}\llap{\parbox[b]{54mm}{(b)\\\rule{0ex}{44mm}}}
  \caption{
  \label{fig:BL-LD}
  The section-wise (a) pressure lift and (b) pressure drag coefficients along the span
  for all the geometries of this study. 
  Line styles and colors are the same as those in figure~\ref{fig:BL_betaX-delta99}.
  \st{The additional thin solid lines in panel (a)
  are from inviscid simulations of finite-span wings for the same configurations as RWT-0, RWT-5, and RWT-10.
  The nearly identical variation of pressure lift coefficient (approximately proportional to $\alpha_{\rm eff}$)
  emphasizes the inviscid nature of the phenomenon. }
%  and the observed differences in figure~\ref{fig:BL_betaX-delta99}.}
%%  impact of the effective angle of attack
%%  on the boundary layers.
%  The main cause of the difference between turbulent and inviscid finite-span wings,
%  especially near the wing tip,
%  is due to their difference in vortex formation process and its location on the wing.}
  Refer to figure~\ref{fig:intro-wingtip} (b) and its caption for \st{a schematic} explanation of the 
  \st{induced drag which explains the}
  increased drag coefficient of the finite-span wings.
%  \st{More details can be found in related references given in section~\ref{sec:intro}.} 
  } 
\end{figure}

There is a large difference between the 
boundary layers formed on the finite-span wings
compared to %those of 
the periodic ones at the same angle of attack.
Namely, the boundary layers on the suction sides are thinner 
and encounter lower adverse pressure gradients 
(lower $\beta_{x_{\rm BL}}$ values),
while they are thicker on the pressure side and
encounter a larger Clauser parameter
(interesting to note that $\beta_{x_{\rm BL}}>0$ over the majority of the pressure side as well). 
Furthermore, the discrepancies are larger for higher angles of attack 
and nearly vanish for $\alpha=0^\circ$.
%The observed differences are \
These are all consequences of
the lower effective angle of attack $\alpha_{\rm eff}$
of the finite-span wings compared to their geometric angle of attack $\alpha$
(a well-known inviscid effect
\st{caused by the induced downwash of the wing-tip vortex,}
explained in figure~\ref{fig:intro-wingtip}). 
This is further
demonstrated by figure~\ref{fig:BL-LD}
which shows the section-wise 
pressure component of the
lift and drag coefficients of all cases defined as, 
\beq \nonumber
\begin{aligned}
c_{L,p}(z_0) =  \frac{2}{\rho U_\infty^2 c} \oint {(\avg{p(x_{\rm BL},0,z_0)} \cdot \mathbf{e}_{\rm n}) {\rm d}{x}_{\rm BL}}, \\
c_{D,p}(z_0) = \frac{2}{\rho U_\infty^2 c} \oint {(\avg{p(x_{\rm BL},0,z_0)} \cdot \mathbf{e}_{\rm t}) {\rm d}{x}_{\rm BL}},
\end{aligned}
\eeq 
where 
$\mathbf{e}_{\rm n}$ is a unit vector normal to the wall and opposite to $y_{\rm BL}$,
$\mathbf{e}_{\rm t}$ is a unit vector tangential to the wall in the same direction as $x_{\rm BL}$,
and $\rho$ and $c$ are the fluid density and chord length.
%an arbitrary length, 
%and the sectionwise planform area of the wing, respectively.
The integrals are taken on the wing surface ($y_{\rm BL}=0$)
over both the suction and pressure sides
at a fixed spanwise location $z_{\rm BL}=z_0$.
Assuming a linear variation of the lift coefficient with angle of attack
in infinite-span wings~\citep[which is approximately true for relatively low angles of attack; cf.][]{Houghton:book,phak}
we could make the approximation that 
$3^\circ \leq \alpha_{\rm eff} \leq 4.5^\circ$ for RWT-10 
(with it being closer to $4.5^\circ$ at the root and around $3^\circ$ close to the tip)
and $1^\circ \leq \alpha_{\rm eff} \leq 2^\circ$ for RWT-5
(the lifting-line theory~\citep[cf.][]{Houghton:book} leads to similar approximated values for $\alpha_{\rm eff}$
over the majority of the span).
This justifies the fact that
in figure~\ref{fig:BL_betaX-delta99} 
both the boundary layer thickness and the Clauser parameter on the suction side of
RWT-10 are very close to those of the P-5,
or why the RWT-5 curves
(corresponding to $\alpha_{\rm eff} \approx 2^\circ$ at the root)
are closer to P-2.

It is important to
emphasize that the observed similarity based on effective angles of attack
is in fact a result of an approximate similarity of the pressure gradient
and boundary layer histories.
Therefore, our reliance on $\alpha_{\rm eff}$ in the next sections
is meant as an approximate equivalence of pressure gradients and history effects. 
In fact, there are additional (inviscid) effects caused by 
a positive (i.e., away from the wall) and variable wall-normal velocity along the span,
especially near the tip at locations earlier in the development of the wing-tip vortex,
that are not captured by the simplified use of $\alpha_{\rm eff}$.
These effects will be discussed further in section~\ref{sec:3D-vs-2D}.

\st{
Figure~\ref{fig:BL-LD} (a) also compares 
the turbulent finite-span wings, RWT-0, RWT-5, and RWT-10, with 
inviscid (incompressible Euler) simulations of the same setups (without the tripping)
performed using the open-source solver SU2~\citep{su2}. 
The nearly identical variation of the lift (and effective angle of attack) 
over a large portion of the span 
%is due to 
emphasizes
the inviscid nature of the phenomenon. 
The larger observed differences near the tip should be mainly attributed to the 
difference in vortex formation and the resulting change 
in both the vertical and spanwise location of the vortex in inviscid flows. 
These differences are discussed briefly in appendix~\ref{app:general-vortex}.
Additional viscous effects are present near the tip and are expected to contribute 
to the observed differences in that region; 
however, they will not be discussed here. 
}

%
%\st{
%It is also important to 
%note that the majority of the difference between the finite and infinite span wings
%observed in figures~\ref{fig:BL_betaX-delta99} and~\ref{fig:BL-LD}
%}
%

%\subsection{Collateral flow and simplification by transformation into wall-shear coordinates \label{sec:BL-2D3D}}
%\subsection{Transformation into wall-shear coordinates \label{sec:BL-2D3D}}
\subsection{Collateral flow and transformation into wall-shear coordinates \label{sec:BL-2D3D}}

The skewed velocity profile and 
the variable deflection angle of the streamlines 
(figure~\ref{fig:BL_streamline})
result in additional non-zero velocity gradient components,
and consequently, the activation of
additional production terms
in the transport equations of $\tilde{R}_{13}$, $\tilde{R}_{23}$, and $\tilde{R}_{33}$
(for $\tilde{R}_{ij}=\avg{u'_i u'_j}_{\rm BL}$ denoting the Reynolds stress
expressed in $(x_{\rm BL},y_{\rm BL},z_{\rm BL})$ coordinates)
with the largest impact on the production of $\tilde{R}_{13}$. 

The main goal of this section is to use the flow characteristics to
find a coordinate system that simplifies the analysis.

\begin{figure}
  \centering
 \includegraphics[angle=0,origin=c,width=44mm,clip=true,trim=0mm 0mm 10mm 1mm]{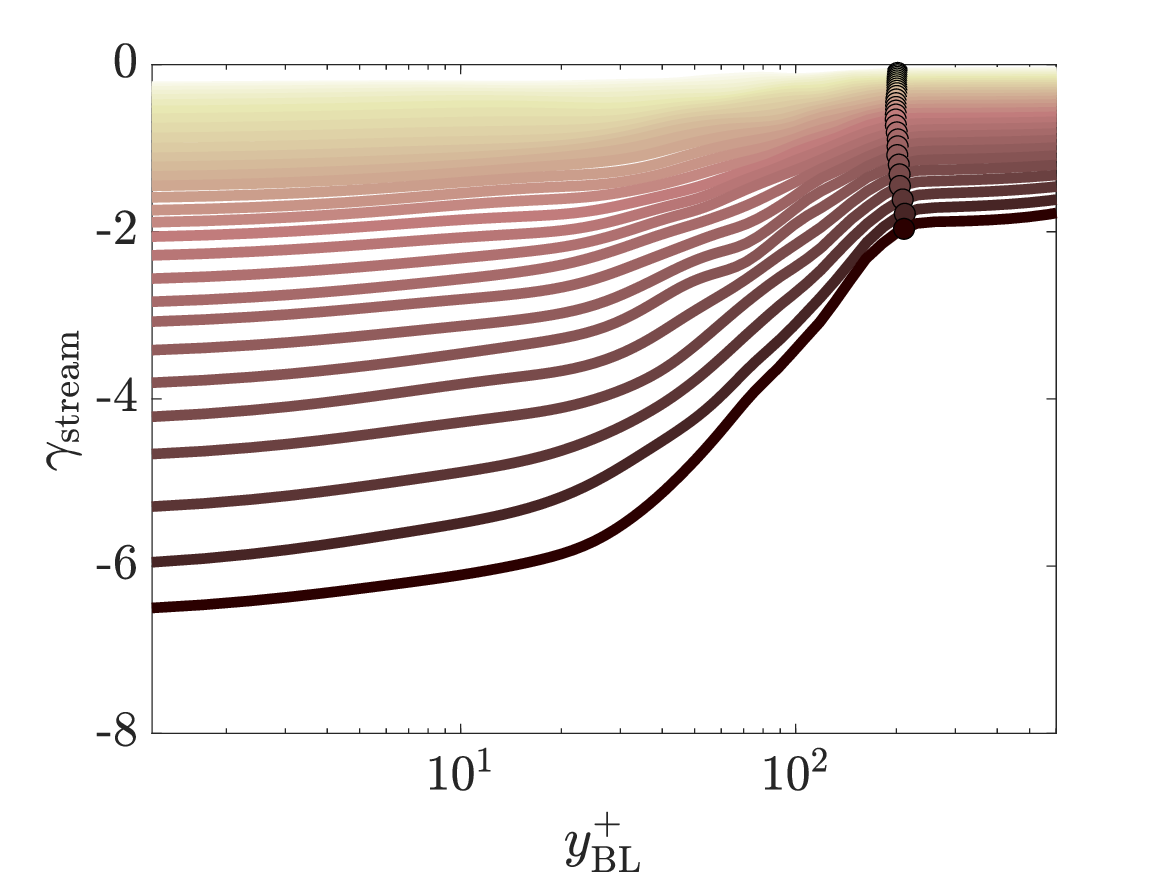}\llap{\parbox[b]{7mm}{(a)\\\rule{0ex}{7mm}}}
 \includegraphics[angle=0,origin=c,width=44mm,clip=true,trim=0mm 0mm 10mm 1mm]{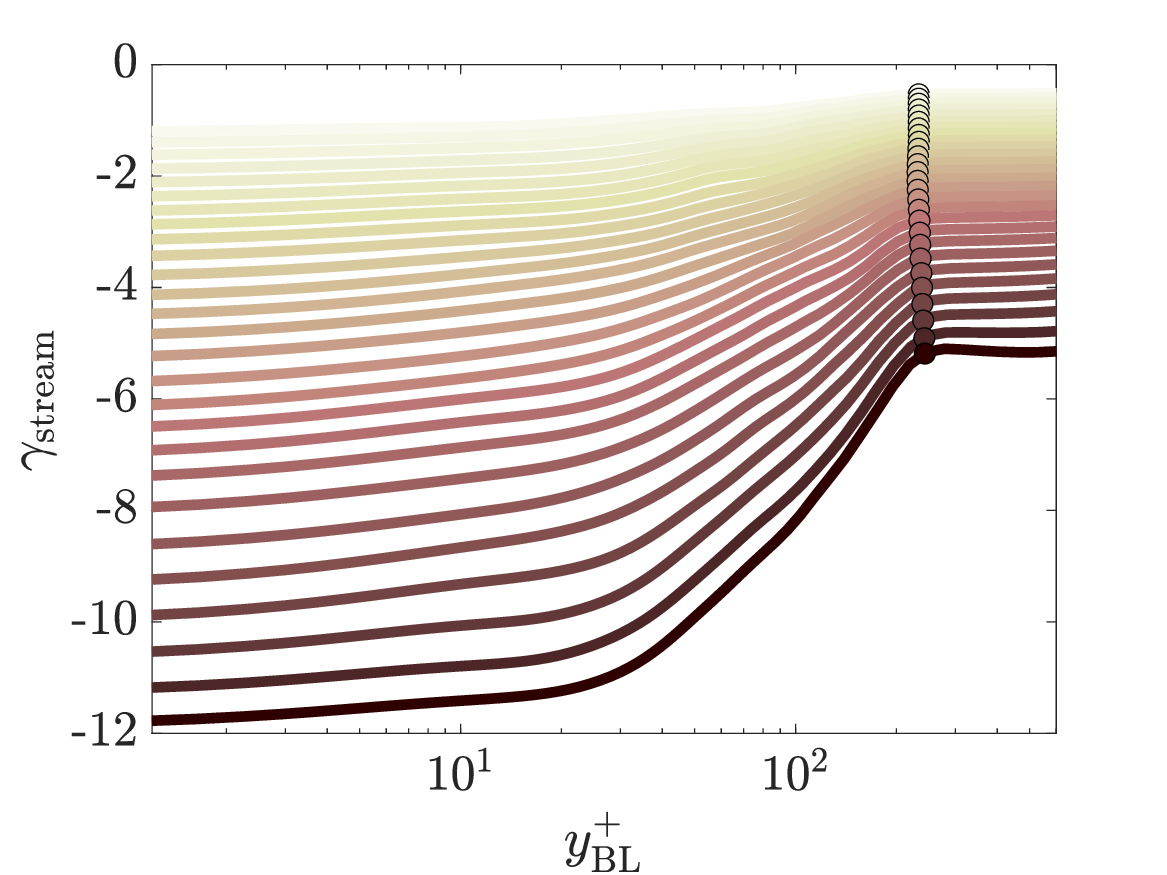}\llap{\parbox[b]{7mm}{(b)\\\rule{0ex}{7mm}}}
  \includegraphics[angle=0,origin=c,width=44mm,clip=true,trim=0mm 0mm 10mm 1mm]{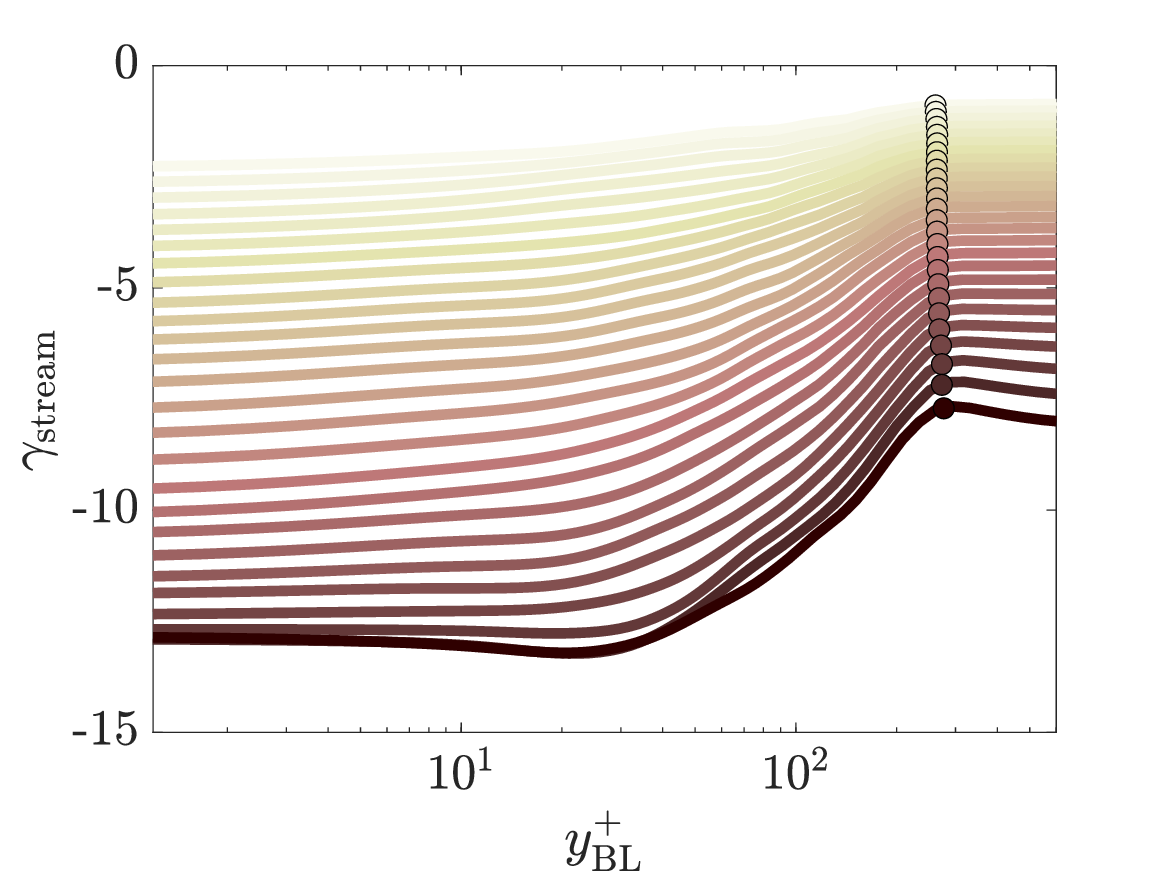}\llap{\parbox[b]{7mm}{(c)\\\rule{0ex}{7mm}}} \\
 \includegraphics[angle=0,origin=c,width=44mm,clip=true,trim=0mm 0mm 10mm 1mm]{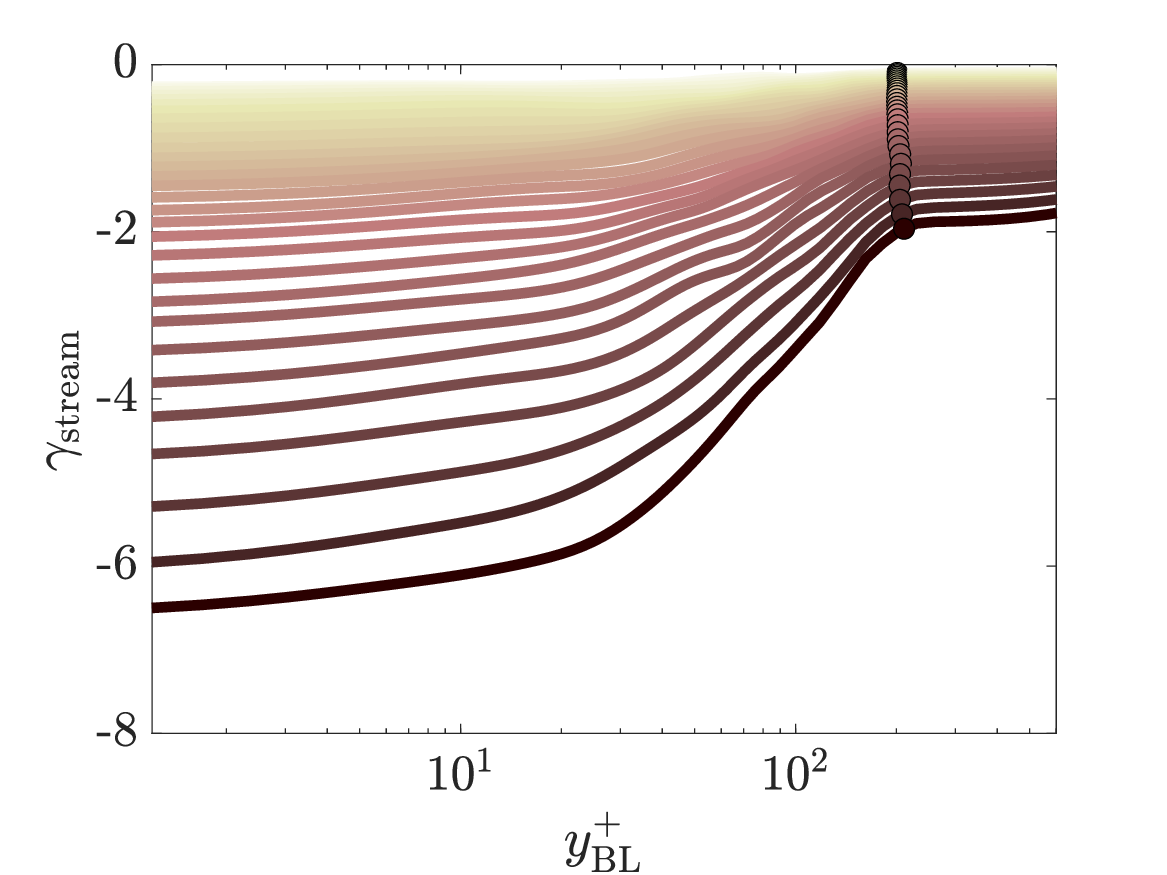}\llap{\parbox[b]{7mm}{(d)\\\rule{0ex}{7mm}}}
 \includegraphics[angle=0,origin=c,width=44mm,clip=true,trim=0mm 0mm 10mm 1mm]{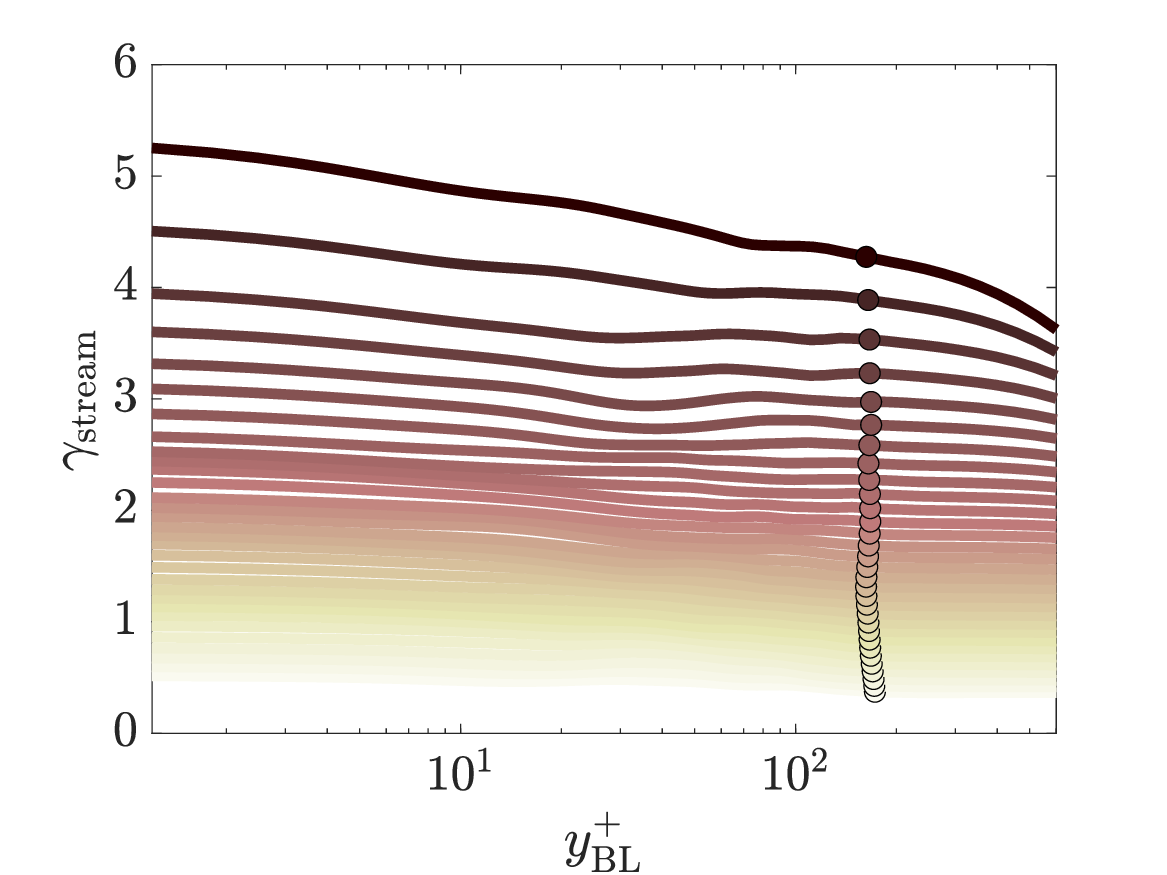}\llap{\parbox[b]{7mm}{(e)\\\rule{0ex}{7mm}}}
 \includegraphics[angle=0,origin=c,width=44mm,clip=true,trim=0mm 0mm 10mm 1mm]{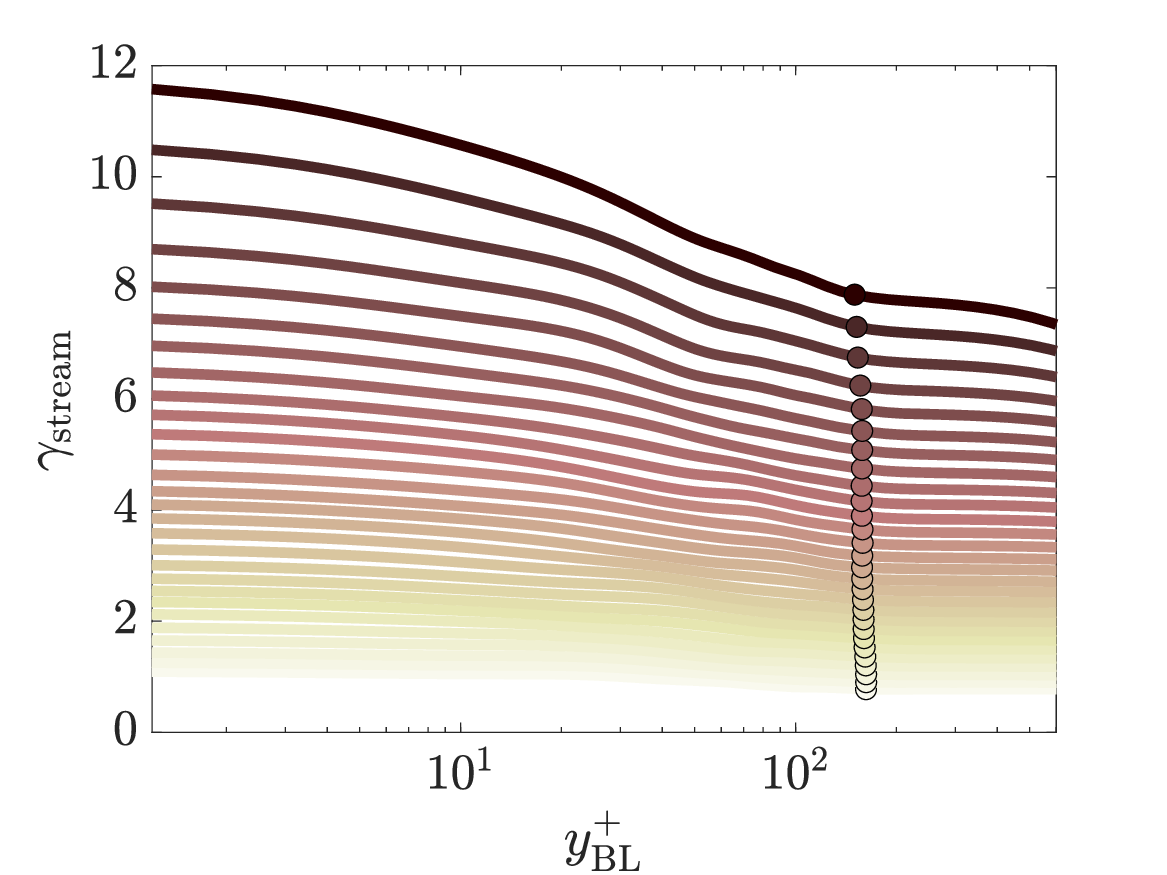}\llap{\parbox[b]{7mm}{(f)\\\rule{0ex}{7mm}}}
  \caption{
  \label{fig:BL_deflection}
  The deflection angle 
  (in degrees)
  at $x'/c=0.7$ for 
  multiple spanwise locations from 
  $z'/c=0.1$ to 0.6 (with 0.02 increments) 
  on the suction side (a, b, c)
  and the pressure sides (d, e, f) of
  (a, d) RWT-0, (b, e) RWT-5, and (c, f) RWT-10.
  Darker colors are sampled closer to the tip. 
  Circles show the edge of the boundary layer, $\delta_{99}$. 
  Negative (positive) $\gamma_{\rm stream}$ values correspond to
  deflection towards the root (tip). 
  Note the relatively constant value of $\gamma_{\rm stream}$ for $y_{\rm BL}^+ \leq 30 $.
  }
\end{figure}

Figure~\ref{fig:BL_deflection} plots the deflection angle $\gamma_{\rm stream}$
defined as the angle between the wall-parallel part of the velocity vector and the 
wall-parallel direction $x_{\rm BL}$ 
(with the wall-normal component of the velocity vector $\avg{u_2}_{\rm BL}$ excluded). 
Visually, this is the angle that the streamlines of figure~\ref{fig:BL_streamline}
make with the chord-wise lines (free-stream direction).
While $\gamma_{\rm stream}$ is highly variable across the boundary layer thickness, 
it exhibits an %remarkably 
important quality:
it is approximately constant in 
the most active and important region of wall turbulence, $y_{\rm BL}^+ \leq 30$.
This region of \emph{collateral flow} is a common feature of many three-dimensional
boundary layers~\citep[cf.][]{johnston:60,perry:65,pirece:83,olcmen:95,devenport:22}.
In the context of our discussion,
this means that the most active part of the near-wall region
can potentially be considered nearly two-dimensional 
in a rotated coordinate system that is aligned with the direction of the streamlines
in the $y_{\rm BL}^+ \leq 30$ region.
The region above $30\delta_\nu$ still experiences a variable shear,
but since both the Reynolds stresses and mean velocity gradients are significantly 
smaller in this region,
the contribution from the new production terms
(which are the product of the Reynolds stresses and the mean velocity gradient)
will remain low for a large portion of the boundary layer.

In the rest of this paper
the boundary layers are studied in a
coordinate system $(x_\tau,y_\tau,z_\tau)$,
which is defined by rotating $(x_{\rm BL},y_{\rm BL},z_{\rm BL})$ 
(figure~\ref{fig:wing-geometry})
around its wall-normal axis $y_{\rm BL}$
in such a way that $x_\tau$ becomes aligned with
the direction of the wall shear stress (the streamlines).
Note that the orientation of this coordinate system changes %is in fact variable 
across both the span and the chord of the wing
%with lower rotation angles closer to the root or the leading-edge
%and higher angles close to the tip or the trailing edge 
(see the streamlines of figure~\ref{fig:BL_streamline}). 
This does not affect the analysis in the rest of the paper 
which is performed in local coordinates.

The rotated Reynolds-stress components
exhibit a structure that is very close to a two-dimensional 
boundary layer, where $R_{13} \approx R_{23} \approx 0$
(for $R_{ij}=\avg{u'_i u'_j}_\tau$ denoting the Reynolds stress
expressed in $(x_\tau,y_\tau,z_\tau)$ coordinates).
This simplification is possible for these boundary layers mostly because of the
relatively small change in the deflection angle across the boundary layer thickness
(less than $10^\circ$; see figure~\ref{fig:BL_deflection})
and is not necessarily possible for all three-dimensional boundary layers.
The spanwise component of the mean velocity
is by definition close to zero for $y_\tau^+\leq30$ 
in this coordinate system
and only increases closer to the edge of the boundary layer.
This results in a Reynolds-stress production term close 
to that of a statistically two-dimensional boundary layer
for similar conditions.
The obtained simplification by rotation into the $(x_\tau,y_\tau,z_\tau)$ coordinate system 
cannot remove %all the three-dimensional effects. %especially 
the less local effects.
%This fact will be further discussed in the next section.
Such effects are discussed in the next section.

\subsection{Additional departures from infinite-span wings
\label{sec:3D-vs-2D}}

Two other notable effects are associated with the finite span of the wing
and the three-dimensional flow field: %induced by the wing-tip vortices 
%include:
\begin{enumerate}
\item 	Flow acceleration from the pressure side towards the suction side,
during the formation of the wing-tip vortices,
results in a wall-normal velocity that
is different along the chord and the span of the wing.
This variation is such that the
locations closer to the tip and earlier in the development of the wing-tip vortex 
(i.e., closer to the leading edge)
experience higher positive wall-normal velocities.
The wall-normal velocity impacts the development of the boundary layers,
such that increased values of this quantity lead to a faster growth rate.
As a result, 
there is a slight increase in $\delta_{99}$ on the suction side 
for locations closer to the tip,
and similarly thinner boundary layers closer to the tip on the pressure side.
This is primarily an inviscid effect.
\item 	Variation of the deflection angle with normal distance from the wall 
(a result of variable momentum across $y_\tau$, mainly due to viscous effects)
means that
streamlines crossing a wall-normal line have converged from different spanwise locations on the wing. 
Since the effective angle of attack, and thus the streamwise pressure gradients, are different at
different spanwise locations (mostly an inviscid effect), 
fluid particles at different wall-normal locations 
have experienced different histories.
This effect is larger 
closer to the tip region, where both the variation in the deflection angle (see figure~\ref{fig:BL_deflection})
and the variation in the pressure gradient 
(approximately characterized by the effective angle of attack; see figure~\ref{fig:BL-LD})
are larger.
\end{enumerate}

%In the following we investigate and characterize 
This section investigates and characterizes
these effects
by comparing the %three-dimensional 
boundary layers formed on the finite-span wings 
with their equivalent %two-dimensional boundary layers 
from the infinite-span wings.
Equivalence is quantified here 
as closeness in terms of the local values
of $Re_\tau$ and $\beta_{x_\tau}$,
as well as similarity of their history. 
Appendix~\ref{app:sec-BL-params} explains in more detail 
the procedure used to match each of these quantities.
The importance of this procedure is discussed briefly at the end of this section.

\begin{figure}
  \centering
  \begin{overpic}[angle=0,origin=c,width=65mm,clip=true,trim=0mm 0mm 0mm 0mm]{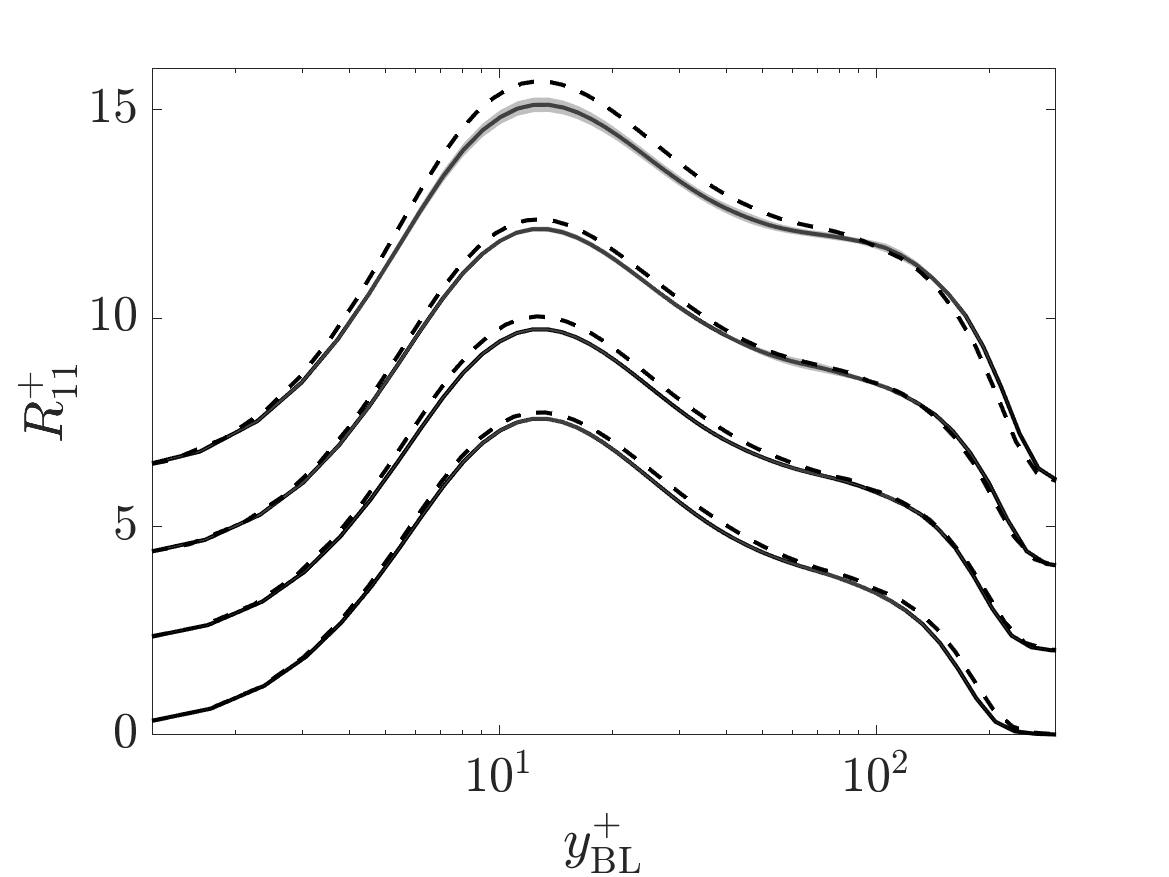}
  \put(60,25){\vector(1,4){10}} 
%  \put(40,3){$\overrightarrow{U}_\infty$}
  \put(72,59){$x'\uparrow$}
%  \put(40,3){flow direction}
  \put(15,65){(a)}
  \put(39,15){$z'_{\rm start}/c=0.1$}
  \end{overpic}
    \begin{overpic}[angle=0,origin=c,width=65mm,clip=true,trim=0mm 0mm 0mm 0mm]{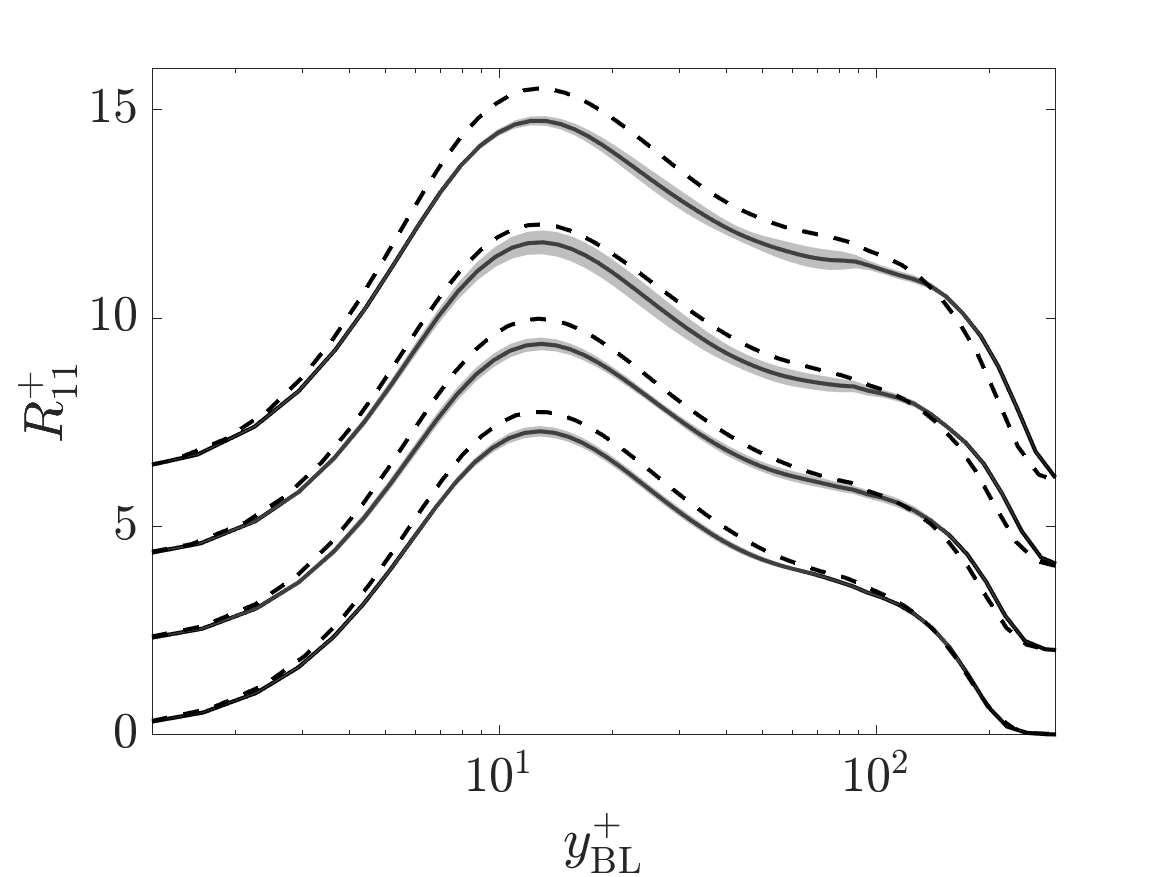}
  \put(60,25){\vector(1,4){10}} 
%  \put(40,3){$\overrightarrow{U}_\infty$}
  \put(72,59){$x'\uparrow$}
%  \put(40,3){flow direction}
  \put(15,65){(b)}
  \put(39,15){$z'_{\rm start}/c=0.72$}
  \end{overpic}
    \begin{overpic}[angle=0,origin=c,width=65mm,clip=true,trim=0mm 0mm 0mm 0mm]{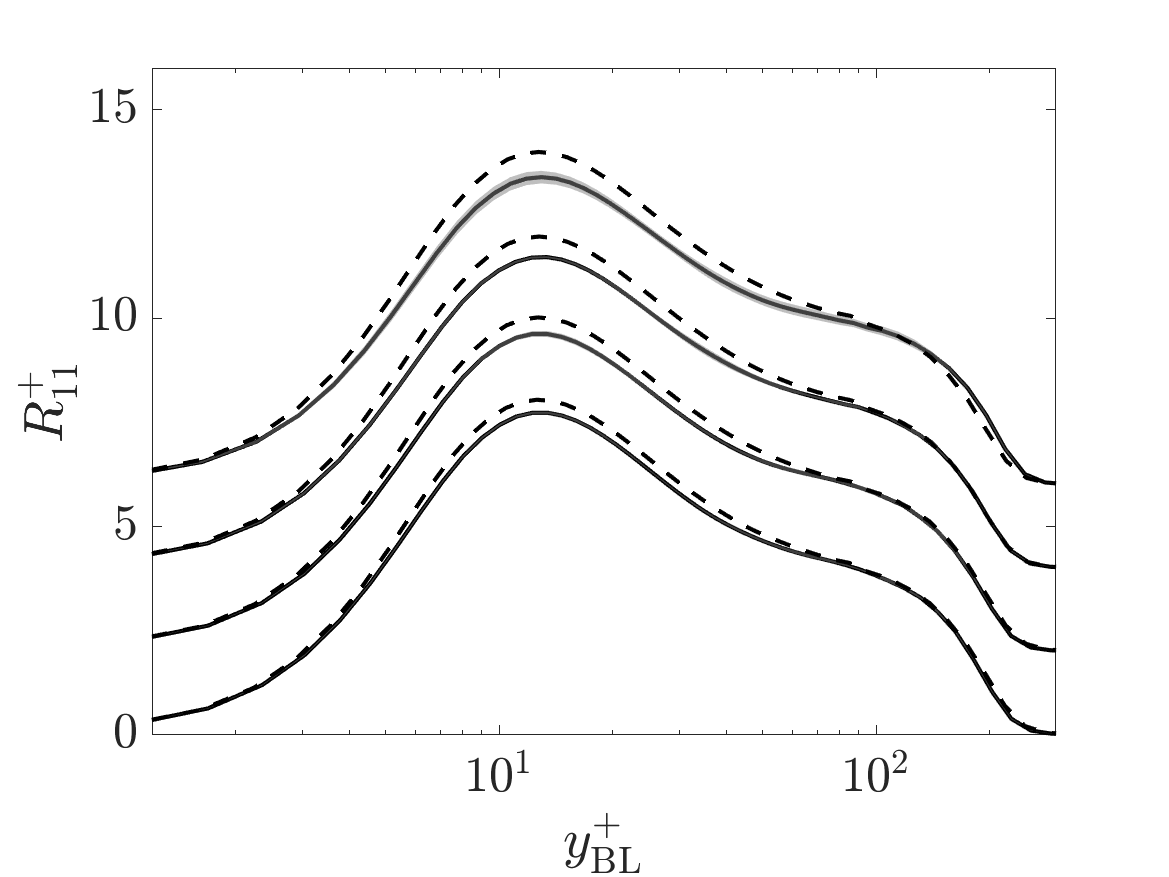}
  \put(60,25){\vector(1,4){10}} 
%  \put(40,3){$\overrightarrow{U}_\infty$}
  \put(72,59){$z'\uparrow$}
%  \put(40,3){flow direction}
  \put(15,65){(c)}
  \put(40,15){$x'/c=0.7$}
  \end{overpic}
    \begin{overpic}[angle=0,origin=c,width=65mm,clip=true,trim=0mm 0mm 0mm 0mm]{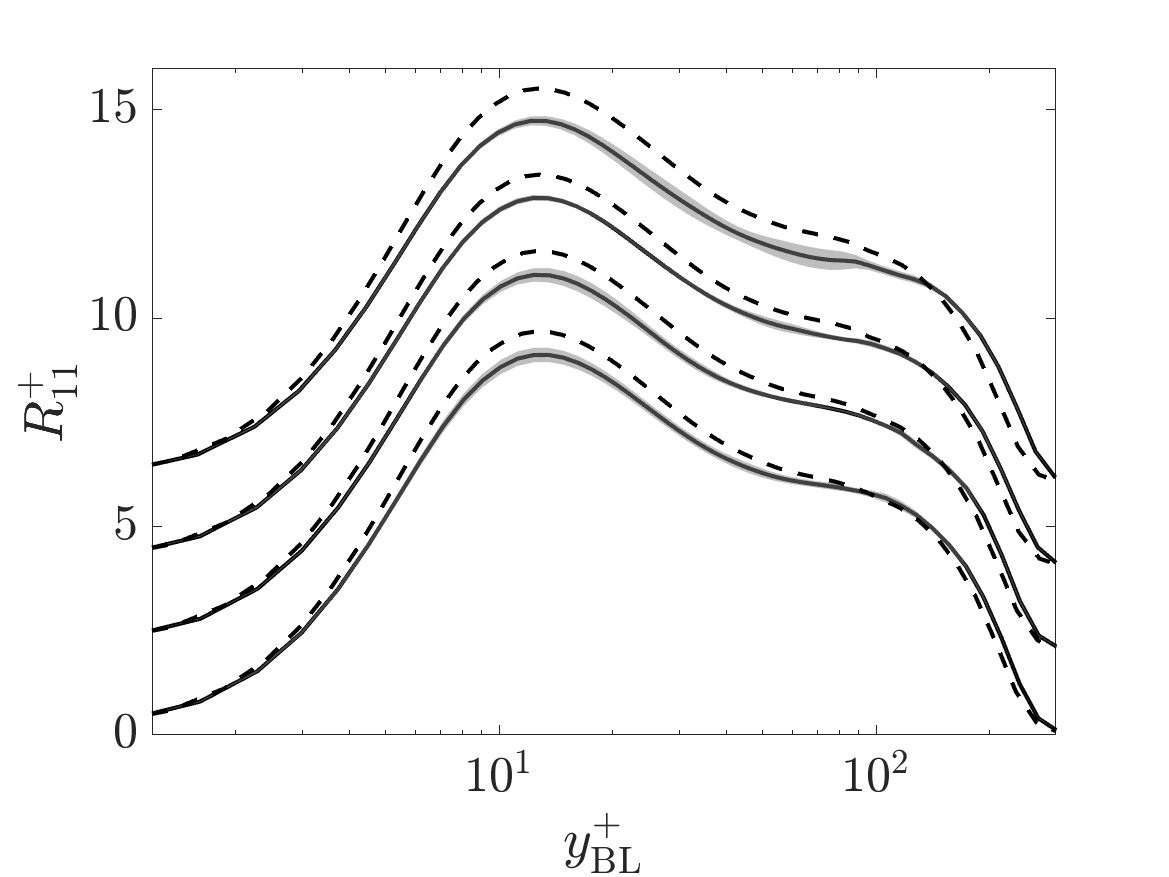}
  \put(60,25){\vector(1,4){10}} 
%  \put(40,3){$\overrightarrow{U}_\infty$}
  \put(72,59){$z'\uparrow$}
%  \put(40,3){flow direction}
  \put(15,65){(d)}
  \put(40,15){$x'/c=0.9$}
  \end{overpic}
 \caption{
  \label{fig:pattern-xz}
  Variation of the inner-scaled Reynolds stress $R_{11}^+$ with 
  (a,b) streamwise distance from the leading edge, 
  and (c,d) spanwise distance from the root.
  Panels (a) and (b) correspond to the evolution 
  along streamlines released at 
  $(x'_{\rm start},z'_{\rm start})/c=(0.12,0.1)$
  and $(x'_{\rm start},z'_{\rm start})/c=(0.12,0.72)$, respectively.
  Panels (c) and (d) show the locations corresponding to the intersection of four streamlines
  released at  $x'_{\rm start}/c=0.12$ at spanwise locations
  $z'_{\rm start}/c=0.1$, $0.31$, $0.51$, $0.72$,
  with planes at (c) $x'/c=0.6$ and (d) $x'/c=0.9$.
  Profiles are shifted vertically by $2$ units with increasing $x'$ or $z'$ for visual clarity. 
  The profiles are from RWT-5 (solid lines) and P-2 (dashed lines), 
  and are a subset of those plotted in figure~\ref{fig:rwt5-RS-allXZ}.
  All streamlines are released in the near-wall (approximately collateral) region.
  See text and appendix~\ref{app:sec-BL-params} for more details. 
  Shaded regions correspond to 80\% confidence intervals of the inner-scaled Reynolds stress profiles (due to finite time averaging),
  computed from the non-overlapping batch method~\citep[cf.][]{conway:63}.
  P-2 has extremely narrow confidence intervals which are not plotted. 
  }
\end{figure}
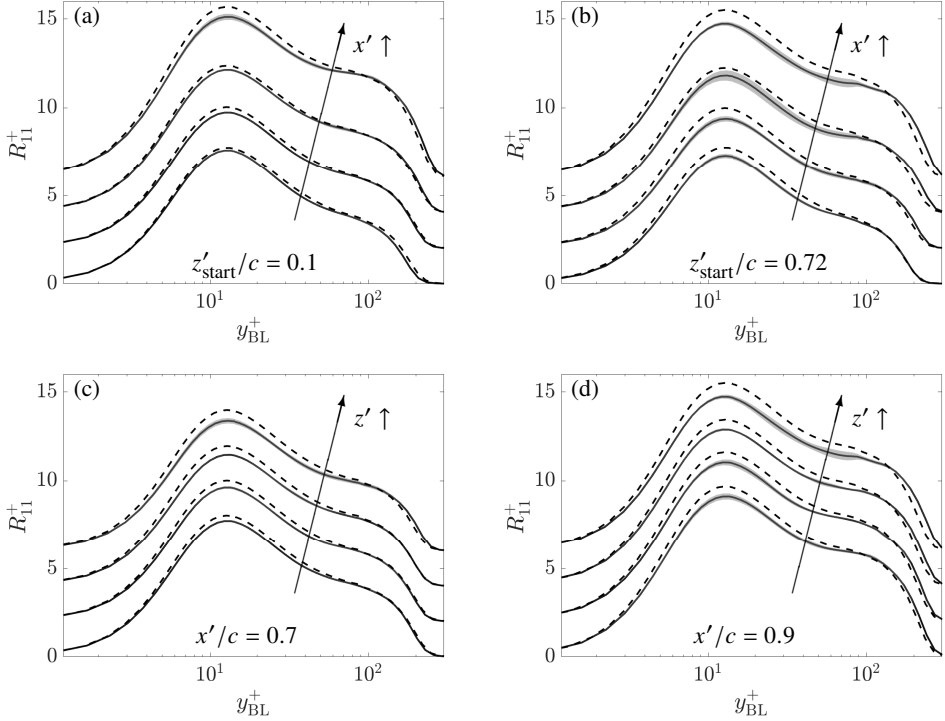

Figure~\ref{fig:pattern-xz} depicts the variation of the normal streamwise Reynolds stress, 
$R_{11}=\avg{u'_1 u'_1}_\tau$, 
in RWT-5 as a function of the streamwise and spanwise locations. 
Additional Reynolds stress components, additional locations, and other angles of attack are included
in figures~\ref{fig:rwt0-RS-allXZ},~\ref{fig:rwt5-RS-allXZ}, and~\ref{fig:rwt10-RS-allXZ} in appendix~\ref{app:sec-BL-params}.
Only a subset of those profiles which is representative of the overall trends and behaviors is shown here.
We focus solely on the suction side of the wings, both here and in the appendix. 
This choice is motivated by the higher Reynolds number and more interesting behavior exhibited on the suction side. 
Information concerning the pressure side of the wings, as well as the suction side, 
including all turbulence statistics, is accessible in the simulation database (please refer to the data availability statement).

The initial pattern identified in figure~\ref{fig:pattern-xz} (a) and (b)
reveals an overall increase in 
$R_{11}^+=R_{11}/u_\tau^2$
in the direction of development of the streamlines (\ie, increasing $x'$)
and the emergence of a more active outer region. 
This is observed in both finite-span and infinite-span profiles
and exhibits considerable similarity across both types.
This phenomenon is attributed to the adverse-pressure-gradient effects, 
a topic extensively studied in the literature~\citep[cf.][]{spalart:93,perry:02,aubertine:05,monty:11,harun:13,bobke:17,bross:19,pozuelo:22,devenport:22}. 
Despite its significance, this phenomenon is not the primary focus of this study. 
Instead, our attention is devoted exclusively to exploring the differences 
between the finite- and infinite-span boundary layers.

Figure~\ref{fig:pattern-xz} (b) demonstrates
a reduction in the near-wall peak of $R_{11}^+$,
which appears to magnify downstream.
This discrepancy mainly arises
due to the milder adverse pressure gradient experienced by the boundary layer
closer to the tip. 
This is a consequence of the lower effective angle of attack in that region.

\begin{figure}
  \centering
   \begin{overpic}[angle=0,origin=c,width=65mm,clip=true,trim=0mm 0mm 0mm 0mm]{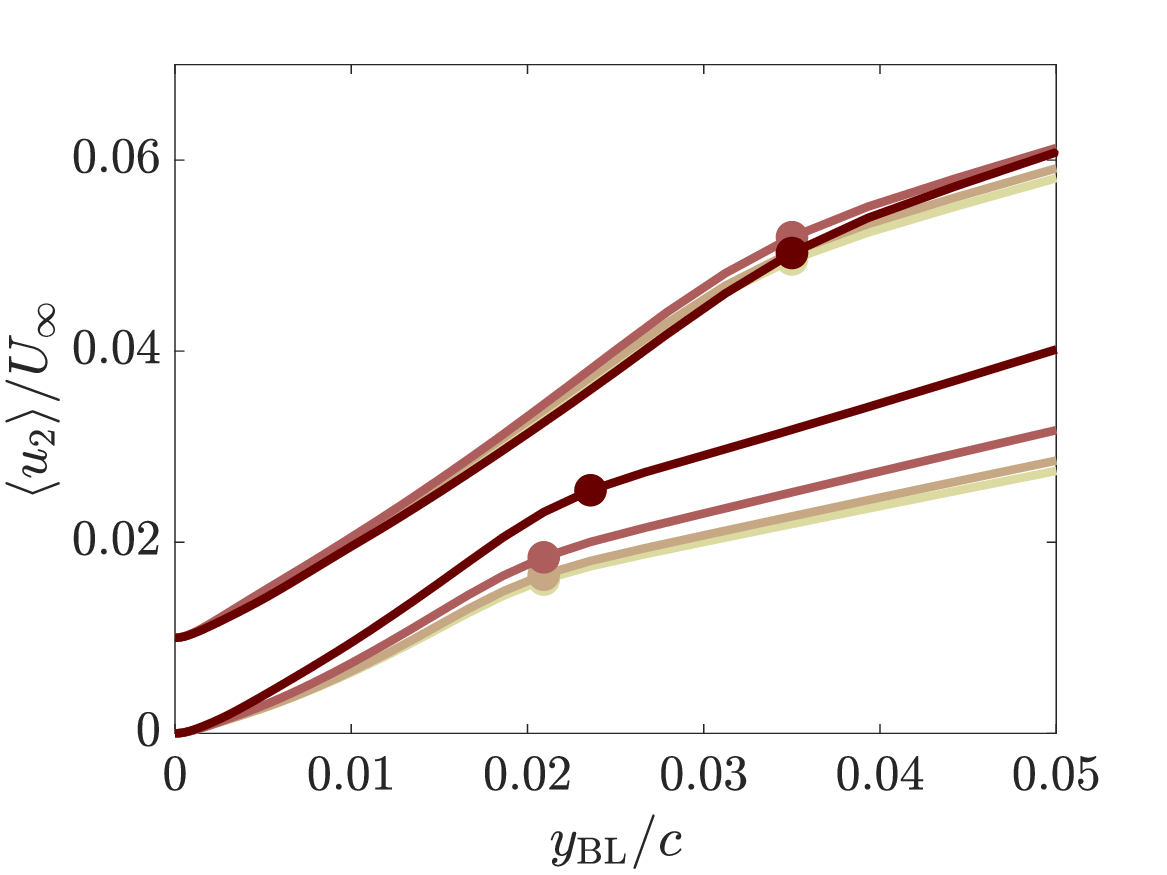}
  \put(83,65){(a)}
  \put(50,22){$x'/c=0.6$}
  \put(16,35){$x'/c=0.9$}
%  \put(39,15){$z'_{\rm start}/c=0.72$}
  \end{overpic}
  \begin{overpic}[angle=0,origin=c,width=65mm,clip=true,trim=0mm 0mm 0mm 0mm]{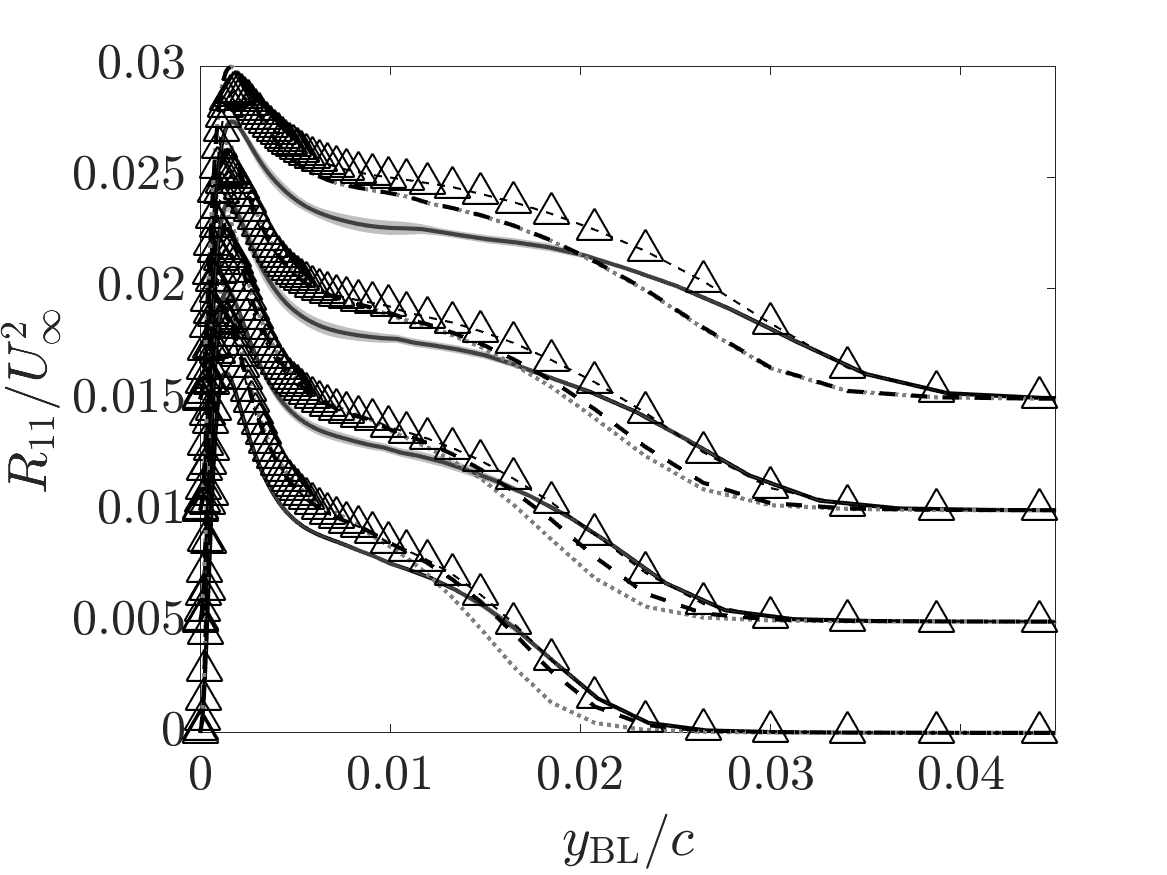}
  \put(40,16){\vector(1,4){12}} 
%  \put(40,3){$\overrightarrow{U}_\infty$}
  \put(55,62){$x'\uparrow$}
  \put(83,65){(b)}
  \end{overpic}
  \caption{
  \label{fig:RS-matched-BL}
  The variation of wall-normal velocity along the span at two streamwise planes (a) 
  and
  the variation of the outer-scaled $R_{11}$ in the streamwise direction (b).
  All solid lines correspond to RWT-5.
  Darker colors in panel (a) denote locations closer to the tip at the intersection of the same streamlines 
  used in figure~\ref{fig:pattern-xz}.
  Profiles shows in panel (b) are the same plotted in figure~\ref{fig:pattern-xz} (b),
%  with solid lines corresponding to the finite-span boundary layers,
  with dotted lines corresponding to matched $x'$ locations from P-2,
  and dashed lines corresponding to 
%  their respective two-dimensional profiles at 
  matched $Re_\tau$ and $\beta_{x_\tau}$ (simple dashed lines)
  and matched boundary layer thickness $\delta_{99}$ (dashed lines with triangles),
  both from P-2. 
  Profiles are shifted vertically by $0.01$ units in panel (a) and
  $5\times10^{-3}$ units in panel (b) for visual clarity. 
  Shaded regions in panel (b) correspond to 80\% confidence intervals in the outer-scaled Reynolds stress profiles (due to finite time averaging),
  computed from the non-overlapping batch method.
  Panel (a) has narrow confidence intervals which are not plotted.
  }
\end{figure}
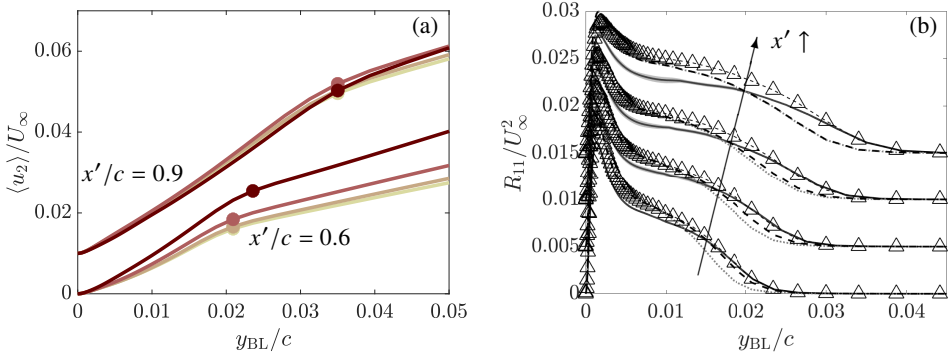

Another trend evident in figure~\ref{fig:pattern-xz} (b) is the 
increased Reynolds stress level in the outer region of the boundary layer,
particularly near its edge.
This effect becomes more pronounced in spanwise locations closer to the tip
(compare figures~\ref{fig:pattern-xz} (a) and (b)),
and it displays a growing spanwise influence as one moves downstream 
(compare figures~\ref{fig:pattern-xz} (c) and (d)).
This phenomenon can be attributed to 
the increased wall-normal velocity %, denoted as $\avg{u_2}_\tau$, 
and its wall-normal gradient %, $\pdline{\avg{u_2}_\tau}{y_\tau}$, 
closer to the wing's tip.
This is illustrated in figure~\ref{fig:RS-matched-BL} (a) for two streamwise locations.
The increased wall-normal velocity results in increased boundary layer growth 
(enhanced $\pdline{\avg{u_1}_\tau}{x_\tau}$ due to the continuity equation)
and an increased boundary layer thickness,
manifesting as a non-zero Reynolds stress at greater distances from the wall.
It is worth noting that the spanwise variation in wall-normal velocity 
diminishes as one moves downstream.
Such observations suggest that the widening spanwise influence of this effect
originates from the upstream state of the boundary layer, 
its flow development towards the root, 
and propagation of non-zero fluctuations in the spanwise direction
(due to mixing and transport).

At these low Reynolds numbers, the development length available for the boundary layers, 
proportional to $c/\delta_{99}$, is quite limited. 
This results in a transient response from the boundary layers 
to the change in their thickness, 
primarily characterized as 
a shift in the location of the outer structures away from the wall,
which propagates throughout the boundary layer thickness as it develops.
This hypothesis is examined in figure~\ref{fig:RS-matched-BL} (b), 
where the profiles from figure~\ref{fig:pattern-xz} (b) are shown in outer units. 
%In this figure, %we contrast 
These profiles are contrasted in the figure
with their infinite-span counterparts 
at matched $(Re_\tau, \beta_{x_\tau})$ 
and matched $\delta_{99}$. 
A shift in the location of the outer structures away from the wall would manifest itself 
as a coincident Reynolds stress profile near the boundary layer edge, 
which is indeed the observed behavior in figure~\ref{fig:RS-matched-BL} (b).
The wall-normal extent of agreement around the boundary layer edge
decreases downstream.
This phenomenon may be attributed to the development of the boundary layers 
and their gradual response to the changes induced by the increased boundary layer thickness. 
Further implications of this behavior are discussed later.

The spanwise variation of $R_{11}^+$ is illustrated in figure~\ref{fig:pattern-xz} (c) and (d). 
A trend of diminishing magnitude in the near-wall peak of the Reynolds stress is evident, particularly when approaching the tip.
This phenomenon can be attributed to a few underlying mechanisms.

First, the effective angle of attack, which varies in the spanwise direction, diminishes closer to the tip. 
As a result, milder adverse pressure gradients occur near the tip (characterized by lower \(\beta_{x_\tau}\) values). 
This leads to a lower increase in the inner-scaled Reynolds stress profile 
and the turbulent kinetic energy due to adverse pressure gradients
(in their development in the streamwise direction). 
Consequently, the overall inner-scaled Reynolds stress levels decrease, resulting in a diminished near-wall peak. 
In RWT-10 (see figure~\ref{fig:rwt10-RS-allXZ}), 
there is a larger variation in \(\alpha_{\rm eff}\) and \(\beta_{x_\tau}\) across the span
compared to RWT-5,
which can partially explain the steeper decline of the near-wall peak
in RWT-10.

Closer to the tip, the increased deflection angle near the wall 
results in laminar flow entrainment into the turbulent boundary layers
at their outmost spanwise extent. %of the boundary layers. 
This entrainment is more pronounced in the collateral region $y_{\rm BL}^+\leq 30$
with the highest deflection angles. %, peaking when $y_{\rm BL}^+\leq 30$. 
Consequently, diminished near-wall turbulence activity can be expected close to the spanwise edge
of the turbulent boundary layers.
%especially near the wall. 
Such reductions in turbulence can influence the near-wall peak of $R_{11}^+$ outside areas directly subjected to flow entrainment.
This could be another reason for the observed decline of the near-wall peak for locations plotted in figure~\ref{fig:pattern-xz}.
Further investigations are necessary to evaluate the significance of this mechanism in comparison to others. 

The other mechanism that contributes to attenuation of the near-wall peak of $R_{11}^+$ is linked to the discussed 
increase in growth rate of the boundary layers near the tip. 
This, on the one hand, reduces the production term near the wall through decreased velocity gradients
in the wall-normal direction, 
and on the other, leads to additional transport away from the wall.
Both of these mechanisms contribute towards reducing %the turbulence levels near the wall and
the peak in $R_{11}^+$ profiles.

Another element to consider is the spanwise pressure gradient present within the boundary layers,
quantified by $\beta_{z_\tau}$ in tables~\ref{tab:app-rwt0-vs-p0},~\ref{tab:app-rwt5-vs-p2}, and~\ref{tab:app-rwt10-vs-p5}. 
To take advantage of the findings of~\cite{ALD:20},
an equivalent expression for their parameter $\Pi$
can be established by considering the relation $\pdline{\avg{p}}{x}=\tau_w/\delta$ in fully-developed channel flows 
(where $\delta$ is the channel half-height).
This leads to a definition of $\Pi$ in turbulent boundary layers as:
\beq \nonumber
\Pi \approx \frac{\delta_{99}}{\tau_w} \pd{P_e}{z_\tau} = \frac{\delta_{99}}{\delta^*}\beta_{z_\tau}
\, .
\eeq
This parameter was found to be always below $0.015 Re_\tau$ 
at all locations reported here and in figures~\ref{fig:rwt0-RS-allXZ},~\ref{fig:rwt5-RS-allXZ}, and~\ref{fig:rwt10-RS-allXZ}. 
This is well below the range of the approximate value of $\Pi > 0.03 Re_\tau$ suggested by~\cite{ALD:20}
as the regime in which the effect of spanwise pressure gradients become significant. 
As a result, it is unlikely for this mechanism to have a significant impact
on the attenuation of turbulence and the near-wall peak of the Reynolds stress profile.
It should be noted that the recommended criterion was derived for %under specific conditions: 
higher $Re_\tau$, %(implying a larger separation of scales), 
in the absence of adverse pressure gradients, 
and for an initially fully-developed channel flow.
Whether a violation of these conditions leads to significant departures from the proposed criterion is not entirely clear
and merits further investigation.

%To conclude this section, 
%Before concluding this section,
It is important to acknowledge certain observed departures 
in figures~\ref{fig:rwt0-RS-allXZ},~\ref{fig:rwt5-RS-allXZ}, and~\ref{fig:rwt10-RS-allXZ}
from the trends described above. 
Firstly, RWT-0 exhibits a profile almost identical to P-0 near the root at $x'/c=0.6$ (figure~\ref{fig:rwt0-RS-allXZ} (a)), 
yet later evolves to display lower levels of fluctuations for $y_{\rm BL}^+ \geq 30$ at downstream locations as early as $x'/c=0.7$,
despite having the same boundary layer thickness.
While this is most likely an artifact caused by differences in the trip or different response of the boundary layers to the trip
(e.g., a history effect), 
further investigations are needed to ascertain whether this might be a consequence 
of additional unidentified mechanisms.
Another noticeable departure is found in the case of RWT-10, particularly at $x'/c=0.8$ and $0.9$ 
(figures~\ref{fig:rwt10-RS-allXZ} (k), (l), (o) and (p)), which do not display a decrease in the near-wall peak. 
Possible explanations for this behavior could include the larger variation of $\beta_{x_\tau}$ at those locations, 
the reduced deflection angle at $x'/c=0.8$ and $0.9$ (see figure~\ref{fig:BL_streamline} (c)), 
and the stronger three-dimensional effects in RWT-10.

Before concluding this section,
%Lastly, we wish to 
we should emphasize that the reference profiles chosen for the comparisons in this section 
were specifically selected to eliminate as many causes of difference between the finite- and infinite-span profiles as possible. 
This ensures that the observed differences primarily arise from a few key mechanisms. 
However, this selection also results in smaller differences between the compared profiles, 
which may not be fully representative of the real discrepancy. 
This point is partially illustrated in figure~\ref{fig:RS-matched-BL}~(b), 
which shows a larger difference in $R_{11}$ profiles when only the streamwise locations are matched. 
It is worth noting that, at the spanwise location selected for this figure, RWT-5 exhibits a reduced effective angle of attack of approximately \(1^\circ\). 
Given that a decreased effective angle of attack leads to thinner boundary layers and a higher outer-scaled near-wall peak in $R_{11}$ , 
the actual disparity between the finite- and infinite-span wings is likely greater than what is depicted in the figure.

\section{Summary and conclusions \label{sec:conclusions}}

High-fidelity simulations of finite-span wings 
with a symmetric NACA0012 profile and a rounded wing tip geometry
were performed
at a chord-based Reynolds number of 200\,000
in free-flight conditions.
Three angles of attack ($0^\circ$, $5^\circ$, and $10^\circ$) were considered
and supplemented with infinite-span (periodic) wings at
corresponding geometric and effective angles of attack.
Tripping was used to ensure turbulent boundary layers on both
the suction and pressure sides.
The resulting database 
can be used for a number of academic and engineering applications,
including careful studies of the boundary layers and the wake, 
the wing-tip vortex and its formation, development and interaction with 
the surrounding flow,
or improving the turbulence models
such as the Reynolds-averaged Navier--Stokes (RANS) models
or wall models used in LES.
The main focus here was on the effect of wing-tip vortices and their induced three-dimensionality
on the turbulent boundary layers.
Other aspects of the flow will be studied in future works.

The general flow field of the lift-generating wings (RWT-5 and RWT-10)
was characterized by the presence
of a main wing-tip vortex as well as additional secondary vortices
formed on the suction side of the wings.
These vortices were stronger and formed earlier
for higher angles of attack.
A spanwise pressure gradient was present in all finite-span cases
(a well-known inviscid effect), 
including RWT-0 which does not generate lift.
As a result of this pressure gradient
the boundary layers were deflected towards the root on the suction side
and towards the tip on the pressure side.
The deflection angle was different across the span and chord of the wing, 
but also in the wall-normal direction due to
the faster response of the low momentum fluid 
near the wall~\citep[see][and references therein]{devenport:22}.

Despite the non-canonical nature of these boundary layers,
they could be simplified by
first accounting for the effective angle of attack
and its impact on the pressure gradient imposed on the boundary layers,
followed by 
a (tensor) rotation into
a coordinate system aligned with the direction of wall shear.
This could further simplify the structure of the Reynolds stress tensor
and make it similar to a 2D boundary layer,
where $R_{13} \approx R_{23} \approx 0$. 
Two additional effects were observed and discussed.
Firstly, the variable deflection angle across the boundary layer thickness
means that the fluid particles across a wall-normal line have converged
from different spanwise locations, 
and due to the variable pressure gradient across the span
have different histories.
Secondly, the variable 
wall-normal velocity along the span, with higher values closer to the tip,
leads to a higher growth rate for the boundary layer farther from the root.
Each of these mechanisms has additional consequences.
The normal streamwise Reynolds stress $R_{11}$ (as well as the other components)
of the finite-span boundary layers
%expressed in wall-shear coordinates 
were compared with 
their corresponding profiles from infinite-span wings,
the observed differences were explained,
and the mechanisms at play were discussed.
Some of these differences 
could not be fully explained by the simplified representation here
and thus need further investigations.

Different terms in the transport equation of the turbulent kinetic energy
and Reynolds stresses~\citep[\ie, the budget terms; see][]{pope:00} 
were computed (and are available in the simulation database; see the data availability statement), but
were not shown or discussed here.
This was because no additional insight was gained from those terms. 
For example, while a decrease in the production of $R_{11}$
was observed near the wall, 
this was as much a consequence of the 
decreased near-wall peak of $R_{11}$ as it was a cause.
Similarly, the increase in the production term near the boundary layer edge
and a corresponding increase in the dissipation term
could be attributed to the increased $R_{11}$ in that region 
and the required balance between different terms.
This highlights the need for predictive models in wall turbulence.

It is important to emphasize that the assumptions leading to section~\ref{sec:3D-vs-2D},
and the subsequent conclusions, are not valid 
close to the wing-tip vortex or the trailing edge of the wing.
For instance, 
for spanwise locations slightly closer to the tip than those 
discussed here,
while still inside the turbulent region of the boundary layer,
the near-wall streamlines have entered from 
the laminar flow region near the vortex.
At the trailing edge, 
the boundary layer on the suction side with a spanwise velocity towards the root 
(\ie, $\avg{u_3}<0$)
approaches the boundary layer on the pressure side with its spanwise velocity towards the tip 
($\avg{u_3}>0$).
The two boundary layers also have opposite wall-normal and
different streamwise velocities. 
These complex trailing edge effects were not discussed here.

%Additional work is still required to make use of the findings of this study
%and construct predictive models.

Throughout this work we relied on quantities such as the Clauser pressure-gradient parameter,
with little modification in their formulation or
discussion about their application in 
complex or three-dimensional boundary layers.
Here, the three-dimensional effects were somewhat weak and
we mostly relied on qualitative comparisons;
therefore, the current definitions were deemed sufficient.
Going forward, 
careful studies on the role and optimal definition of such parameters
in 3D boundary layers,
and potentially developing new ones, 
are absolutely essential.

Many of the attributes of 3D turbulent boundary 
layers~\citep[summarized in][]{devenport:22}
were not observed here.
This includes effects such as
depressed wake of the mean velocity profile~\citep[cf.][]{spalart:08:ekman},
reduction in the Townsend's structure parameter~\citep[cf.][]{littell:94},
or a significant change in the pressure-strain term~\citep[cf.][]{ALD:20}. 
This is most likely due to the relatively weak 
variation of the deflection angle in the wall-normal direction, 
and the gradual variation of this and other similar parameters along the streamlines,
both of which allow the boundary layers to recover (or approach) their two-dimensional state.
In terms of modeling,
the weak three-dimensionality of these boundary layers
makes it easier to adapt the current turbulence models
to account for these effects.

The wings of this study had a relatively low aspect ratio of 1.5.
This was a deliberate choice, 
motivated by the scope of this study,
the stronger three-dimensional effects near the tip,
and as a measure to save cost.
For a fixed Reynolds number, a
wing with a higher aspect ratio will 
experience a higher effective angle of attack
and a stronger vortex. 
It will also have a slightly different spanwise variation 
in its effective angle of attack
due to the change in the spanwise distance to the mirror vortex.
A potentially more significant impact of the low aspect ratio is
the (global) blocking effect of the symmetry plane on
the spanwise velocity,
which could lead to higher variations in the deflection angle along the span
or a slightly different location of the wing-tip vortex. 
While it is important to keep these in mind when generalizing the findings of this study, such
effects are most likely secondary to 
the mechanisms discussed here,
and are therefore not expected to change any of the conclusions.

Only straight, rectangular geometries with rounded wing tips were considered for finite-span wings. 
This was another deliberate choice to minimize the potential differences
between the finite- and infinite-span wings
and help facilitate our analysis. 
For that reason, several factors present in realistic wing designs were excluded. 
These include 
more realistic airfoil profiles and tip geometries, 
drag reduction devices,
wing sweep and twist, 
or variable chord and thickness.
In addition,
the spanwise pressure distribution on the wings
was different from the (nearly) elliptic distribution
encountered in realistic wings.
Compressible effects were also not considered here.
It is important to be aware of these differences 
when generalizing the findings of this work.

It is also important
to acknowledge that the present simulations 
were performed at relatively low Reynolds numbers,
which could have an impact on some of the conclusions.
In general, increasing the Reynolds number leads to
thinner boundary layers on the wing,
\ie, smaller values for $\delta_{99}$ and $\delta^*$,
and higher values of wall-shear stress. 
This, in principle,
leads to a decrease in the streamwise and spanwise Clauser pressure-gradient parameters
and a subsequent reduction in the effect of 
both the streamwise and spanwise pressure gradients.
It would also reduce the variation of the deflection angle $\gamma_{\rm stream}$
in the wall-normal direction,
which was the source of variation in flow history.
Furthermore, the spanwise variation of boundary layer thickness would occur
over longer distances in terms of $\delta_{99}$.
In other words, most of the non-equilibrium and three-dimensional effects discussed here
tend to diminish with 
increased Reynolds number.
Of course, these predictions and extrapolations need to be confirmed directly
by additional experimental and numerical studies
of these flows at a wider range of Reynolds numbers
and pressure gradients.

From a computational perspective,
one of the main challenges in studying %three-dimensional 
turbulent boundary layers
with no homogeneous direction (such as the ones studied here)
is the excessively long integration times required for accurate statistics
in high-fidelity simulations.
Here, we relied on a filtering method along the streamlines
with a variable filter width in the spanwise direction.
%as a function of variation in flow deflection angle.
This was an \emph{ad hoc} choice made based on our prior experience 
and the observed or expected physics of the flow; 
i.e., relation between variations in $\gamma_{\rm stream}$ and the rate of change of solution statistics. 
Given the importance of 3D boundary layers and their prevalence in engineering applications, 
developing more general, more accurate, and more robust methods
for improving statistical convergence is absolutely necessary.
Additionally, designing an optimal computational grid for complex flows and geometries
is an extremely difficult and time consuming task which,
despite the recent advancements in error estimation and grid specification, %~\citep[see][for a survey of proposed criteria]{toosi:caf:20}.
requires further developments.
Addressing such challenges is essential for future studies of 
more realistic three-dimensional boundary layers.

\backsection[Acknowledgements]{
We acknowledge PRACE for awarding us access to HAWK at GCS@HLRS, Germany.
We also acknowledge the EuroHPC Joint Undertaking for awarding this project access to the EuroHPC supercomputer LUMI, 
hosted by CSC (Finland) and the LUMI consortium
through EuroHPC Regular Access and EuroHPC Extreme Scale Access calls.
Additional computations, data handling, and post-processing were enabled by resources provided by 
the National Academic Infrastructure for Supercomputing in Sweden (NAISS) and 
the Swedish National Infrastructure for Computing (SNIC) at PDC 
partially funded by the Swedish Research Council through grant agreements no. 2022-06725 and no. 2018-05973.
}

\backsection[Funding]{
This work was supported by the European Research Council (ERC), the Swedish Research Council (VR), and the Knut and Alice Wallenberg Foundation. 
RV acknowledges financial support from the ERC Grant No. ``2021-CoG-101043998, DEEPCONTROL''.
Views and opinions expressed are however those of the authors only and do not necessarily reflect those of the European Union or the European Research Council. Neither the European Union nor the granting authority can be held responsible for them.
}

\backsection[Declaration of interests]{The authors report no conflict of interest.}

\backsection[Data availability statement]{
The data that support the findings of this study 
will be made openly available, upon publication of the manuscript, in the following repository:
\url{https://www.vinuesalab.com/databases/}
}

\backsection[Author ORCID]{
S. Toosi, https://orcid.org/0000-0001-6733-9744; 
A. Peplinski, https://orcid.org/0000-0002-7448-3290;
P. Schlatter, https://orcid.org/0000-0001-9627-5903;
R. Vinuesa, https://orcid.org/0000-0001-6570-5499
}

%\backsection[Author contributions]{Authors may include details of the contributions made by each author to the manuscript, for example, ``A.G. and T.F. derived the theory and T.F. and T.D. performed the simulations.  All authors contributed equally to analysing data and reaching conclusions, and in writing the paper.''}

\appendix

\section{Post-processing, filtering, and averaging \label{app:averaging}}

A total of 44 fields are collected during runtime to compute the statistics.
These include the temporal average of $u_i$ and $p$ (velocity and pressure fields),
$u_i u_j$ (six independent terms),
$u_i u_j u_k$ (ten independent terms),
$p u_i$, $p\pdline{u_i}{x_j}$,
$\pdline{u_i}{x_k} \cdot \pdline{u_j}{x_k}$,
and a few other fields~\citep[see][]{osti_1349052}.
These fields are then used to compute the additional fields 
(such as the first and second derivatives of $\avg{u_i}$ and $\avg{u_i u_j}$)
required for computing the Reynolds stress budgets 
and other statistics, resulting in
a total of 99 additional fields.
Everything so far is done on the original grids of table~\ref{table:grids} using Nek5000's numerical operators.
The remaining operations are point-wise and can be performed on a smaller grid.
Therefore, these 143 fields are interpolated from the original unstructured grids used for simulations
onto a structured post-processing grid 
(with an order of magnitude fewer grid points and covering only the near-wing region)
which has a nearly-uniform streamwise and spanwise spacing of 
$(\Delta x, \Delta z)/c= (12.5,7.8)\times10^{-4}$ 
(selected based on an approximate friction length of $\delta_\nu = 10^{-4}$)
over the majority of the turbulent boundary layer regions of the wing.
This is done to facilitate the next post-processing steps, including the spatial filtering.

The spatial filtering is employed for two main reasons:
(i) to reduce some of the artifacts present in the derivatives of the solution
inherent to spectral-element methods, %~\citep[cf.][]{deville:book,karniadakis:sem:book}, 
which are small in our simulations
but could still impact some of the more sensitive budget terms,
and 
(ii) to act as a form of spatial averaging and reduce the required 
integration time of the finite-span configurations.

In the finite-span wings, the employed spatial filter has a two-dimensional Gaussian kernel with its principal axes  
aligned with the streamlines and the spanwise direction. 
In the periodic wings, the filter has a 
one-dimensional kernel along the streamlines. 
The streamwise filter width is small, and comparable to the size of the spectral elements in that direction,
\ie, $2\sigma_x /c \approx 12.5\times 10^{-3}$ 
(equivalent to around $125\delta_\nu$)
for element sizes of around $7.4 \times 10^{-3}c$.
This is to target the potential artifacts present in element boundaries~\citep[cf.][]{massaro:23}
and is common between the finite- and infinite-span wings.
In the finite-span wings, the Gaussian filter has a 
variable spanwise filter width 
which is a function of both $x'$ and $z'$,
and simultaneously 
targets the artifacts at the element boundaries 
(elements are around $5\times 10^{-3} c$ in size in the spanwise direction)
and acts as a form of spanwise averaging to reduce the required integration time. 
In the periodic wings, the fields are already averaged in the spanwise direction and no 
spanwise filtering is required. 
Note that no filtering is used in the wall-normal direction in either of the periodic or finite-span configurations. 
It is worth mentioning that at first 
we tried to remove the artifacts at the element boundaries by using
a median filter, but it became obvious that it introduced additional artifacts in the budget terms
and was therefore quickly abandoned.

The spanwise filter width of the finite-span wings is defined based on the variation 
in the streamline deflection angle $\gamma_{\rm stream}$ at the edge of the boundary layer (see figure~\ref{fig:BL_deflection}). 
The main reason for using the value of $\gamma_{\rm stream}$ at the edge of the boundary layer 
(as opposed to the arguably more important value at the wall, or somewhere inside the boundary layer)
is for its relatively low uncertainties due to time-averaging errors.
Additionally, the values of $\gamma_{\rm stream}$ at the wall and inside the 
boundary layer are directly related to the value at the edge.
These two reasons made $\gamma_{\rm stream}$ at the boundary layer edge 
a suitable quantity for our purpose.
The spanwise filter width is defined by the following procedure: 
(i) identifying the variation of $\gamma_{\rm stream}$ at the boundary layer edge across $z'$ from the root $z'/c=0$ 
up to a location where the flow is identified as a turbulent boundary layer 
(excluding the region strongly affected by the wing-tip vortex 
or laminar flow entrance),
(ii) dividing the variation into 10 equal intervals, finding the location of the dividing boundaries, and computing the size of each interval,
(iii) using a linear fit to the identified interval sizes such that it matches the value exactly at the root and the closest location to the tip
(to ensure a smooth variation of the filter width along the span, and to avoid overshoots and undershoots 
typical of higher-order polynomial fits),
(iv) defining the filter width by requiring that 80\% of the kernel weight
lies within the (interpolated) interval size at that location
(\ie, divide the interpolated interval size by $2 \mathcal{Z}_{0.9}\approx2.56$, 
where $\mathcal{Z}_{0.9}$ is the location of 90th percentile in a normal cumulative distribution function).
With this method the spanwise filter width is different for different flows,
at different $x'$ and $z'$ locations, and on different sides of the wing,
but on the suction side
has typical values of $2\sigma_z /c \approx 12\times 10^{-2}$
at the root
and $2\sigma_z /c \approx 3\times 10^{-2}$
near the tip. 
One could use the approximate relation $\delta_\nu/c \gtrapprox 10^{-4} $ to 
convert these values into wall units.

The implementation of the filter was verified by 
comparing the statistics of P-2 computed using the classical spanwise averaging 
(over the entire span)
to those computed by spanwise filtering,
and making sure that the filtered statistics converge to the spanwise averaged ones
for wider filter widths.

\section{The wing-tip region \label{app:general-vortex}}

\begin{figure}
  \centering
    \includegraphics[height=30mm,clip=true,trim=0mm 29mm 30mm 12mm]{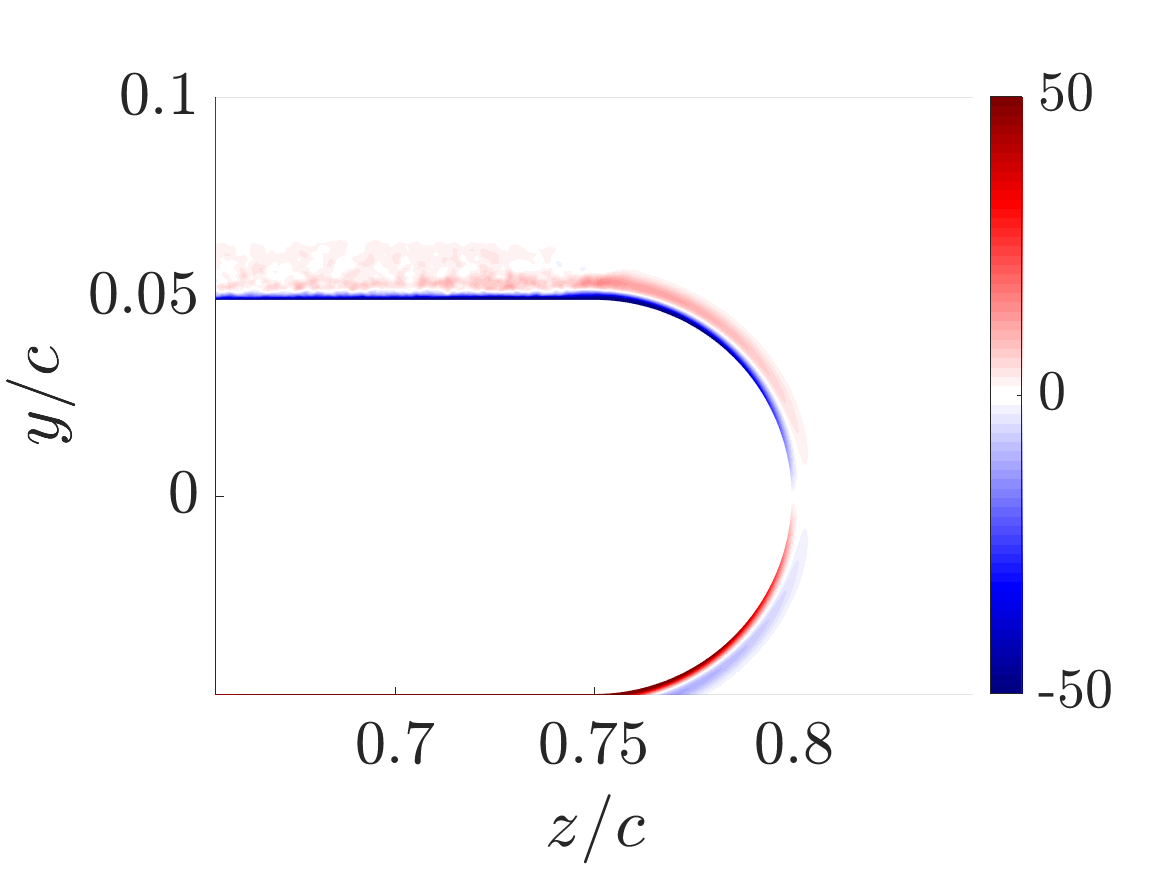}\llap{\parbox[b]{5mm}{(a)\\\rule{0ex}{24mm}}}
    \includegraphics[height=30mm,clip=true,trim=39mm 29mm 30mm 12mm]{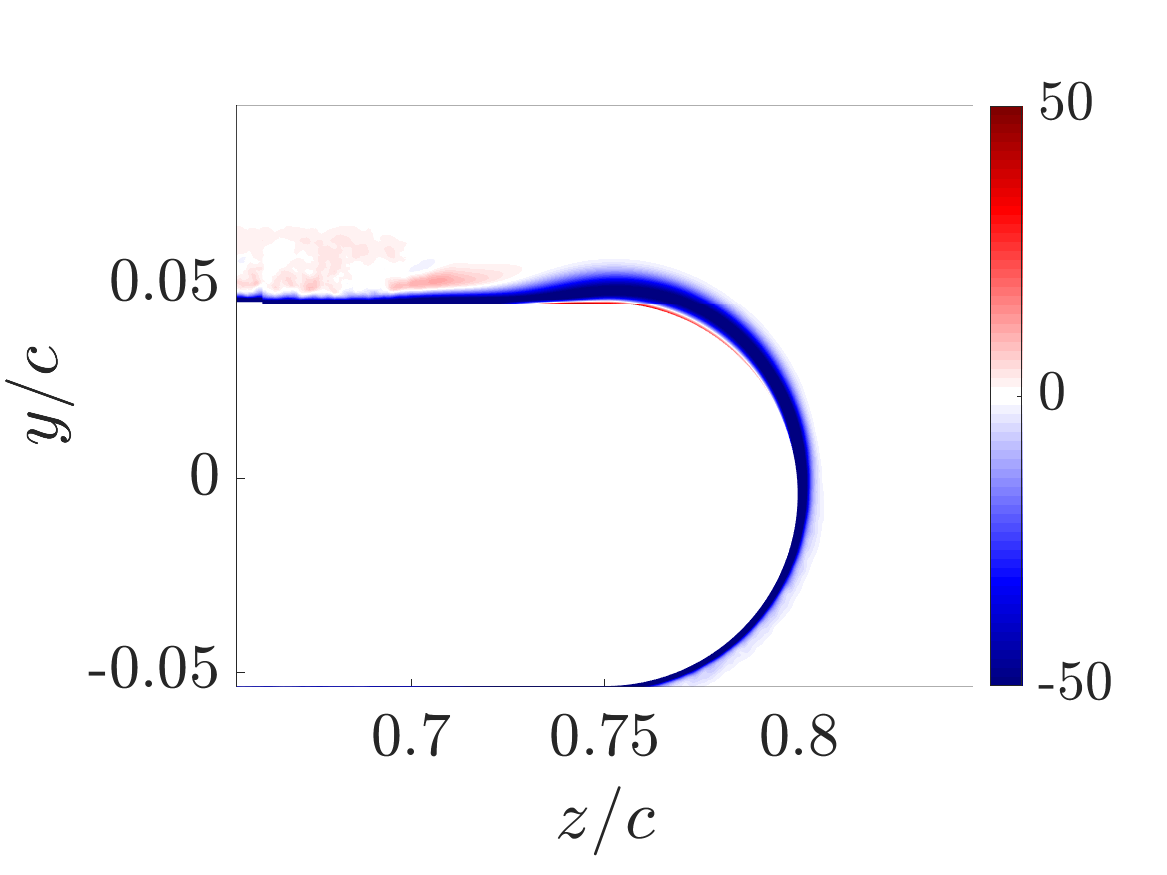}\llap{\parbox[b]{5mm}{(b)\\\rule{0ex}{24mm}}}
    \includegraphics[height=30mm,clip=true,trim=39mm 29mm 30mm 12mm]{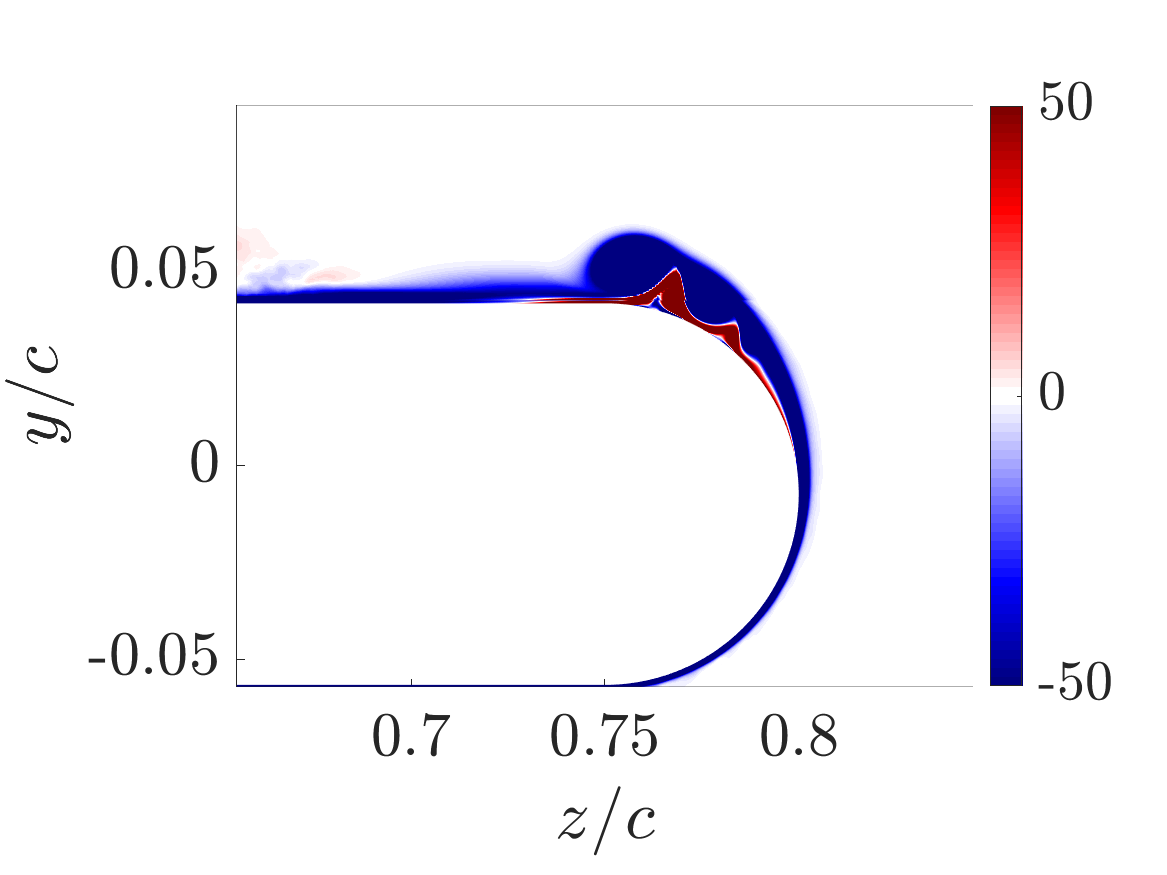}\llap{\parbox[b]{5mm}{(c)\\\rule{0ex}{24mm}}}\\
    \includegraphics[height=30mm,clip=true,trim=0mm 29mm 30mm 12mm]{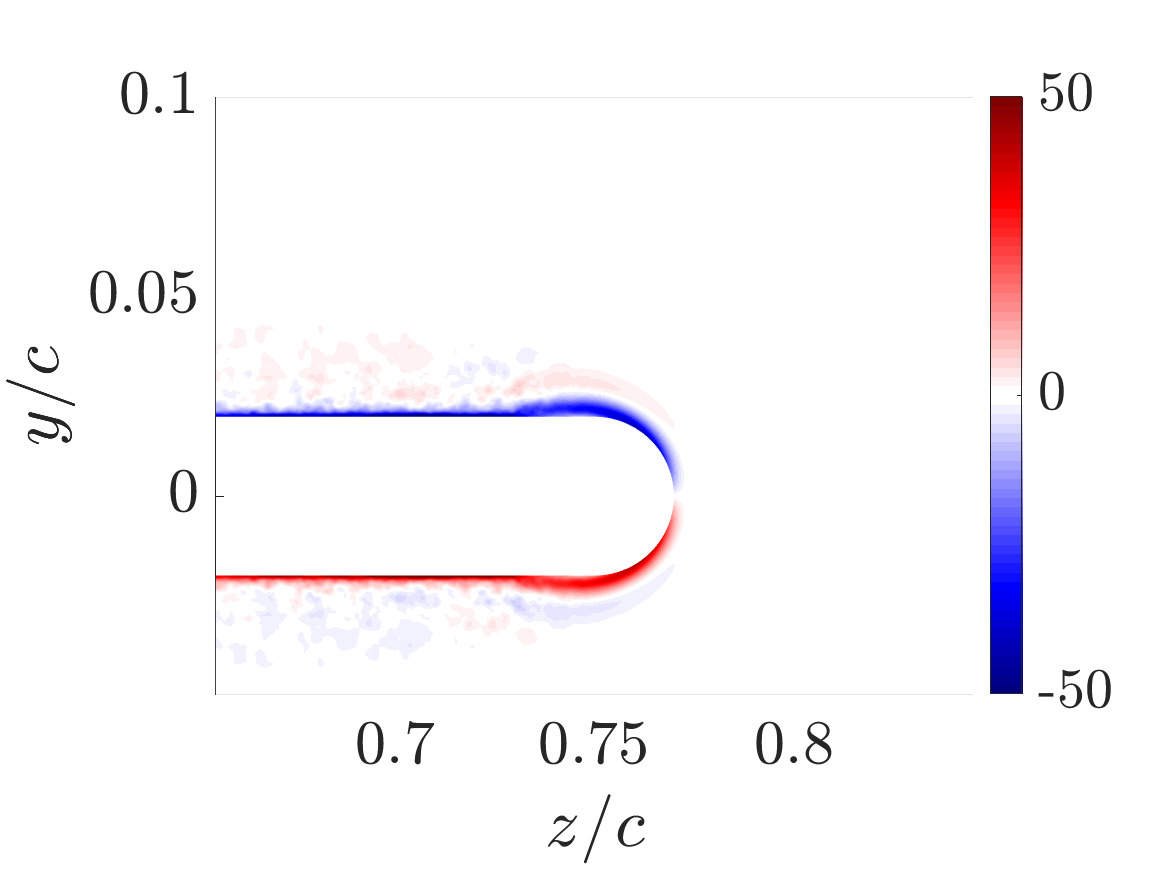}\llap{\parbox[b]{5mm}{(d)\\\rule{0ex}{24mm}}}
    \includegraphics[height=30mm,clip=true,trim=39mm 29mm 30mm 12mm]{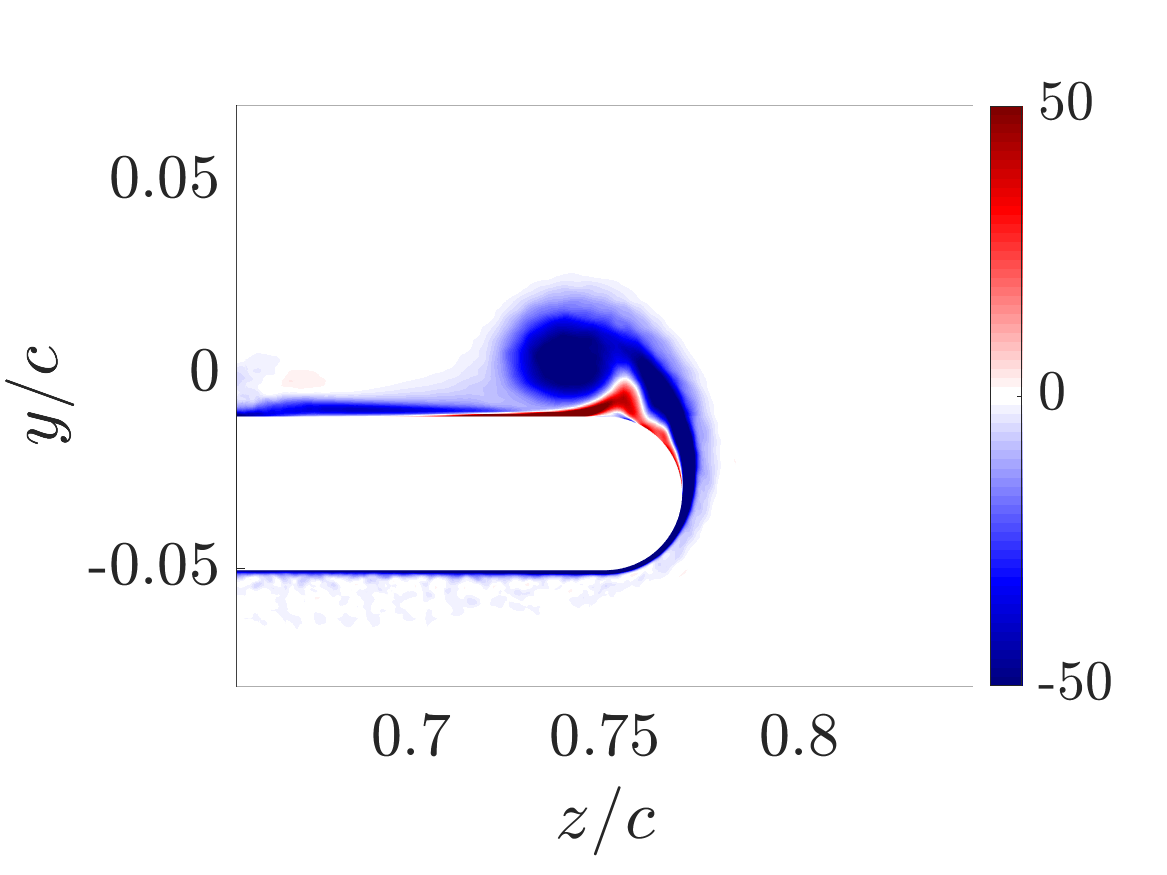}\llap{\parbox[b]{5mm}{(e)\\\rule{0ex}{24mm}}}
    \includegraphics[height=30mm,clip=true,trim=39mm 29mm 30mm 12mm]{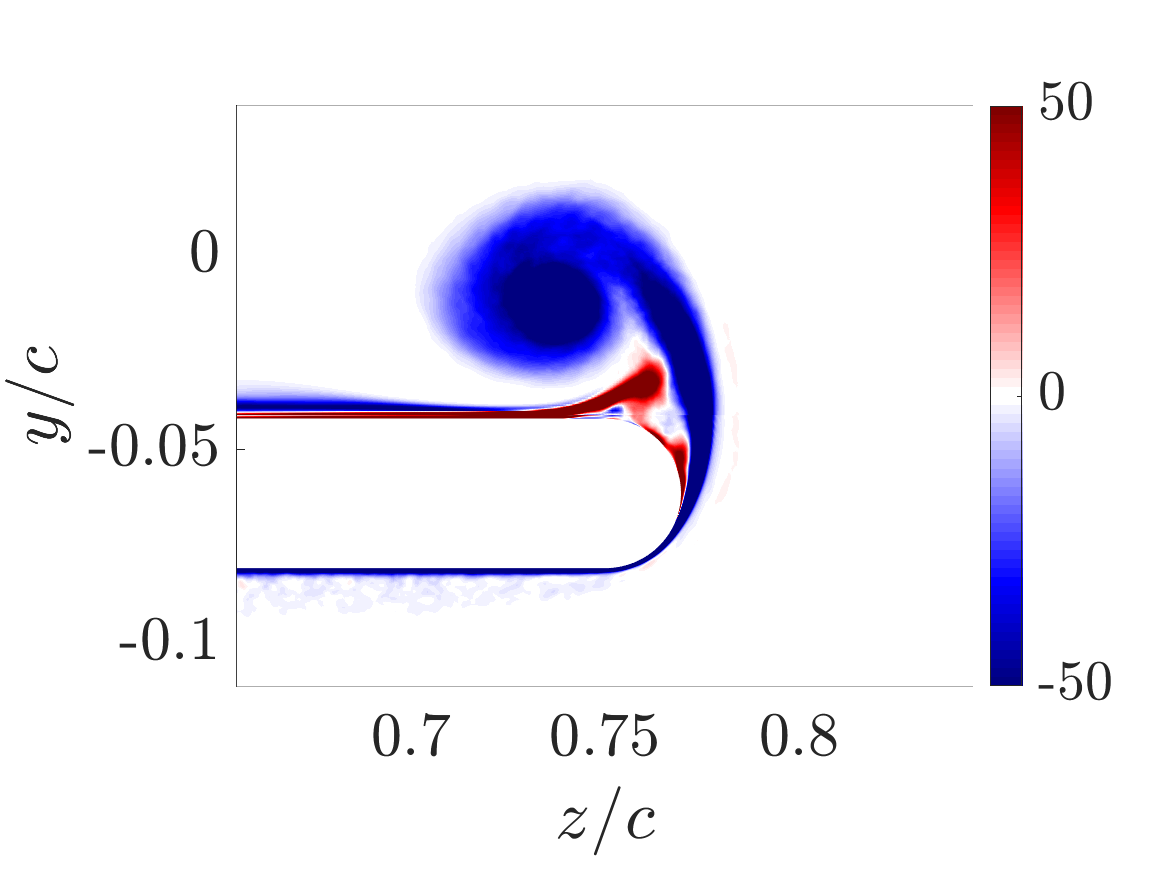}\llap{\parbox[b]{5mm}{(f)\\\rule{0ex}{24mm}}}\\
    \includegraphics[height=38mm,clip=true,trim=0mm 0mm 30mm 12mm]{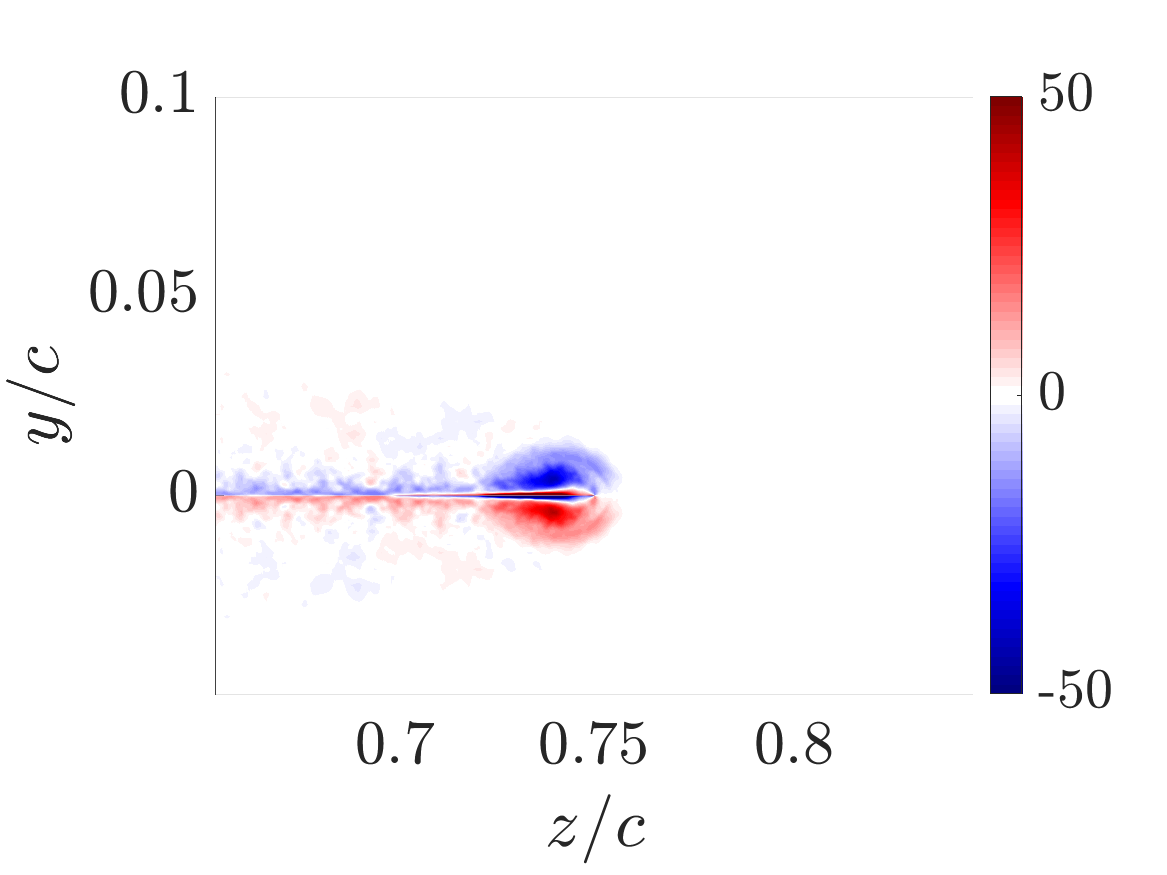}\llap{\parbox[b]{5mm}{(g)\\\rule{0ex}{32mm}}}
    \includegraphics[height=38mm,clip=true,trim=39mm 0mm 30mm 12mm]{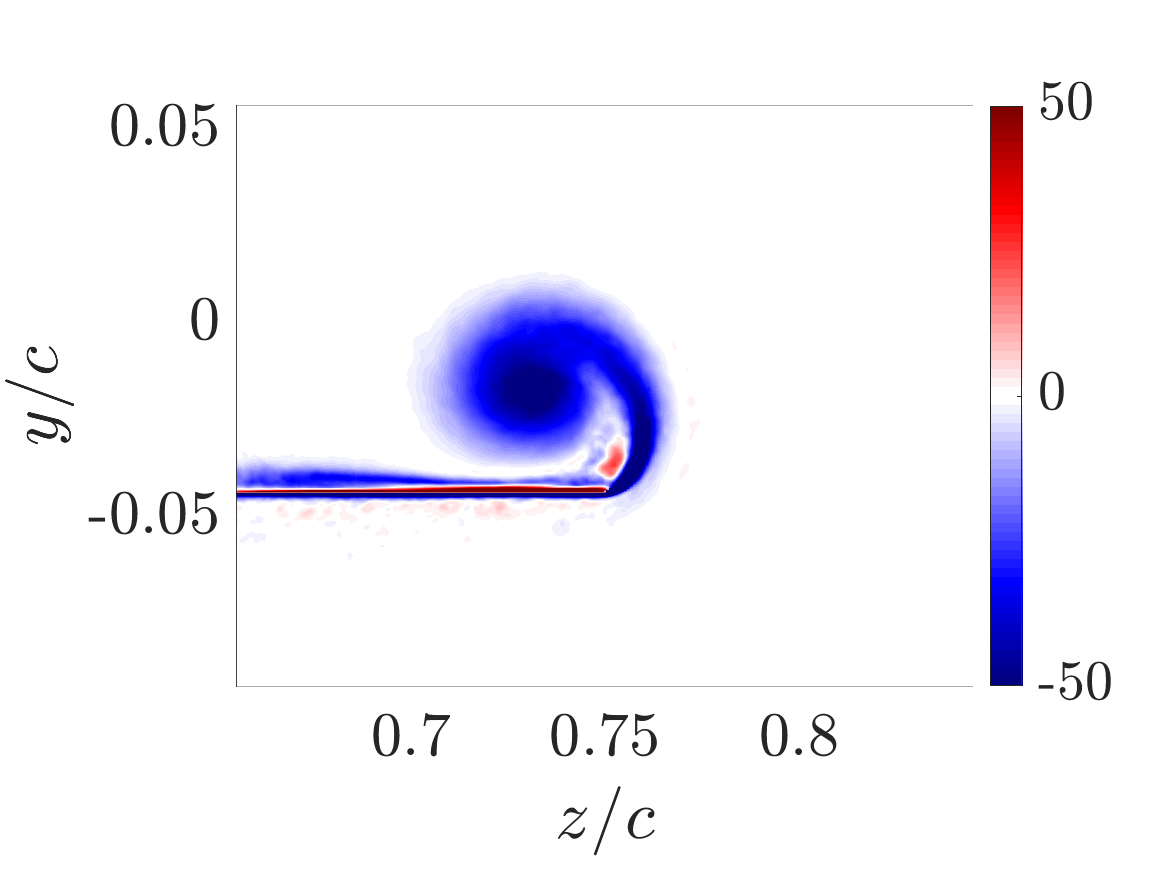}\llap{\parbox[b]{5mm}{(h)\\\rule{0ex}{32mm}}}
    \includegraphics[height=38mm,clip=true,trim=39mm 0mm 30mm 12mm]{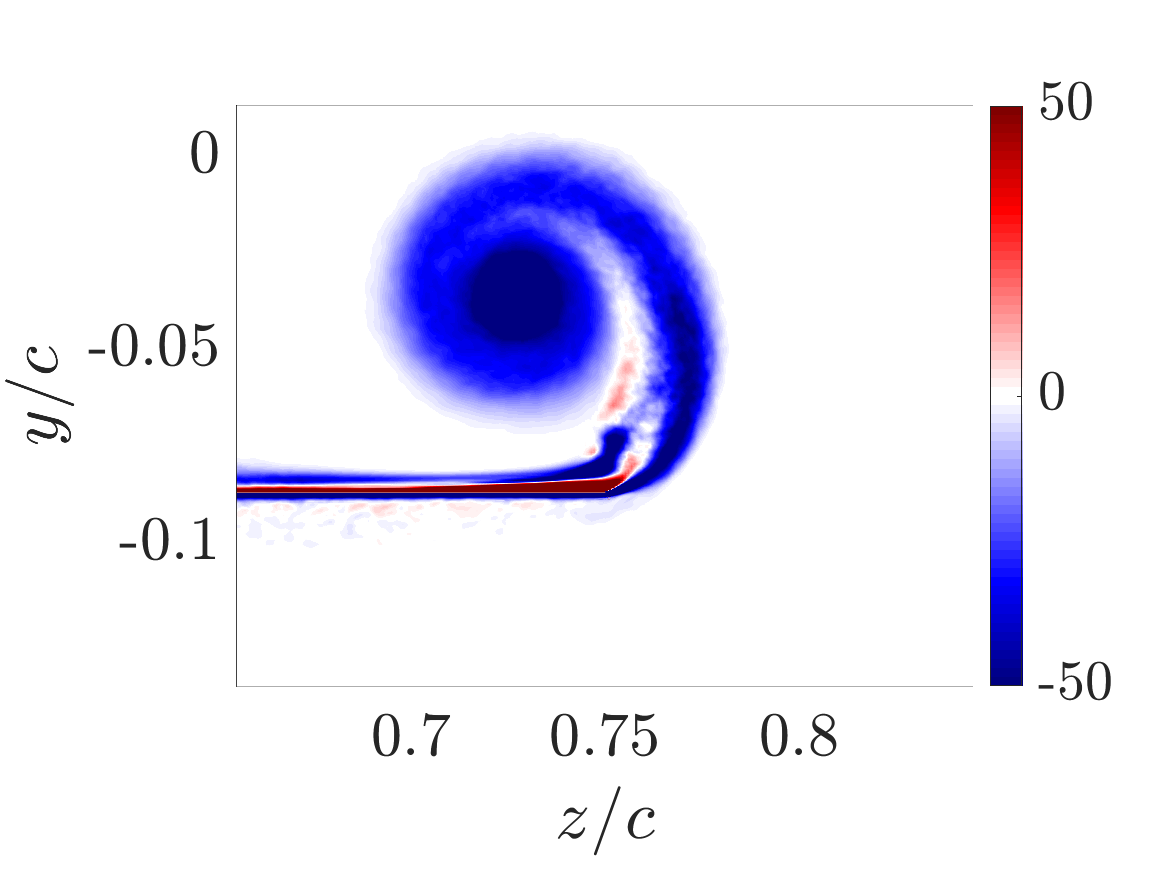}\llap{\parbox[b]{5mm}{(i)\\\rule{0ex}{32mm}}}
  \caption{
  \label{fig:rwt-vorticity}
  Pseudocolor plots of streamwise vorticity in the vicinity of the wing tip for
  (left column) RWT-0, (middle column) RWT-5, and (right column) RWT-10.
  Different rows correspond to
  different locations along the chord at 
  $x'/c=0.55$, $0.85$, and $1.0$, from top to bottom. 
  Colormap is from $\avg{\omega_1}c/U_\infty=-50$ (dark blue) to $50$ (dark red).\\
  }
\end{figure}

Figure~\ref{fig:rwt-vorticity} shows the mean streamwise vorticity of the flow
$\avg{\omega_1}=\avg{\pdline{u_3}{y}}-\avg{\pdline{u_2}{z}}$ 
at a few streamwise locations
near the tip of RWT-0, RWT-5, and RWT-10 wings. 
Here, the wing-tip vortex is characterized by a large negative 
streamwise vorticity region, which is nearly circular in shape
(\ie, approximately homogeneous in the azimuthal direction around the core)
after separation 
from the wing surface (e.g., at the trailing edge, see figures~\ref{fig:rwt-vorticity} (h) and (i)).
In addition to the primary wing-tip vortex,
%which attracts a considerable number of streamlines,
one smaller vortex with an opposite direction of rotation 
can be observed in RWT-5,
and two additional vortices 
(one with opposite rotation, one with the same direction of rotation)
in RWT-10. 
These are the secondary and tertiary vortices 
formed during the formation of the primary (\ie, the wing-tip) vortex
(best visible in figure~\ref{fig:rwt-vorticity} (f)),
and the same vortices identified in figures~\ref{fig:BL_streamline} (b) and (c)
as additional streamline spirals.
The three-dimensionality of RWT-0 and its curved streamlines 
manifest as regions of non-negative streamwise vorticity 
in figures~\ref{fig:rwt-vorticity} (a), (d) and (g).
Interestingly, the streamwise vorticity of RWT-0 near the wall 
(mostly related to the wall-normal variations in the spanwise velocity; \ie, $\pdline{\avg{u_3}}{y}$)
could even exceed that of RWT-5;
for instance, when comparing figures~\ref{fig:rwt-vorticity} (d) and (e).
It is also worth mentioning that 
there is a set of counter-rotating vortices formed at the trailing edge of RWT-0, 
figure~\ref{fig:rwt-vorticity} (g),
as mentioned by~\citet{Giuni:2013}.

% generated by office laptop and display

The pressure gradients 
caused by the pressure difference between the suction and pressure sides of 
the lift-generating wings
lead to a strong flow acceleration from the pressure side towards the suction side.
While this flow remains attached to the surface for lower pressure gradients,
for instance figure~\ref{fig:rwt-vorticity} (b), 
%the presence of viscosity and the introduced vorticity
%eventually lead to flow separation at higher flow accelerations and 
it eventually separates from the wing surface resulting in
the formation of the primary (\ie, wing-tip) vortex,
for instance in figure~\ref{fig:rwt-vorticity} (c).
This in turn leads to the formation of additional (secondary, tertiary, ...) vortices
that could either grow larger, as in figures~\ref{fig:rwt-vorticity} (c) and (f),
or combine with the primary vortex and dissipate, as in figures~\ref{fig:rwt-vorticity} (e) and (h). 

\st{
An important distinction should be made between the vortex formation
described here %compared to 
and that of an inviscid flow. 
In particular, in the absence of viscosity and no-slip boundary conditions at the wall
the wing-tip vortex only separates farther downstream near the tip of the trailing-edge
where the radius of curvature approaches zero~\citep[cf.][]{rizzi:84}.
The change in the location of the vortex 
leads to variations in the induced downwash on the wing. 
As was observed in figure~\ref{fig:BL-LD} 
such variations are rather small away from the vortex core
and become significant only closer to the wing tip. 
Additionally, the formation of secondary and tertiary vortices 
are not observed in inviscid flows.
}

\section{More details on the comparison of boundary layer quantities for finite- and infinite-span wings
\label{app:sec-BL-params}}

The boundary layer histories are approximately matched
by comparing the finite-span wing with a periodic one at a similar 
effective angle of attack, 
and thus pressure-gradient history. 
Here,
the effective angles of attack 
%of the finite-span and periodic wings
are only matched at the root,
and therefore larger differences are expected 
closer to the tip of the wings. 
The remaining differences in tripping,
or the response of the boundary layer to the trip,
can also add to these differences.
%(as observed in section~\ref{sec:BL-alpha}).
These are, however, less significant
on the suction side of the wings considered here.

In order to find an equivalent profile in terms of the local $Re_\tau$ and $\beta_{x_\tau}$,
a location on the periodic wing is found that is closest 
to the $Re_\tau$ and $\beta_{x_\tau}$ values of the
selected location on the finite-span wing
based on a distance on the $(Re_\tau,\beta_{x_\tau})$ plane defined as
\beq \nonumber
d_{(Re_\tau,\beta_{x_\tau})} = 
\sqrt{ 
\left(
\frac{Re_{\tau,{\rm RWT}}-Re_{\tau,{\rm P}}}{Re_{\tau,0}} 
\right)^2
+
\left(
\frac{\beta_{x_\tau,{\rm RWT}}-\beta_{x_\tau,{\rm P}}}{\beta_{x_\tau,0}} 
\right)^2
},
\eeq
where $Re_{\tau,{\rm RWT}}$ and $\beta_{x_\tau,{\rm RWT}}$ are the values
on the finite-span wing for the selected location,
$Re_{\tau,{\rm P}}$ and $\beta_{x_\tau,{\rm P}}$ 
are the values for the periodic wing, %(across the whole chord),
and $Re_{\tau,0}$ and $\beta_{x_\tau,0}$ 
are user-defined values that weight the two quantities 
(which have significantly different values)
when defining the distance.
Here we have chosen  
$Re_{\tau,0}=25$ and $\beta_{x_\tau,0}=0.25$, 
meaning that we expect to see differences between two profiles 
for a variation in $Re_{\tau}$ that is comparable to 25
(in terms of order of magnitude)
and a variation in $\beta_{x_\tau}$ comparable to 0.25.
We effectively assume that
these levels of variation in the two variables
lead to comparable levels of overall departure 
in the studied profiles.

%In addition to the history effects, there are inevitable uncertainties,
%especially in the finite-span wings,
%due to the finite-time averaging of solution statistics
%(see figure~\ref{fig:filter-uncertainty}),
%which should be acknowledged here. 

\begin{figure}
  \centering
 \includegraphics[angle=0,origin=c,height=23mm,clip=true,trim=0mm  22mm 17mm 7mm]{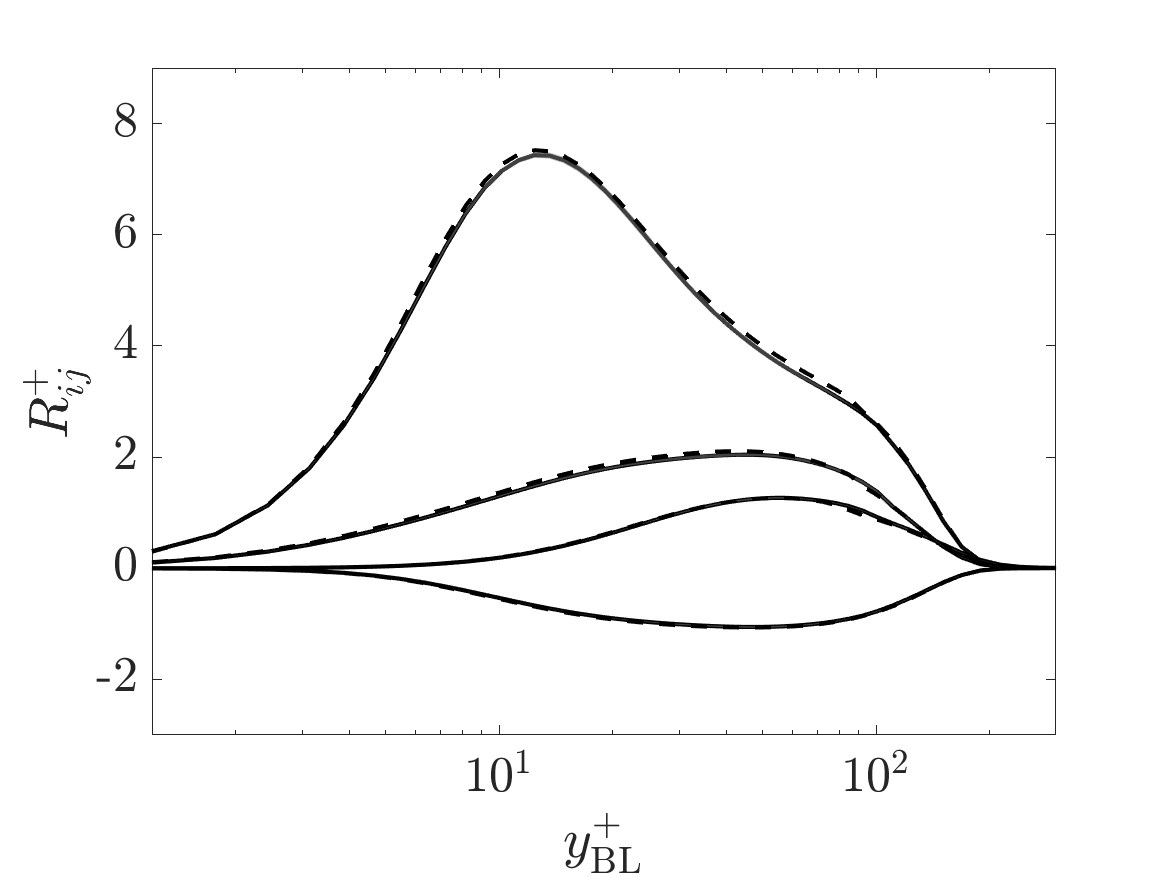}\llap{\parbox[b]{5mm}{(a)\\\rule{0ex}{18.5mm}}}
 \includegraphics[angle=0,origin=c,height=23mm,clip=true,trim=24mm 22mm 17mm 7mm]{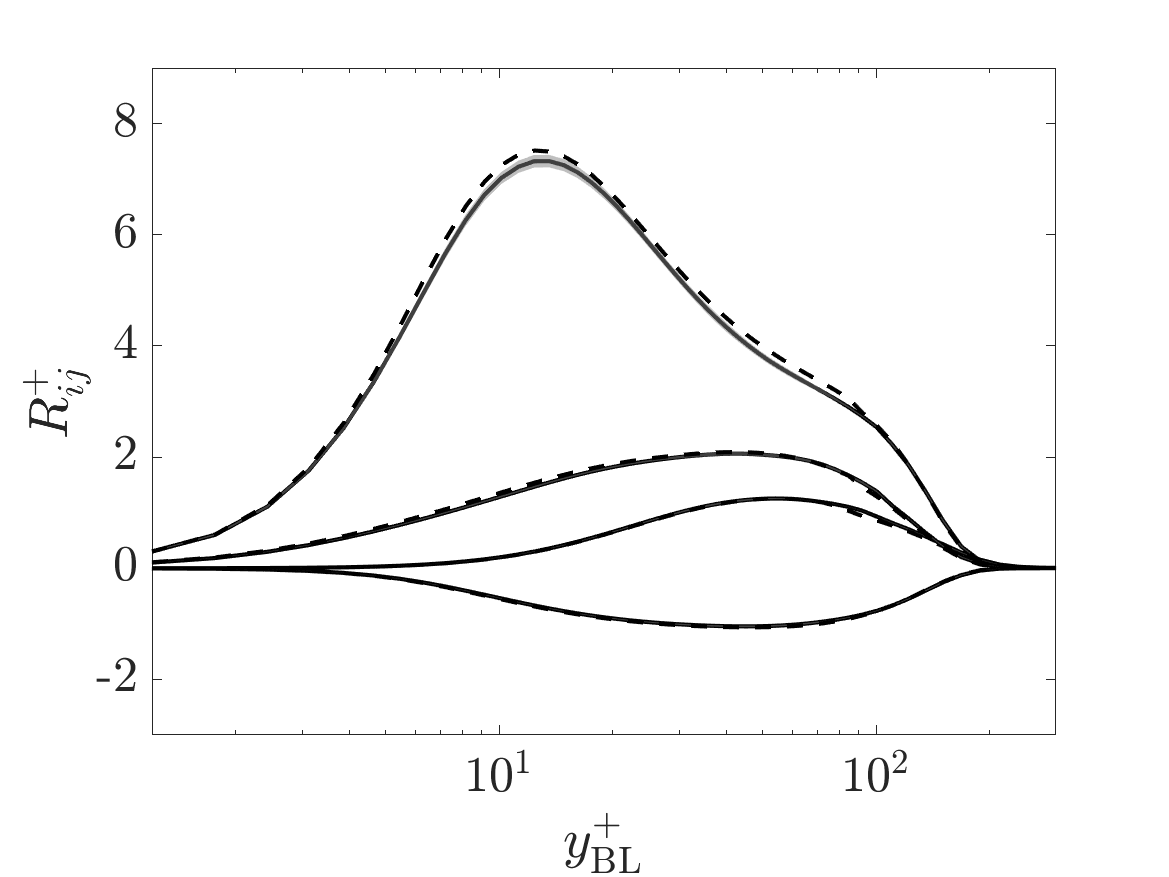}\llap{\parbox[b]{5mm}{(b)\\\rule{0ex}{18.5mm}}}
   \includegraphics[angle=0,origin=c,height=23mm,clip=true,trim=24mm 22mm 17mm 7mm]{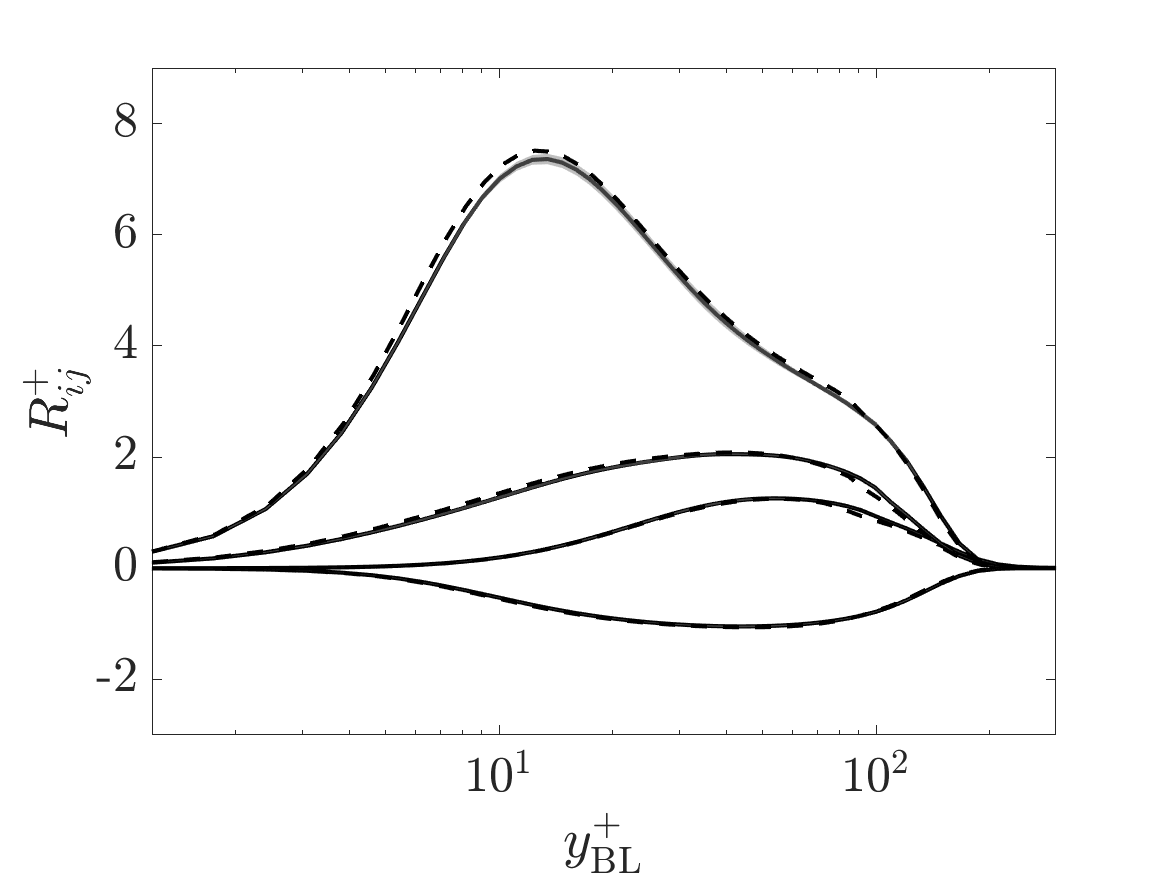}\llap{\parbox[b]{5mm}{(c)\\\rule{0ex}{18.5mm}}}
   \includegraphics[angle=0,origin=c,height=23mm,clip=true,trim=24mm 22mm 17mm 7mm]{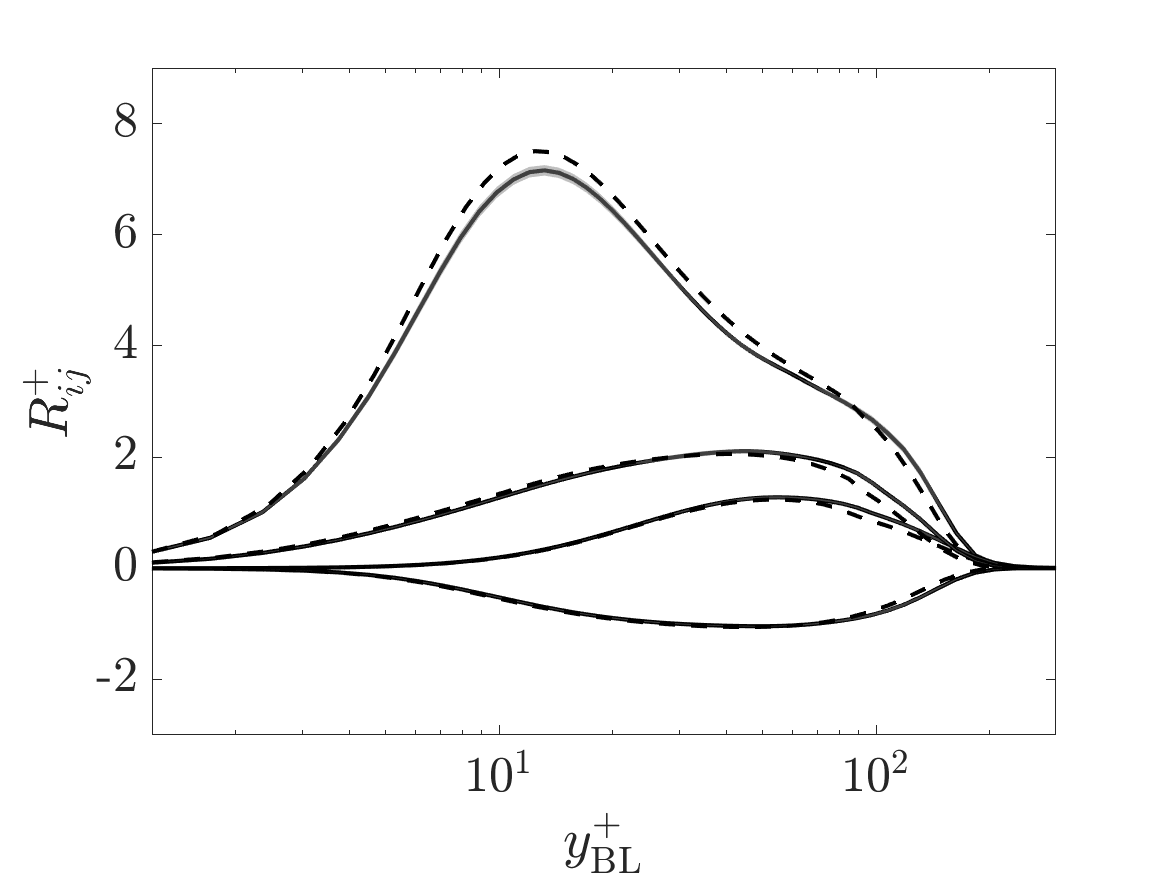}\llap{\parbox[b]{5mm}{(d)\\\rule{0ex}{18.5mm}}}
   \\
    \includegraphics[angle=0,origin=c,height=23mm,clip=true,trim=0mm  22mm 17mm 7mm]{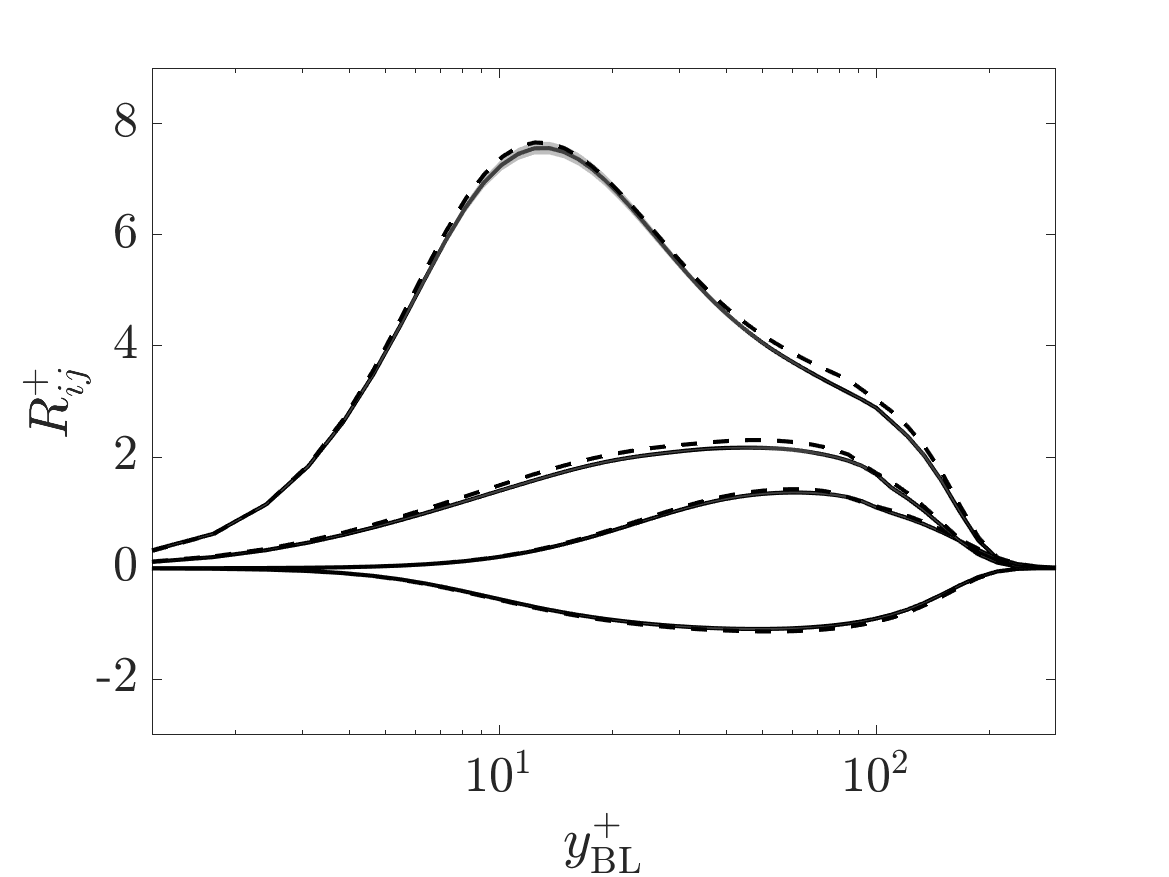}\llap{\parbox[b]{5mm}{(e)\\\rule{0ex}{18.5mm}}}
 \includegraphics[angle=0,origin=c,height=23mm,clip=true,trim=24mm 22mm 17mm 7mm]{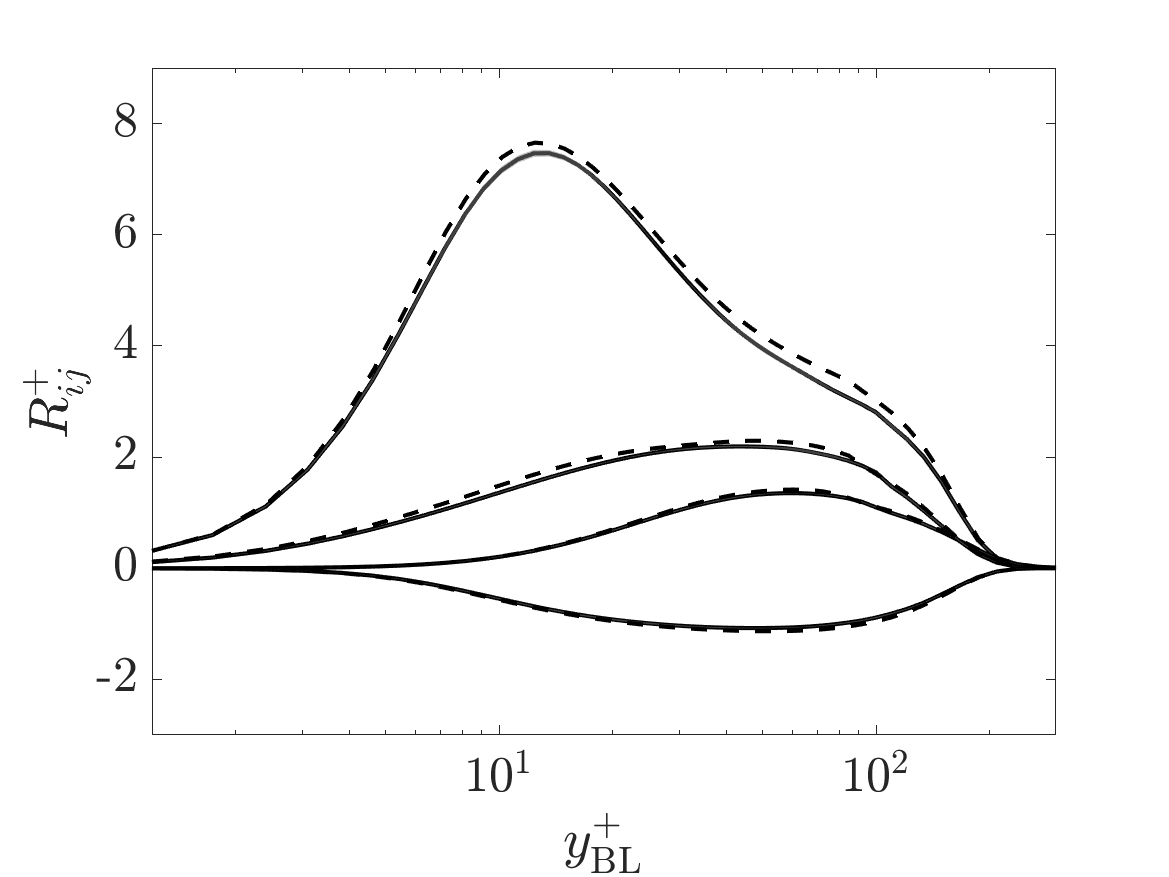}\llap{\parbox[b]{5mm}{(f)\\\rule{0ex}{18.5mm}}}
  \includegraphics[angle=0,origin=c,height=23mm,clip=true,trim=24mm 22mm 17mm 7mm]{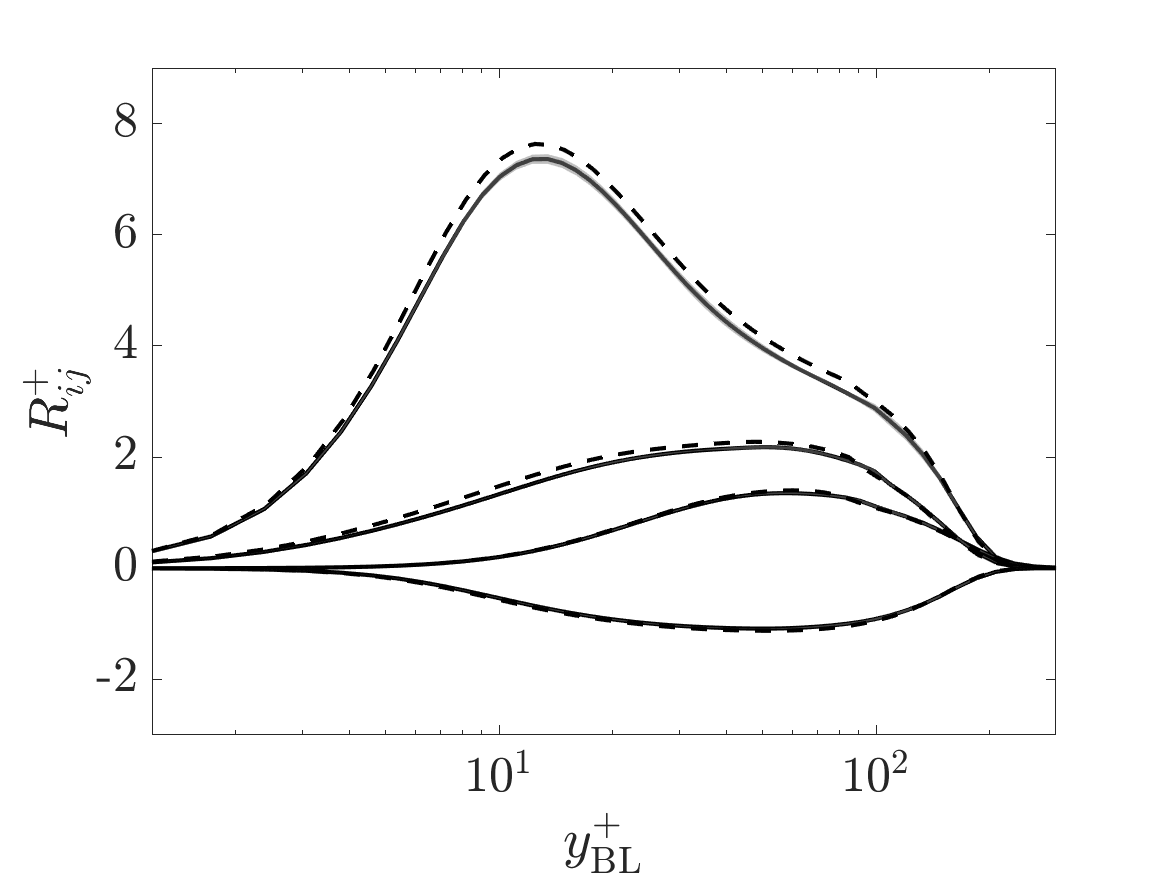}\llap{\parbox[b]{5mm}{(g)\\\rule{0ex}{18.5mm}}}
   \includegraphics[angle=0,origin=c,height=23mm,clip=true,trim=24mm 22mm 17mm 7mm]{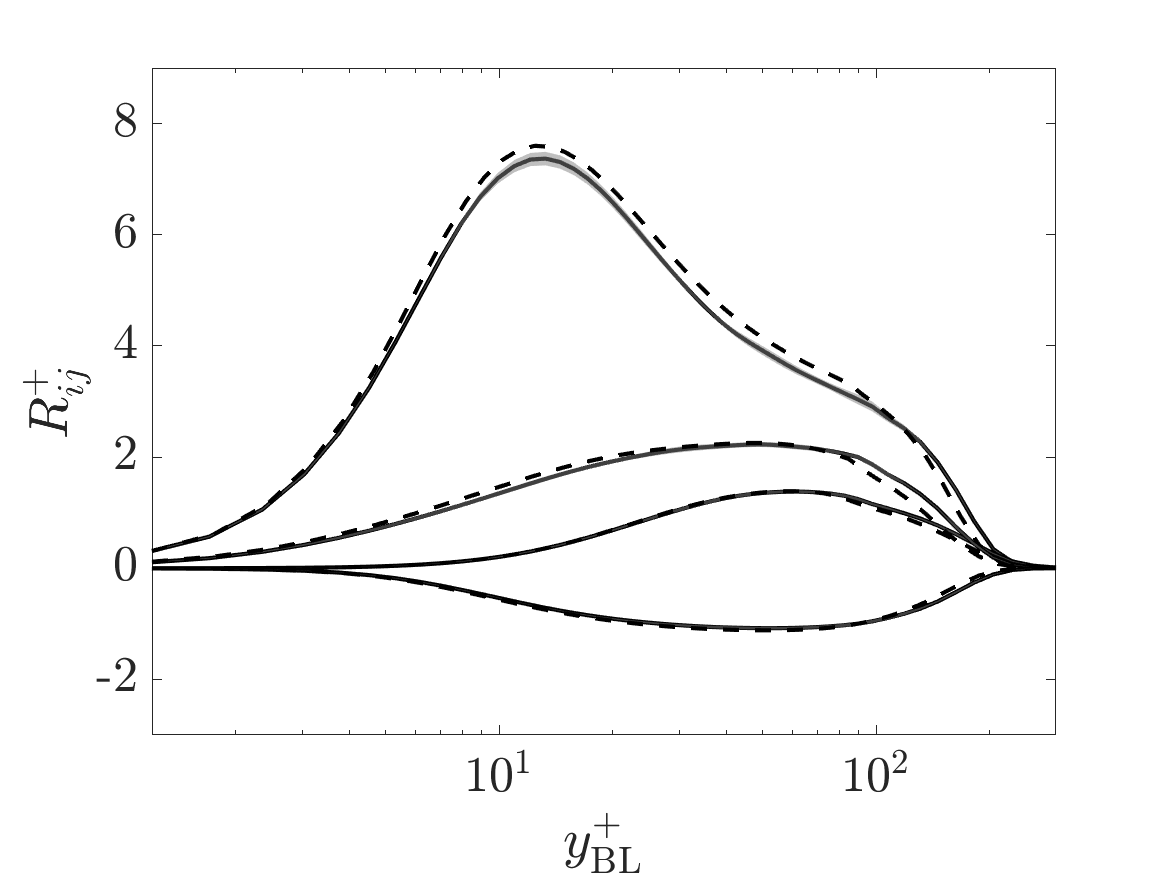}\llap{\parbox[b]{5mm}{(h)\\\rule{0ex}{18.5mm}}}
\\
    \includegraphics[angle=0,origin=c,height=23mm,clip=true,trim=0mm  22mm 17mm 7mm]{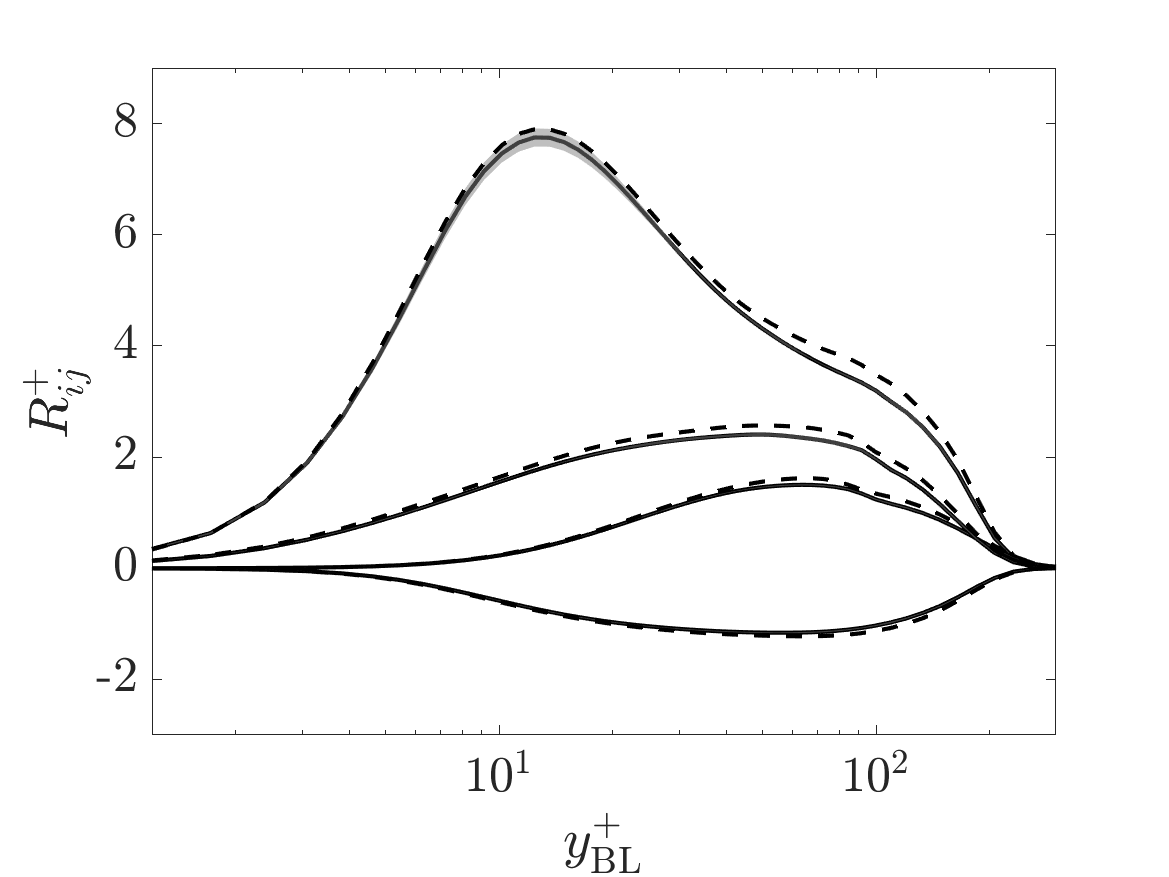}\llap{\parbox[b]{5mm}{(i)\\\rule{0ex}{18.5mm}}}
 \includegraphics[angle=0,origin=c,height=23mm,clip=true,trim=24mm 22mm 17mm 7mm]{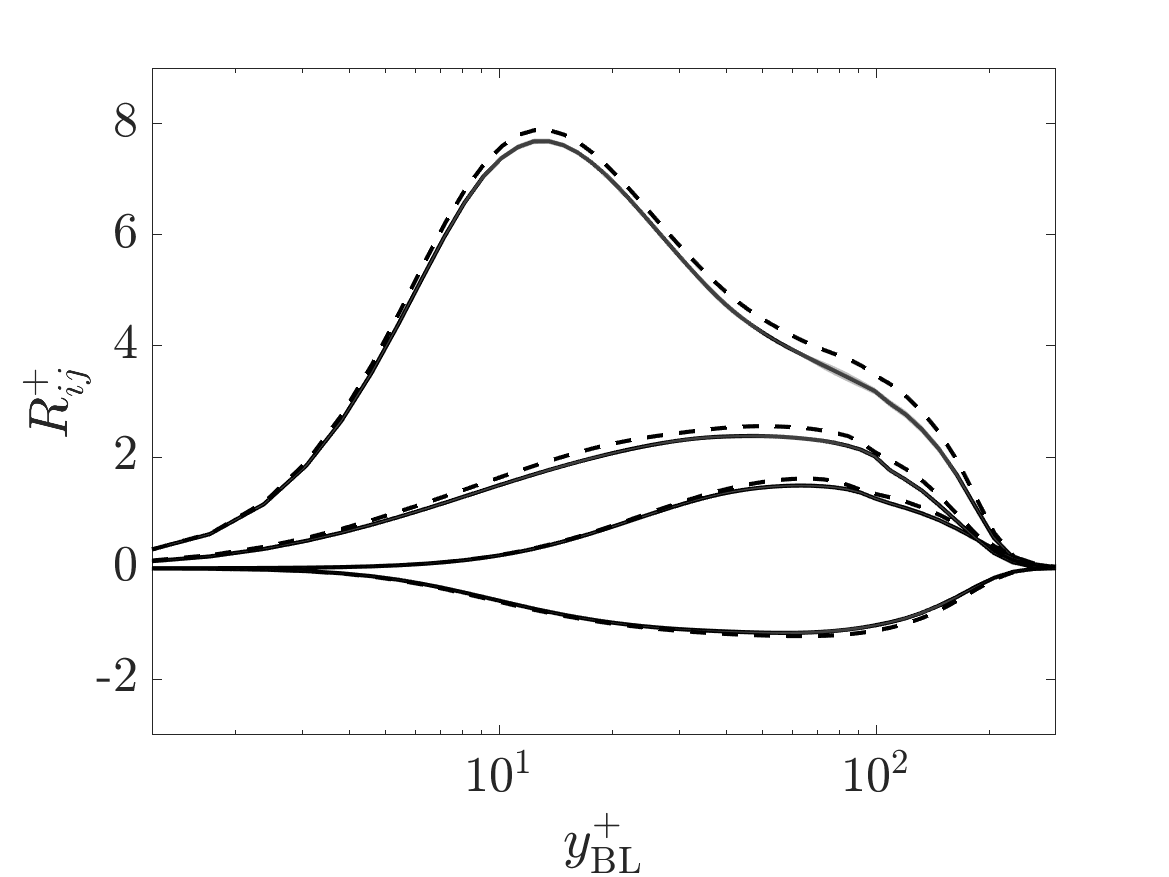}\llap{\parbox[b]{5mm}{(j)\\\rule{0ex}{18.5mm}}}
  \includegraphics[angle=0,origin=c,height=23mm,clip=true,trim=24mm 22mm 17mm 7mm]{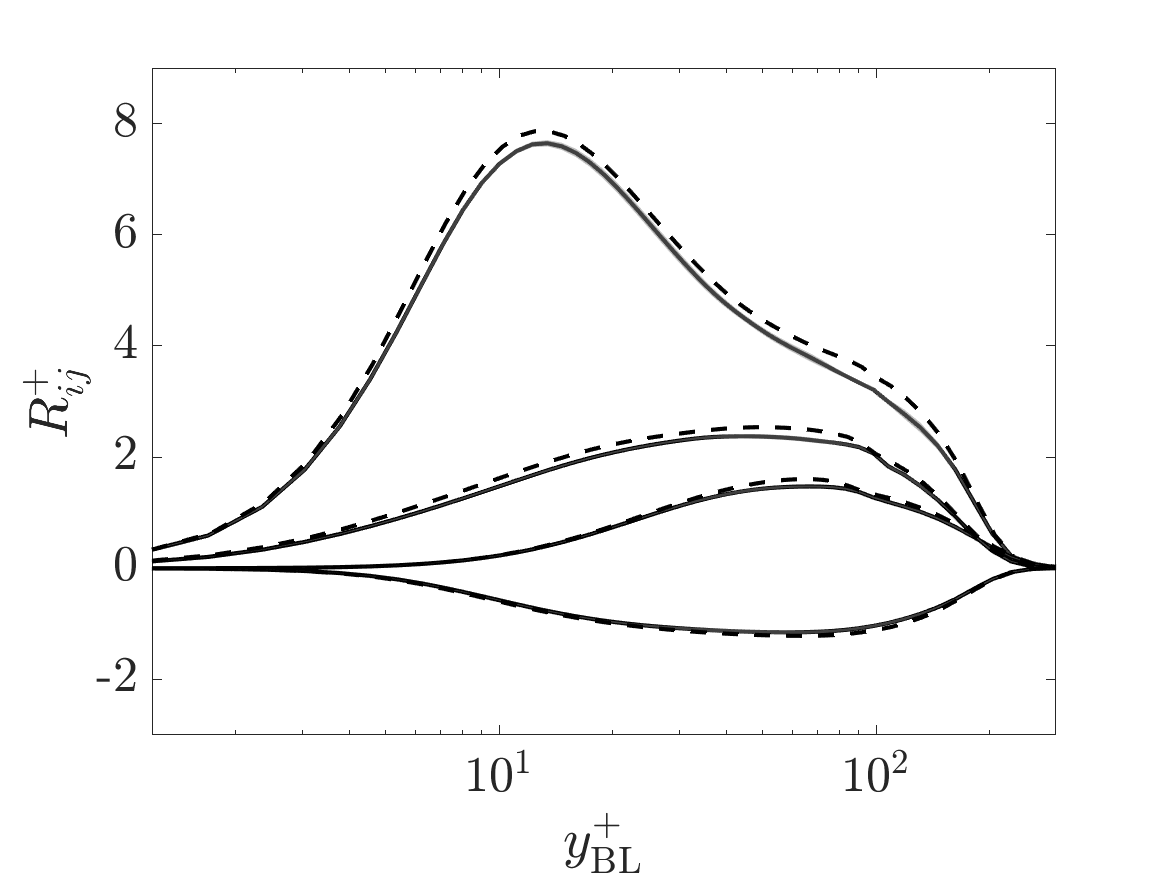}\llap{\parbox[b]{5mm}{(k)\\\rule{0ex}{18.5mm}}}
   \includegraphics[angle=0,origin=c,height=23mm,clip=true,trim=24mm 22mm 17mm 7mm]{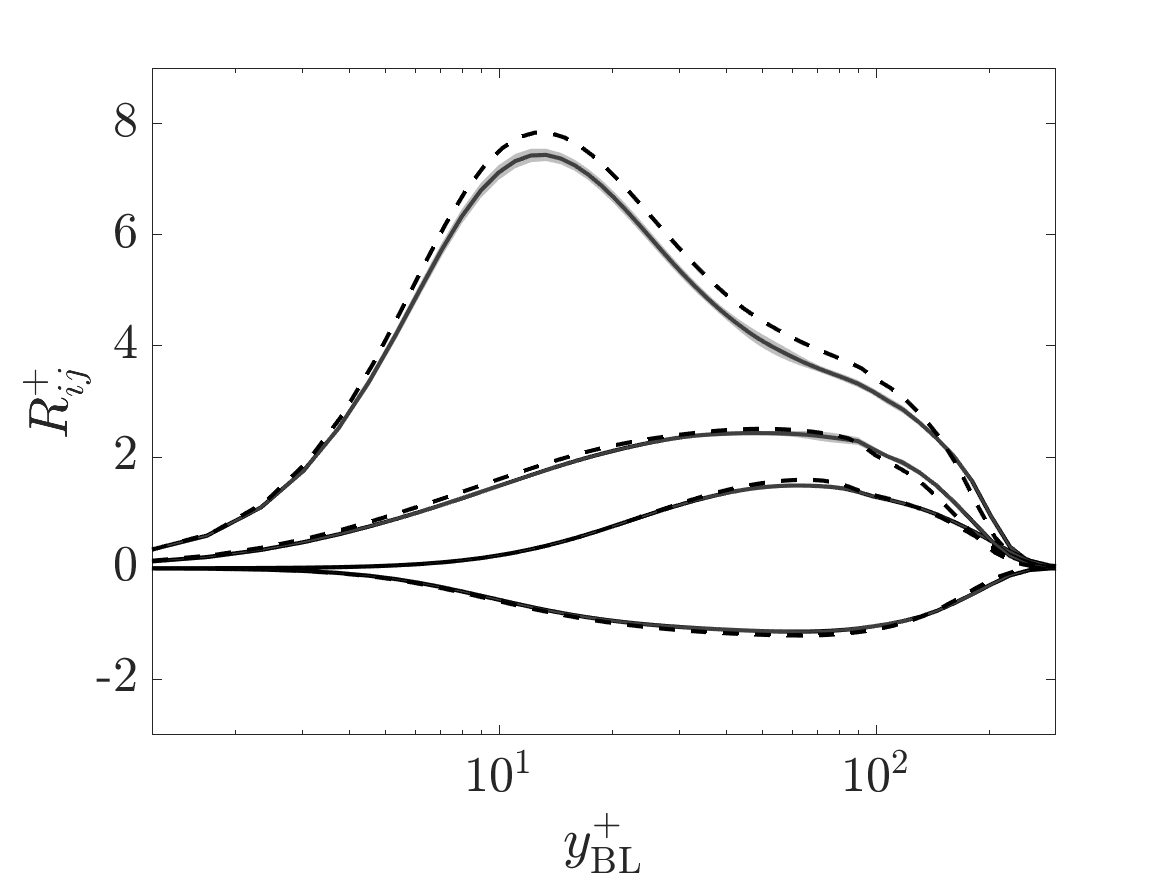}\llap{\parbox[b]{5mm}{(l)\\\rule{0ex}{18.5mm}}}
   \\
    \includegraphics[angle=0,origin=c,height=27.2mm,clip=true,trim=0mm 0mm 17mm 7mm]{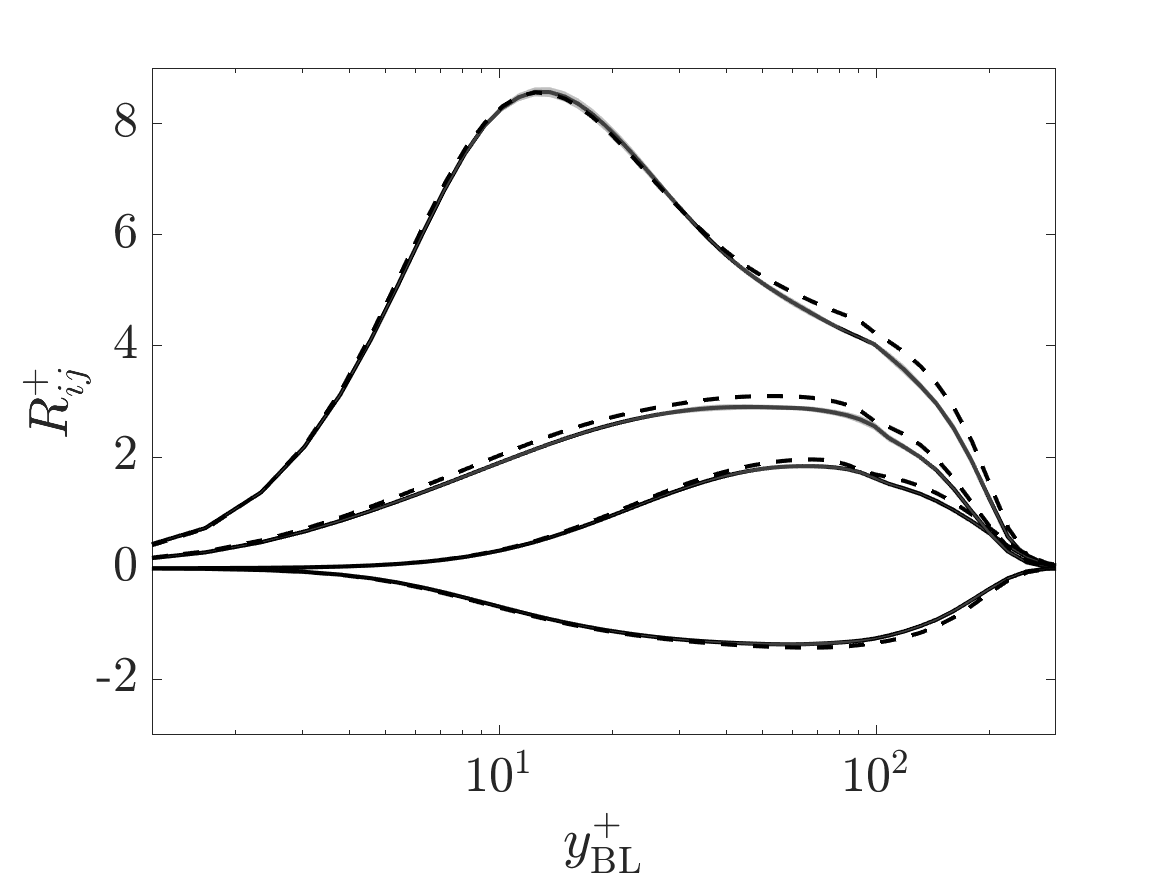}\llap{\parbox[b]{5mm}{(m)\\\rule{0ex}{22.7mm}}}
 \includegraphics[angle=0,origin=c,height=27.2mm,clip=true,trim=24mm 0mm 17mm 7mm]{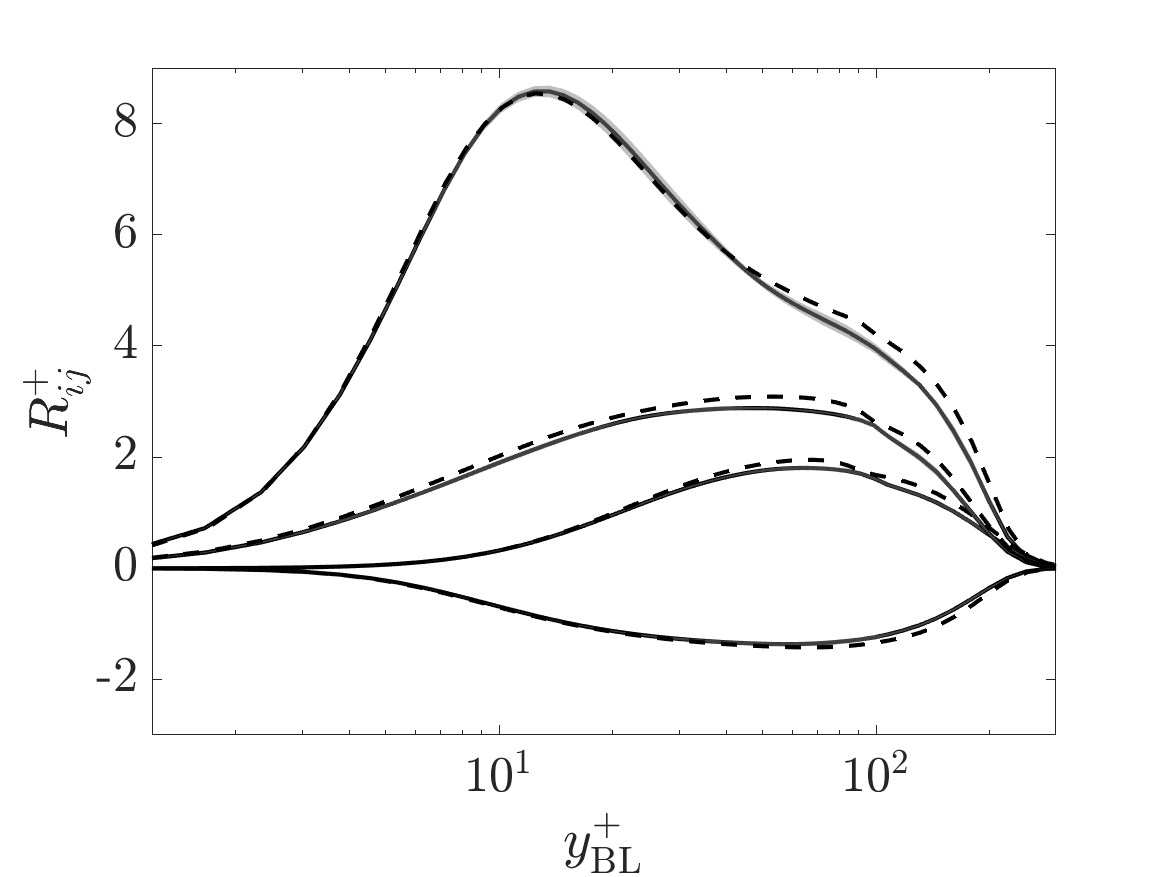}\llap{\parbox[b]{5mm}{(n)\\\rule{0ex}{22.7mm}}}
  \includegraphics[angle=0,origin=c,height=27.2mm,clip=true,trim=24mm 0mm 17mm 7mm]{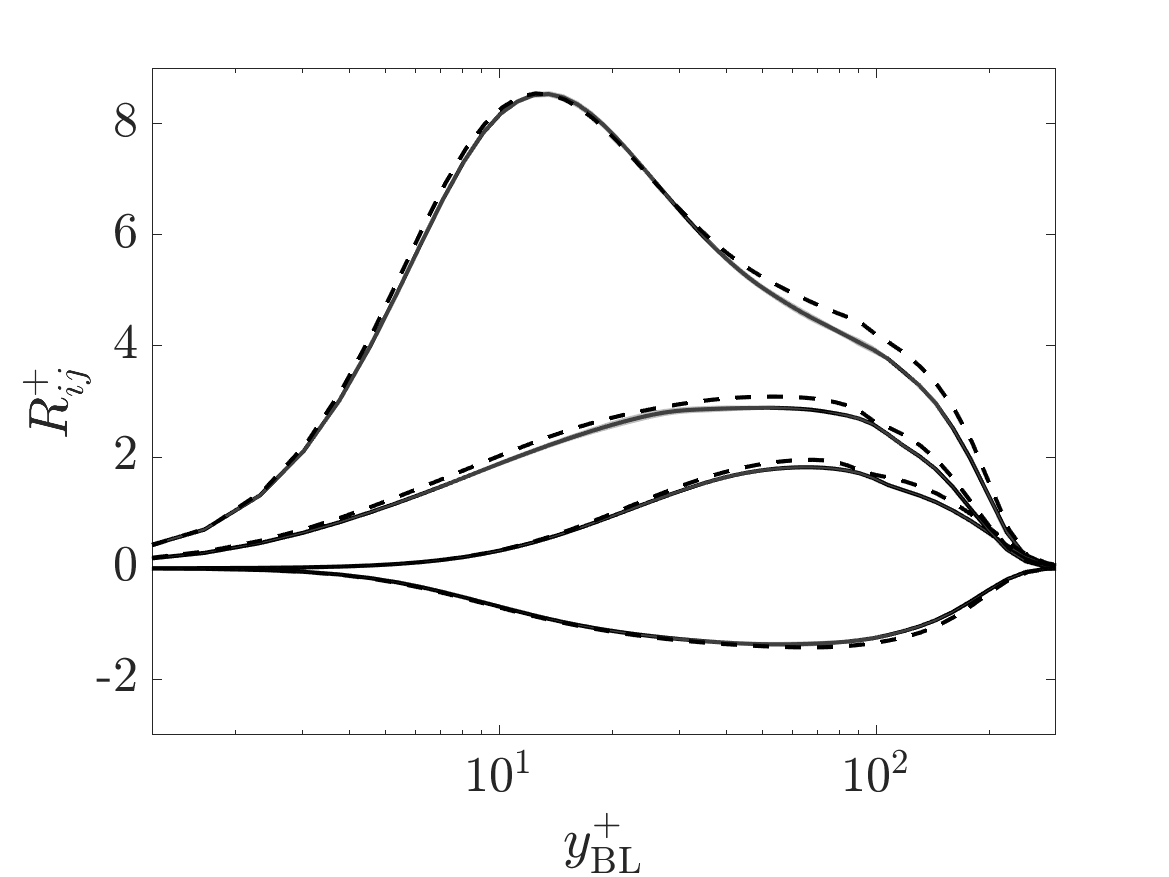}\llap{\parbox[b]{5mm}{(o)\\\rule{0ex}{22.7mm}}}
   \includegraphics[angle=0,origin=c,height=27.2mm,clip=true,trim=24mm 0mm 17mm 7mm]{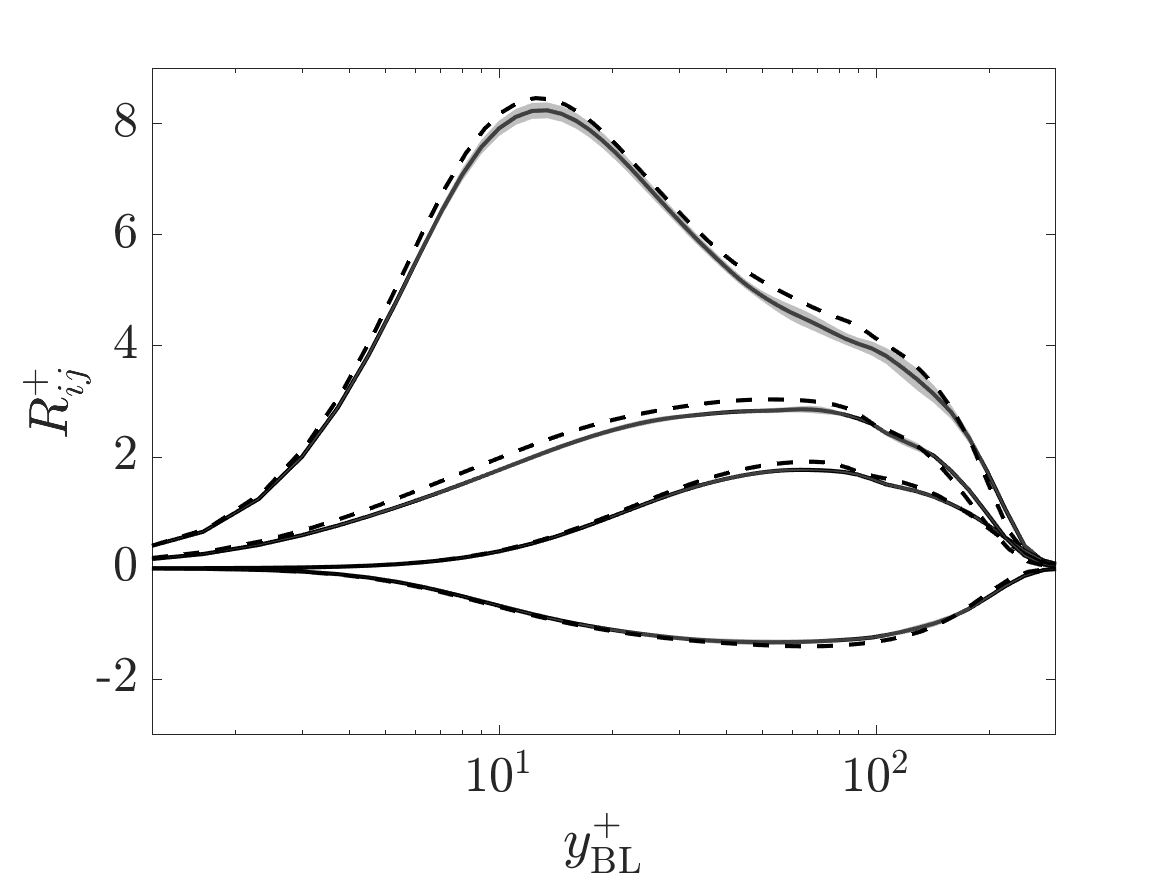}\llap{\parbox[b]{5mm}{(p)\\\rule{0ex}{22.7mm}}}
  \caption{
  \label{fig:rwt0-RS-allXZ}
  Variation of the Reynolds stresses $R_{11}$, $R_{33}$, $R_{22}$, and $R_{12}$
  with distance from the wall
  in RWT-0 (solid lines)
  compared to 
  P-0 (dashed lines) 
  at matched $(Re_\tau,\beta_{x_\tau})$ values.
  Rows from top to bottom correspond to different streamwise locations of
  $x'/c=0.6$, $0.7$, $0.8$, and $0.9$, respectively. 
  Columns from left to right
  correspond to spanwise locations
  of the streamlines near the wall
  (in the collateral region)
  released from $x'_{\rm start}/c=0.12$ (immediately downstream of the trip)
  at $z'_{\rm start}/c=0.1$, $0.31$, $0.51$, and $0.72$, respectively.
  At each location, the Reynolds stress components are expressed in the local
  $(x_\tau,y_\tau,z_\tau)$ coordinate system aligned with the direction of 
  wall-shear stress at that specific location. 
  Shaded regions correspond to 80\% confidence intervals of the inner-scaled Reynolds stress profiles (due to finite time averaging),
  computed from the non-overlapping batch method.
   }
\end{figure}

\begin{figure}
  \centering
 \includegraphics[angle=0,origin=c,height=23mm,clip=true,trim=0mm 22mm 17mm 7mm]{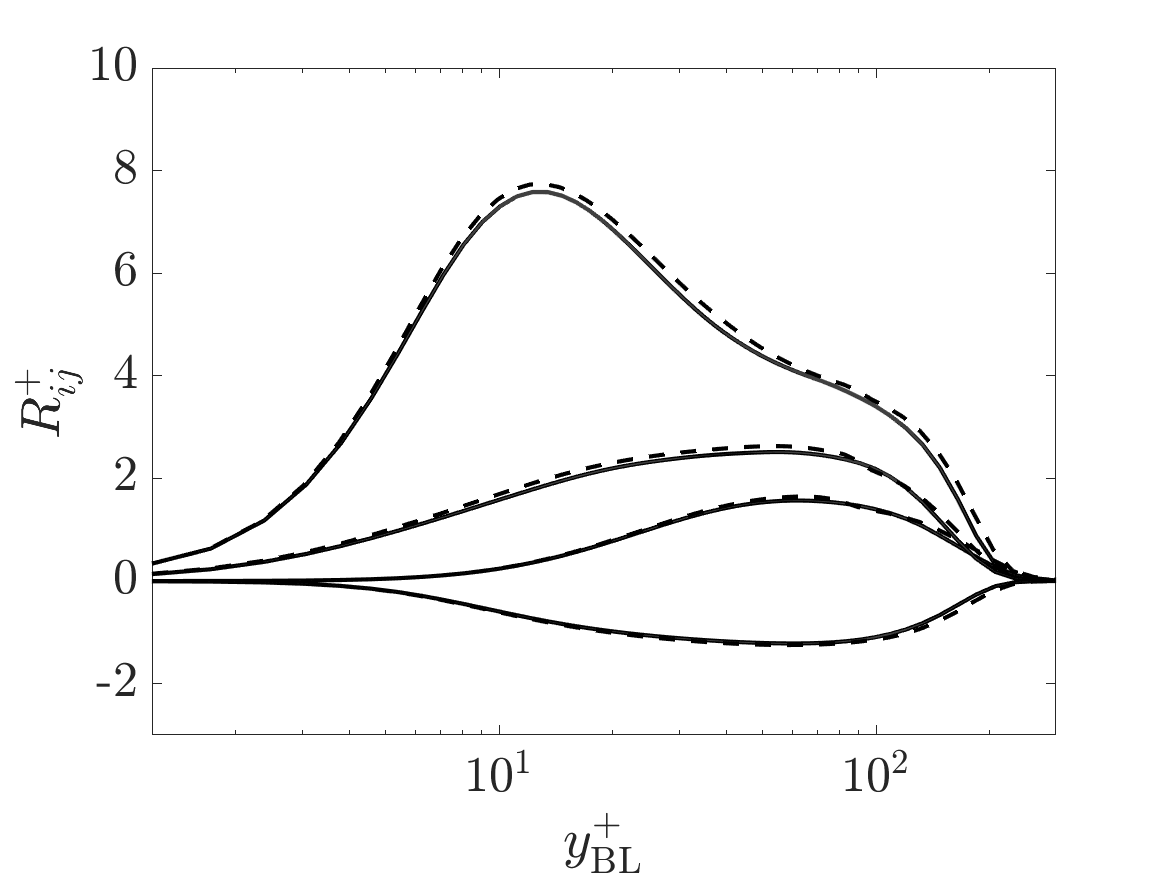}\llap{\parbox[b]{5mm}{(a)\\\rule{0ex}{18.5mm}}}
 \includegraphics[angle=0,origin=c,height=23mm,clip=true,trim=24mm 22mm 17mm 7mm]{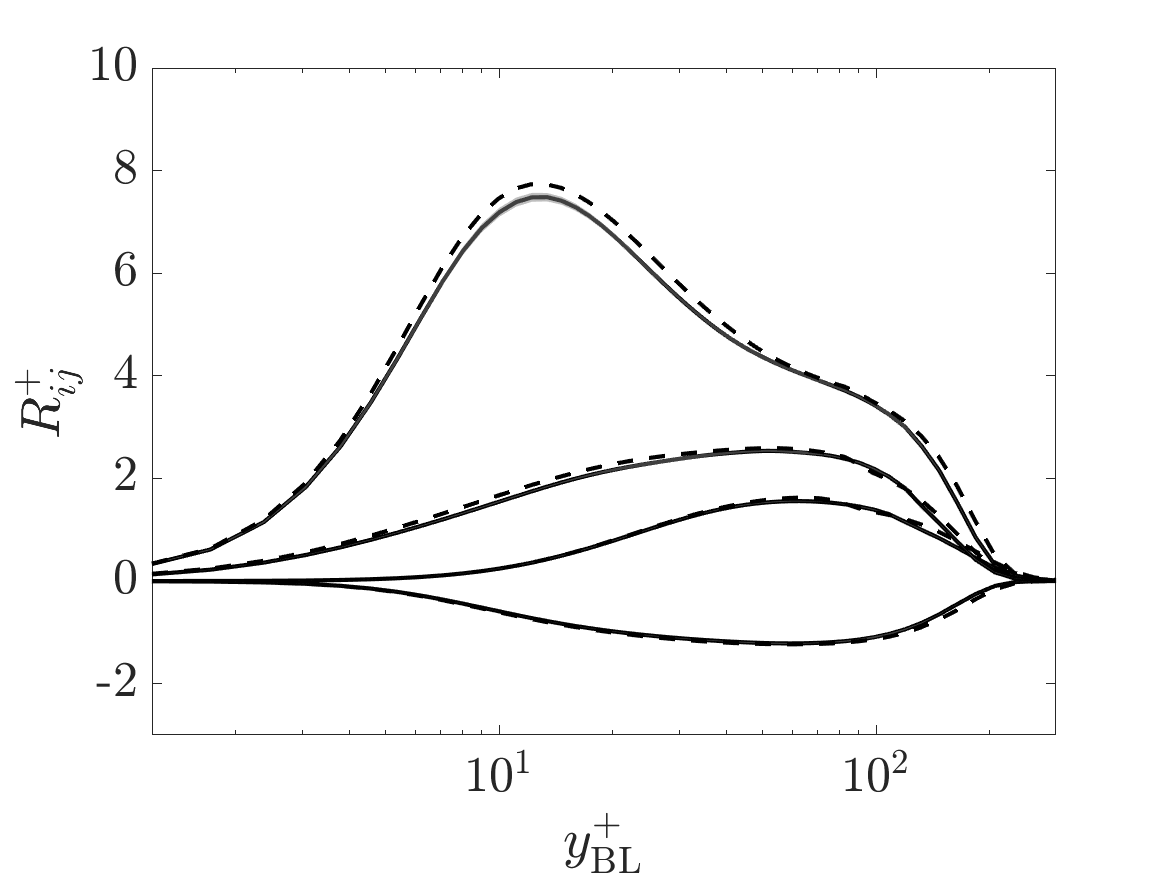}\llap{\parbox[b]{5mm}{(b)\\\rule{0ex}{18.5mm}}}
  \includegraphics[angle=0,origin=c,height=23mm,clip=true,trim=24mm 22mm 17mm 7mm]{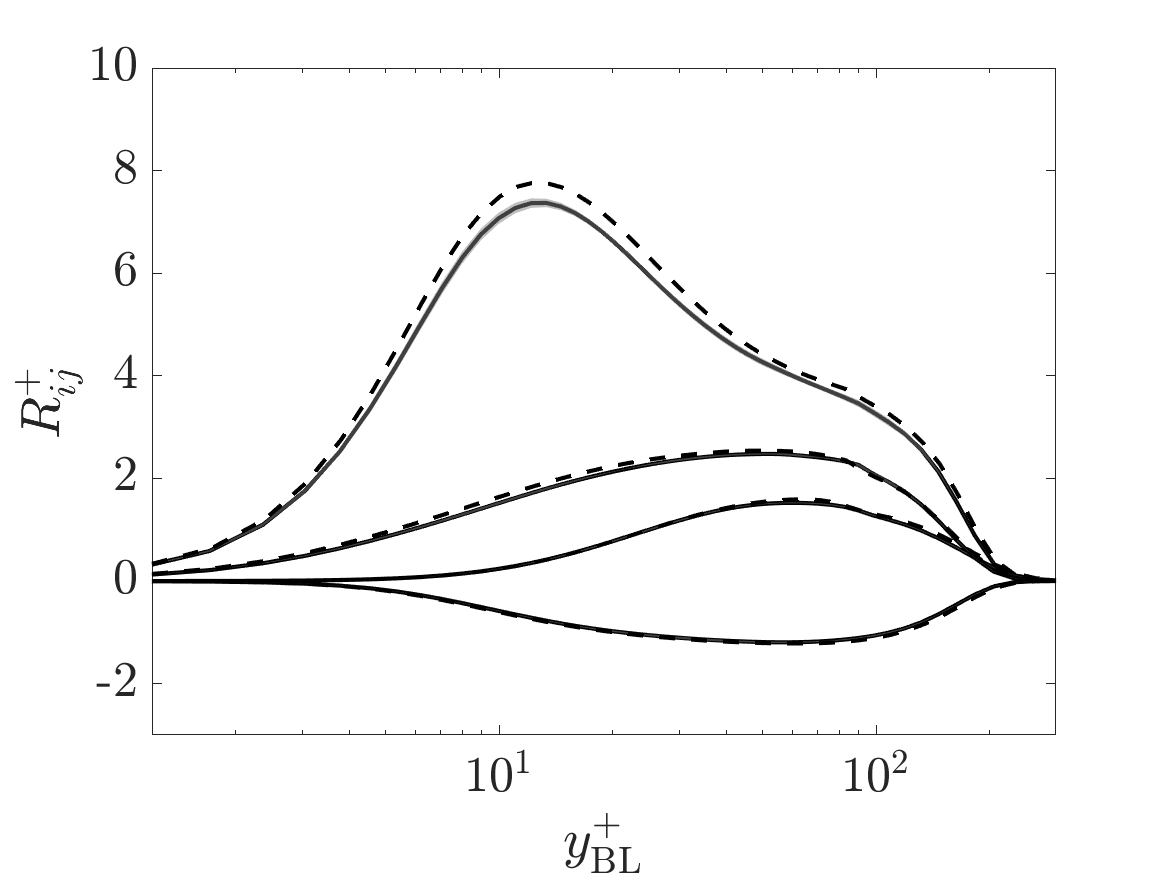}\llap{\parbox[b]{5mm}{(c)\\\rule{0ex}{18.5mm}}}
   \includegraphics[angle=0,origin=c,height=23mm,clip=true,trim=24mm 22mm 17mm 7mm]{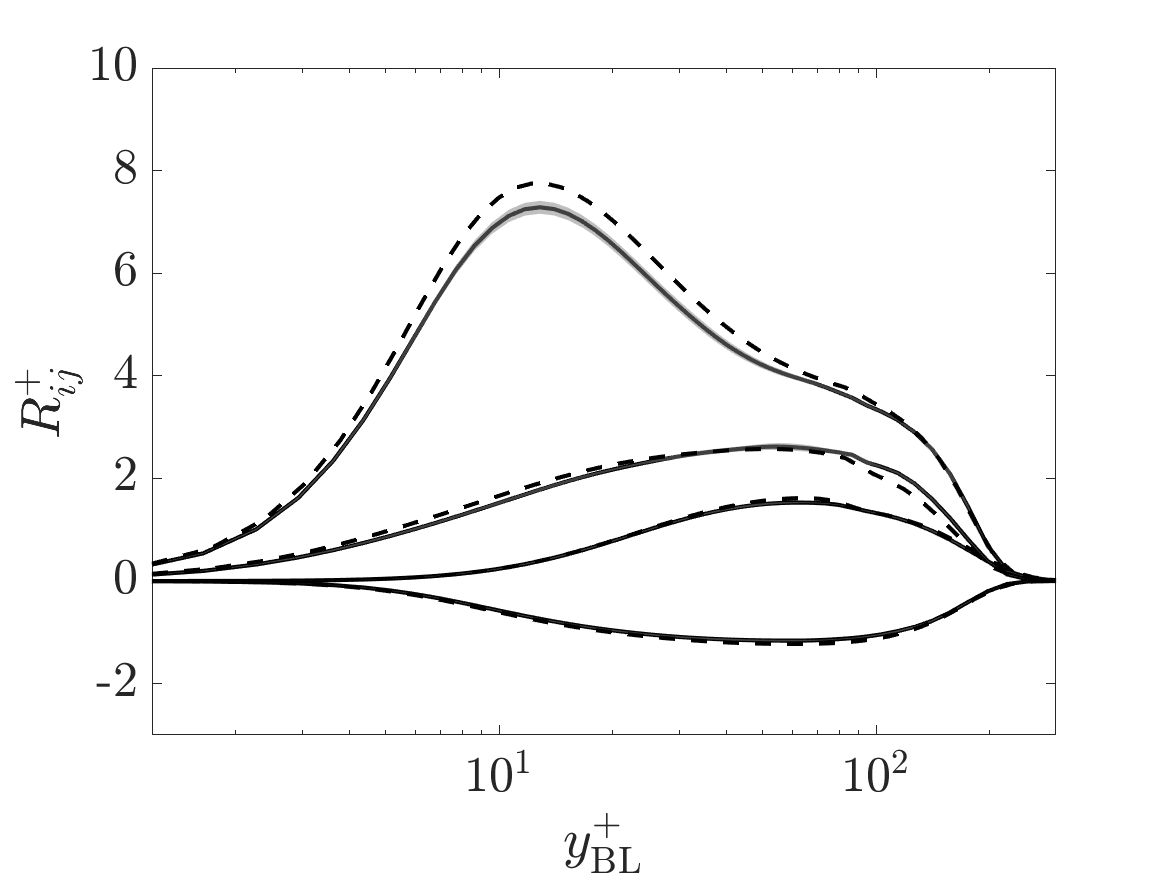}\llap{\parbox[b]{5mm}{(d)\\\rule{0ex}{18.5mm}}}
   \\
    \includegraphics[angle=0,origin=c,height=23mm,clip=true,trim=0mm 22mm 17mm 7mm]{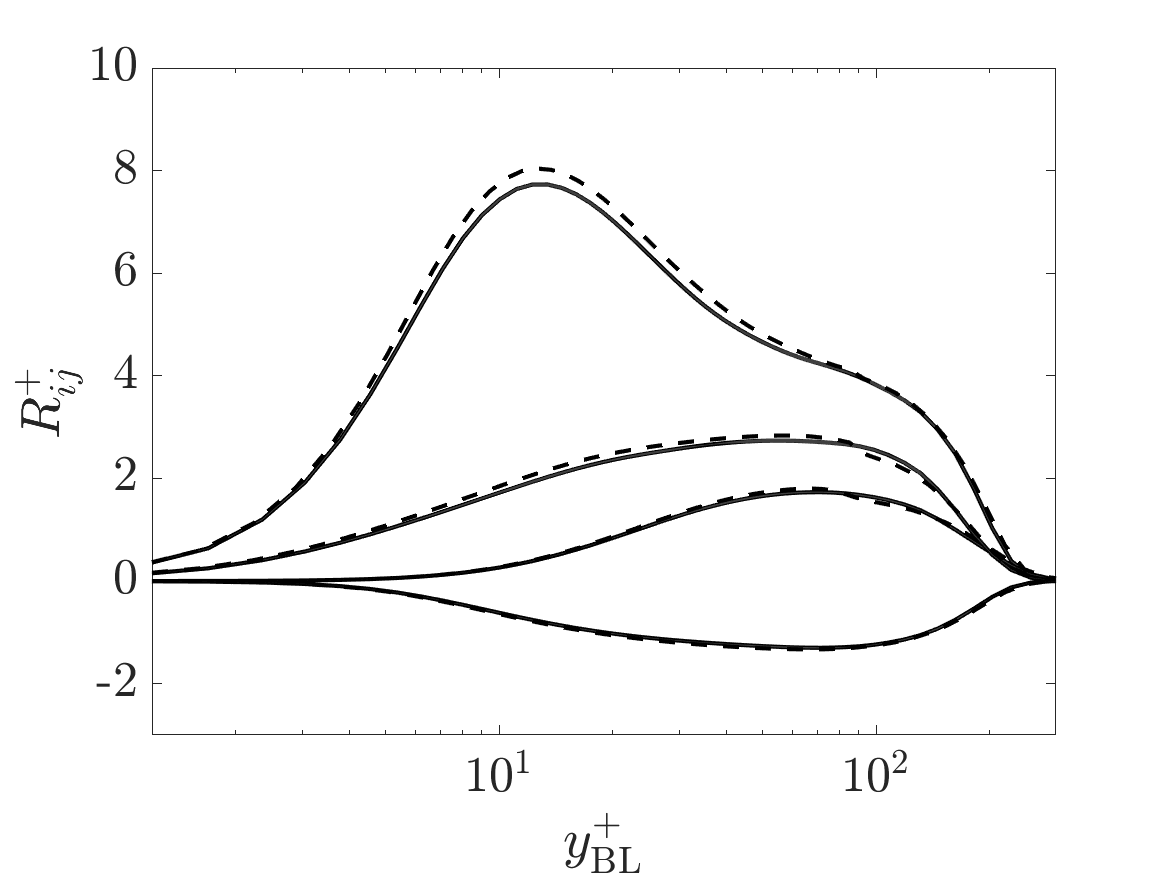}\llap{\parbox[b]{5mm}{(e)\\\rule{0ex}{18.5mm}}}
 \includegraphics[angle=0,origin=c,height=23mm,clip=true,trim=24mm 22mm 17mm 7mm]{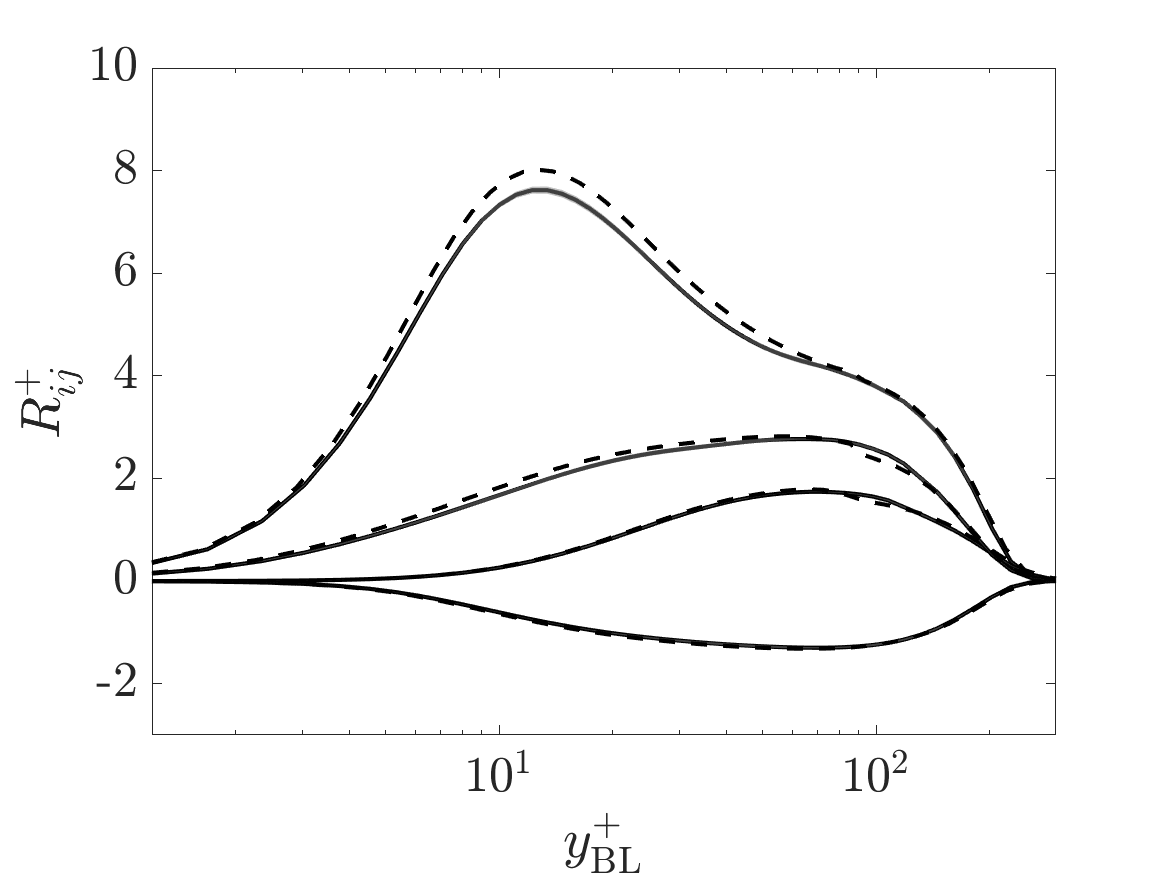}\llap{\parbox[b]{5mm}{(f)\\\rule{0ex}{18.5mm}}}
  \includegraphics[angle=0,origin=c,height=23mm,clip=true,trim=24mm 22mm 17mm 7mm]{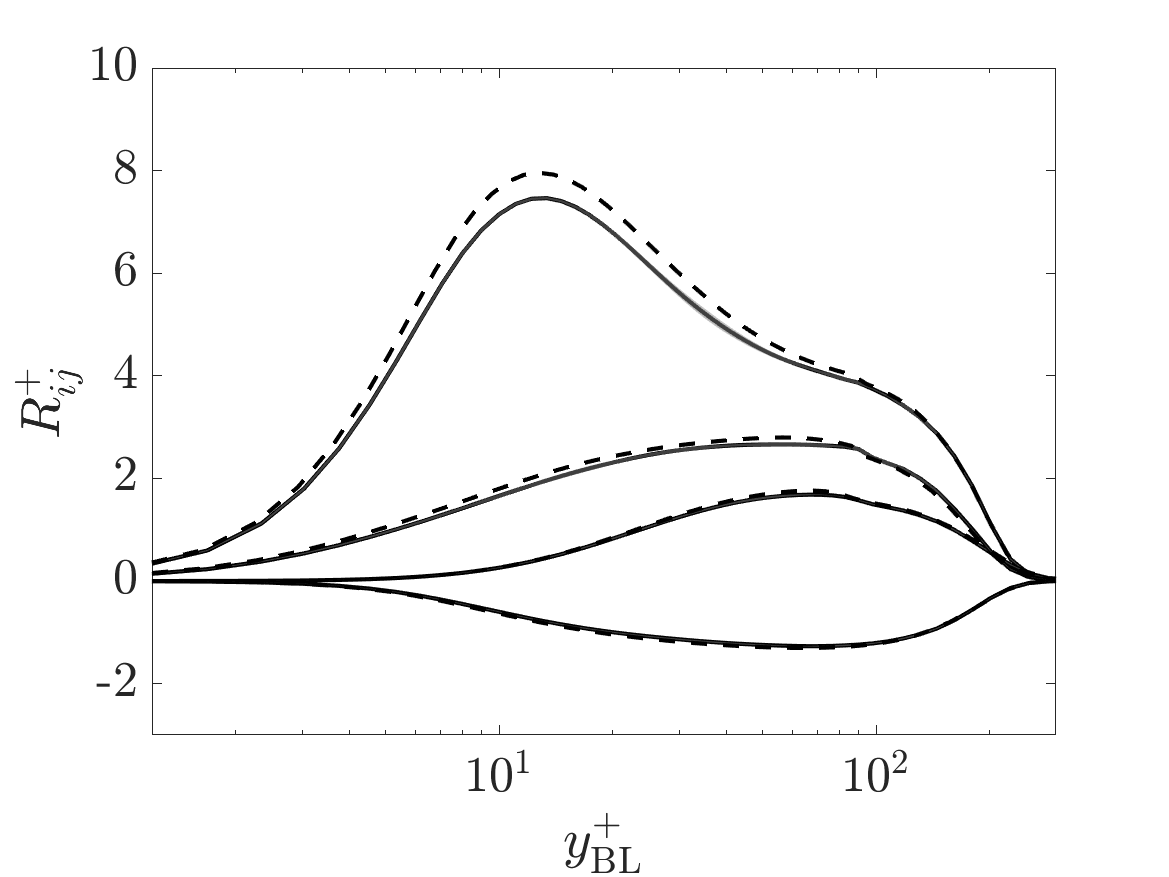}\llap{\parbox[b]{5mm}{(g)\\\rule{0ex}{18.5mm}}}
   \includegraphics[angle=0,origin=c,height=23mm,clip=true,trim=24mm 22mm 17mm 7mm]{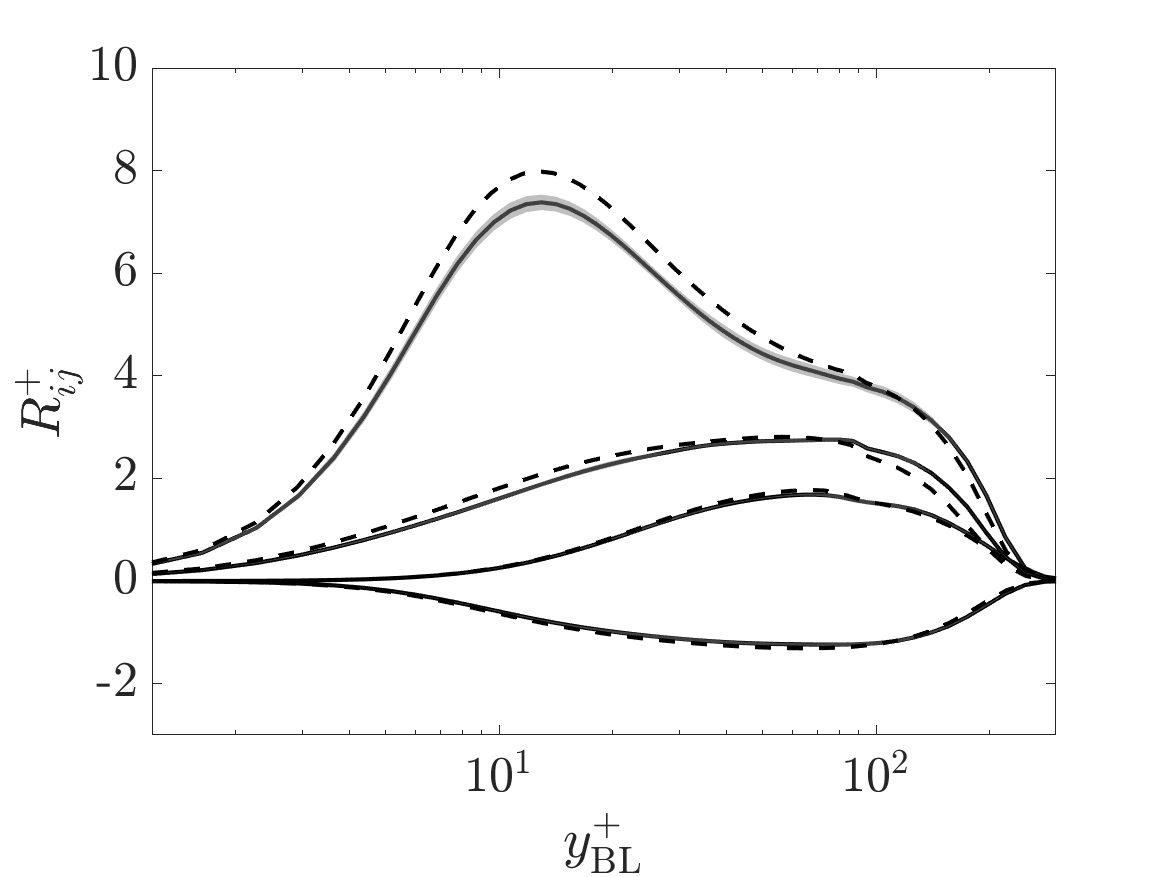}\llap{\parbox[b]{5mm}{(h)\\\rule{0ex}{18.5mm}}}
\\
    \includegraphics[angle=0,origin=c,height=23mm,clip=true,trim=0mm 22mm 17mm 7mm]{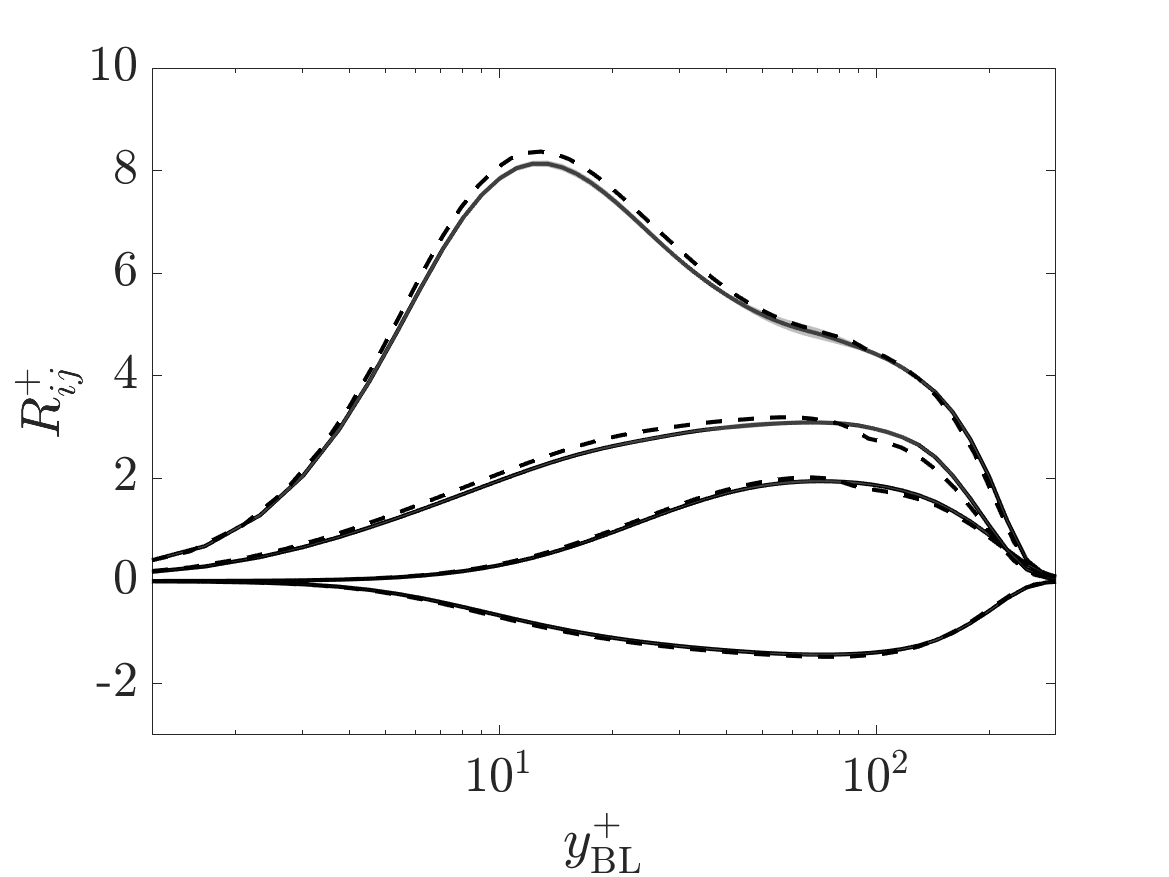}\llap{\parbox[b]{5mm}{(i)\\\rule{0ex}{18.5mm}}}
 \includegraphics[angle=0,origin=c,height=23mm,clip=true,trim=24mm 22mm 17mm 7mm]{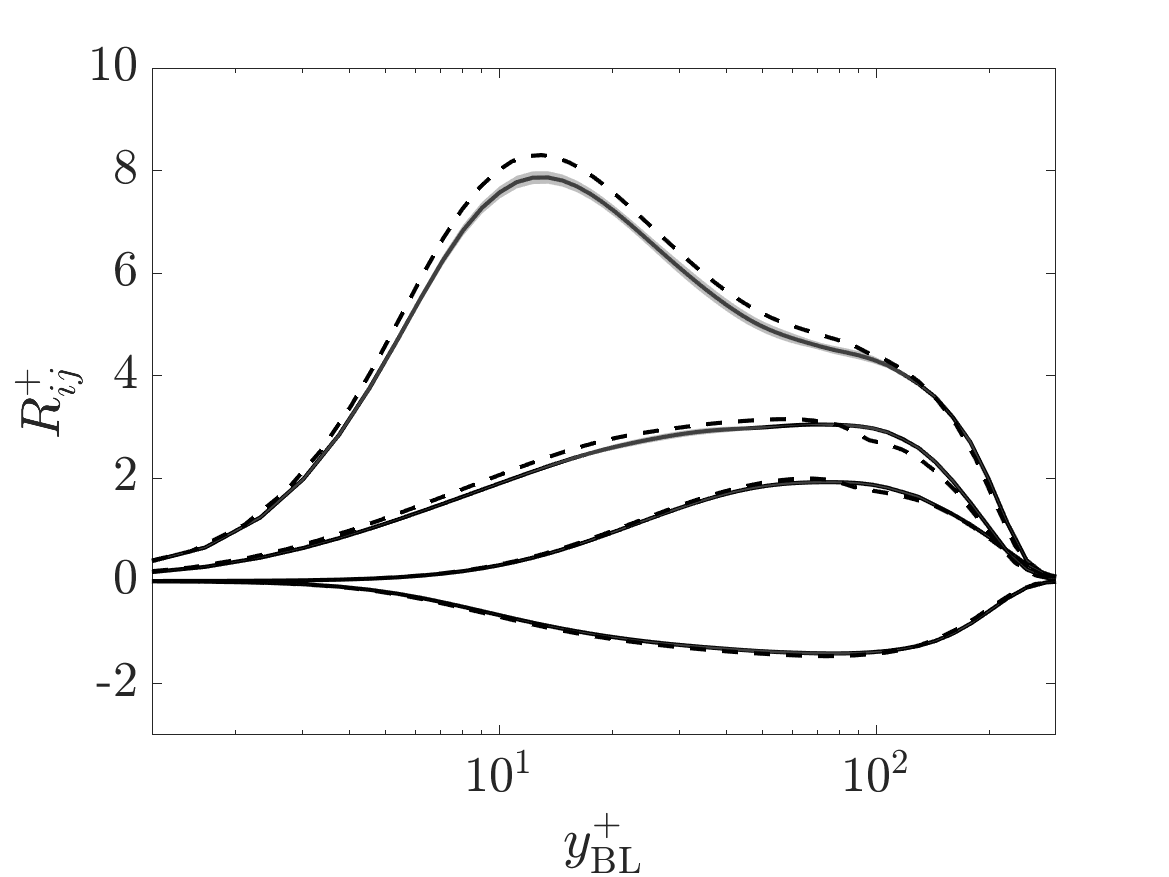}\llap{\parbox[b]{5mm}{(j)\\\rule{0ex}{18.5mm}}}
  \includegraphics[angle=0,origin=c,height=23mm,clip=true,trim=24mm 22mm 17mm 7mm]{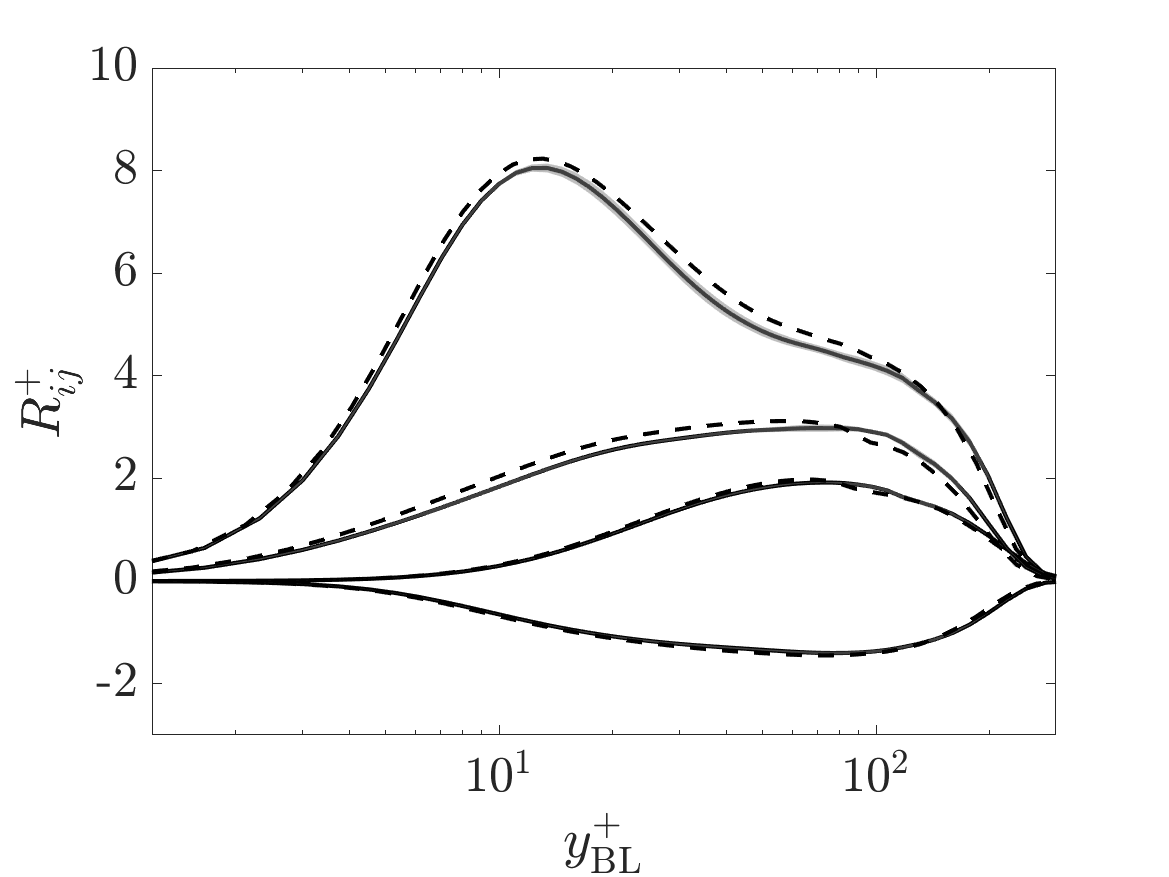}\llap{\parbox[b]{5mm}{(k)\\\rule{0ex}{18.5mm}}}
   \includegraphics[angle=0,origin=c,height=23mm,clip=true,trim=24mm 22mm 17mm 7mm]{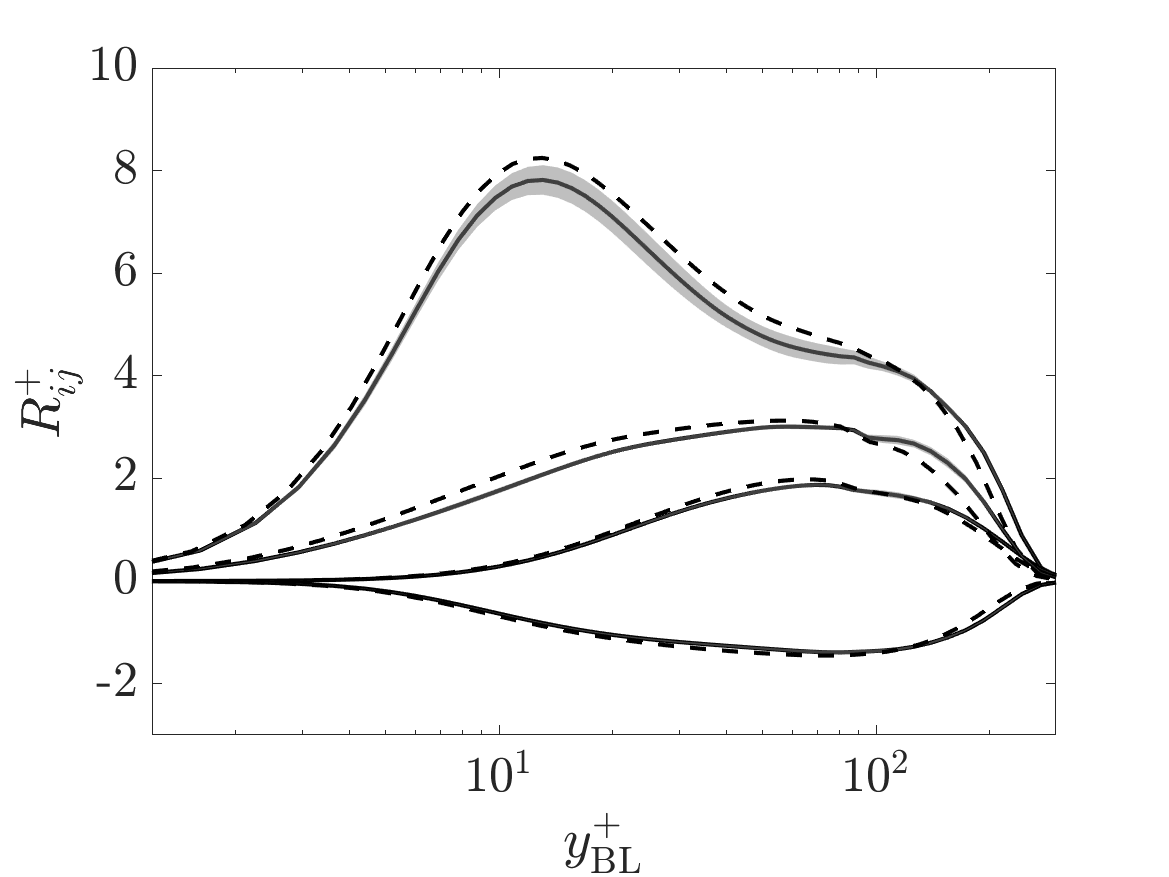}\llap{\parbox[b]{5mm}{(l)\\\rule{0ex}{18.5mm}}}
   \\
    \includegraphics[angle=0,origin=c,height=27.2mm,clip=true,trim=0mm 0mm 17mm 7mm]{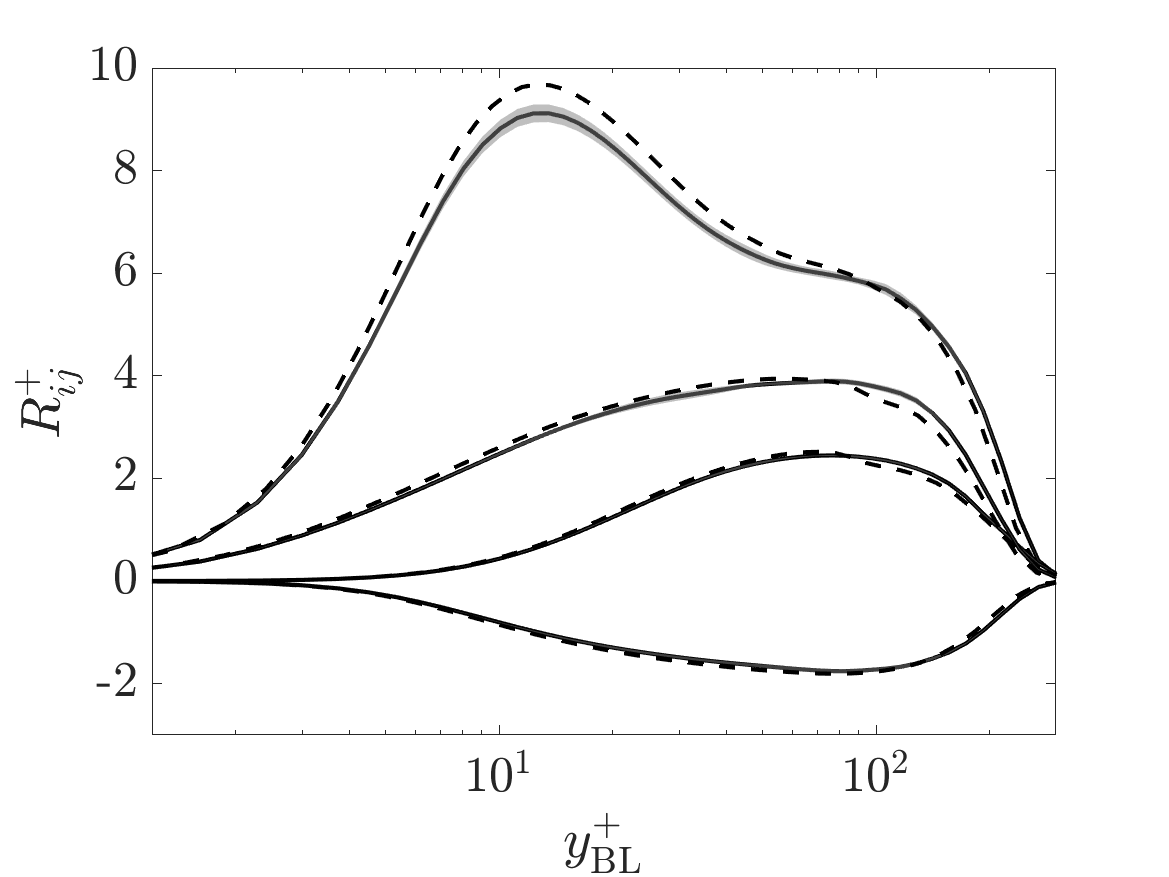}\llap{\parbox[b]{5mm}{(m)\\\rule{0ex}{22.7mm}}}
 \includegraphics[angle=0,origin=c,height=27.2mm,clip=true,trim=24mm 0mm 17mm 7mm]{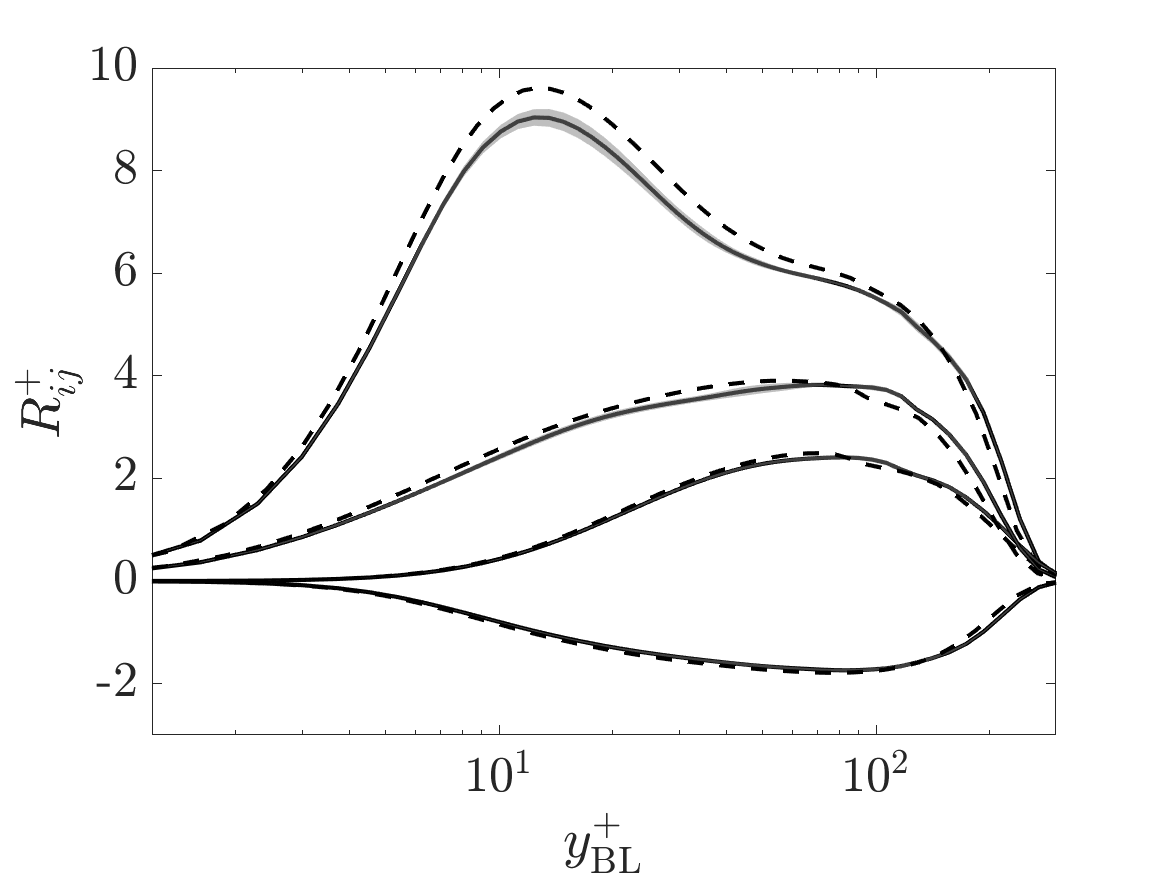}\llap{\parbox[b]{5mm}{(n)\\\rule{0ex}{22.7mm}}}
  \includegraphics[angle=0,origin=c,height=27.2mm,clip=true,trim=24mm 0mm 17mm 7mm]{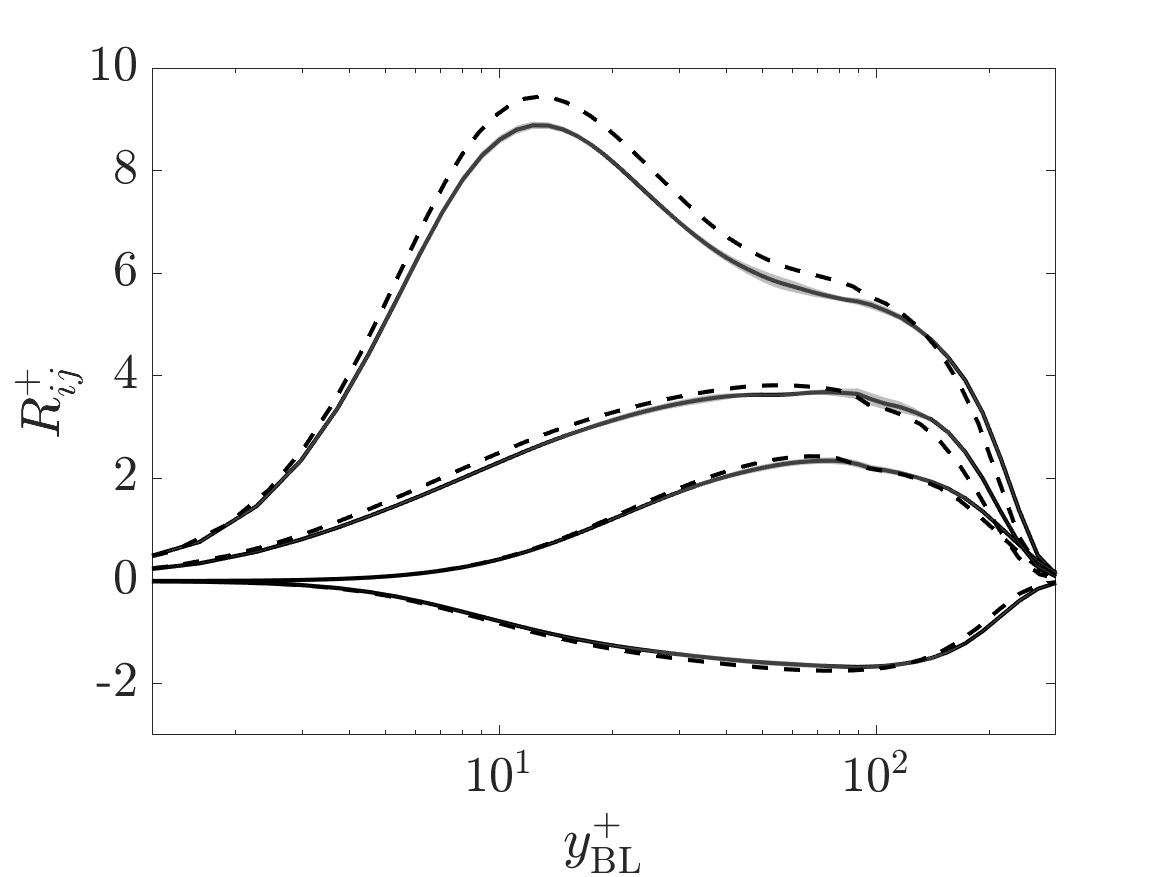}\llap{\parbox[b]{5mm}{(o)\\\rule{0ex}{22.7mm}}}
   \includegraphics[angle=0,origin=c,height=27.2mm,clip=true,trim=24mm 0mm 17mm 7mm]{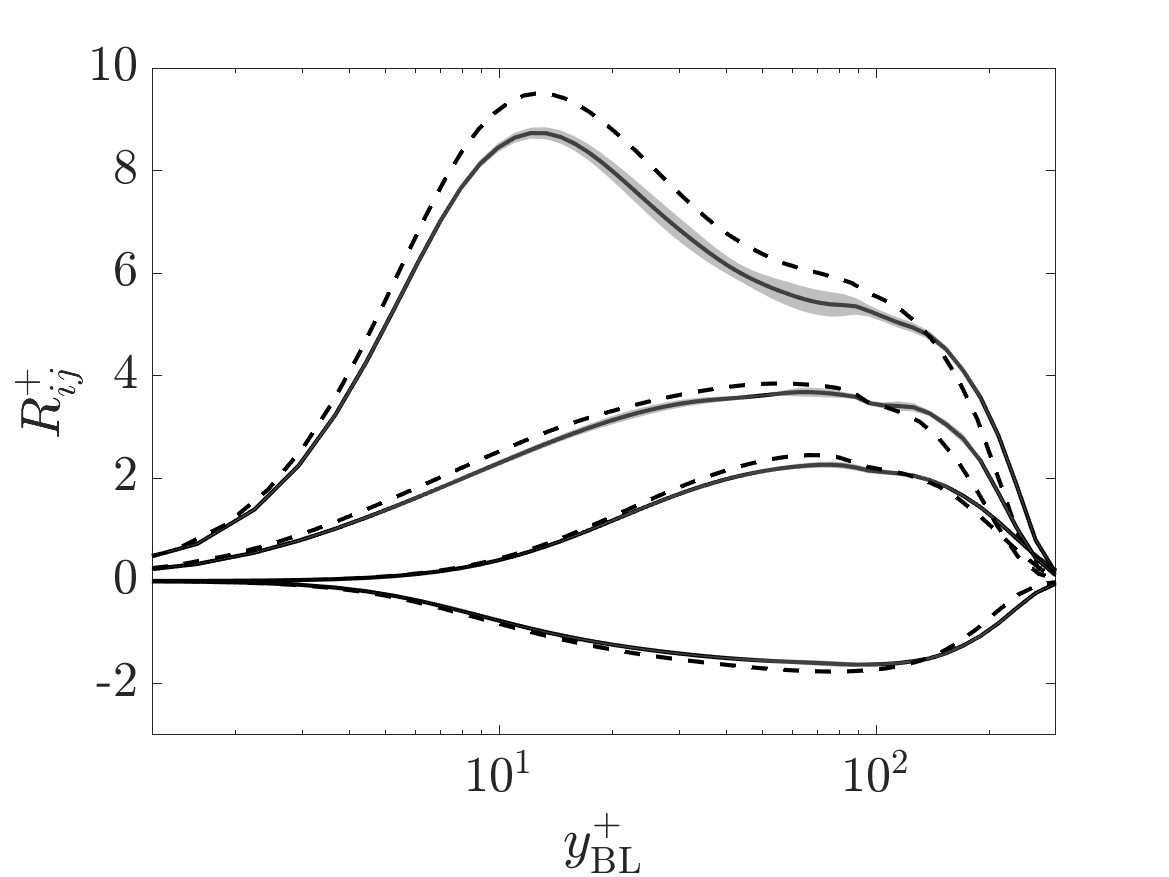}\llap{\parbox[b]{5mm}{(p)\\\rule{0ex}{22.7mm}}}
  \caption{
  \label{fig:rwt5-RS-allXZ}
  Variation of the Reynolds stresses with wall distance 
  at several streamwise and spanwise locations on the suction side of RWT-5 (solid lines)
  compared to P-2 (dashed lines) at matching $(Re_\tau,\beta_{x_\tau})$ values.
  Shaded regions correspond to 80\% confidence intervals of the inner-scaled Reynolds stress profiles due to finite time averaging.
  See the caption of figure~\ref{fig:rwt0-RS-allXZ} for more details.
  }
\end{figure}

\begin{figure}
  \centering
 \includegraphics[angle=0,origin=c,height=23mm,clip=true,trim=0mm 22mm 17mm 10mm]{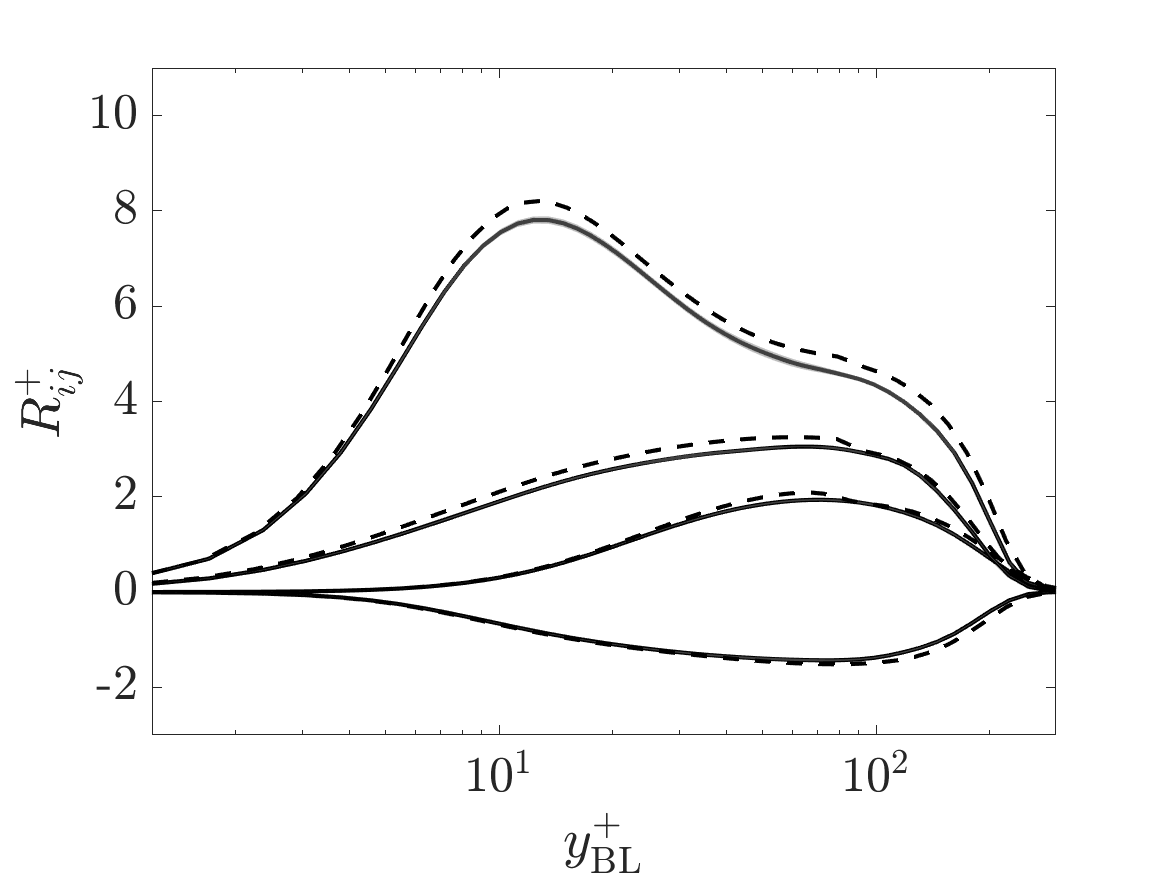}\llap{\parbox[b]{5mm}{(a)\\\rule{0ex}{19mm}}}
 \includegraphics[angle=0,origin=c,height=23mm,clip=true,trim=24mm 22mm 17mm 10mm]{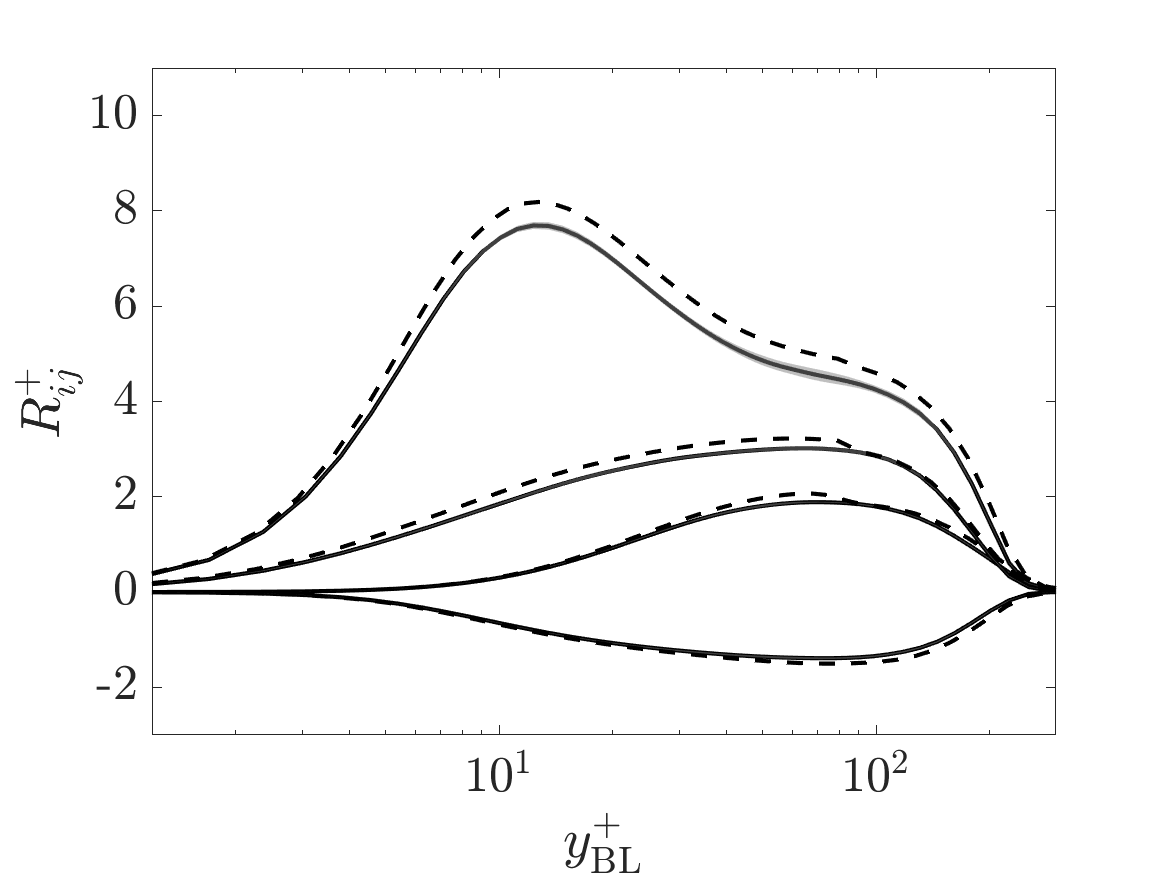}\llap{\parbox[b]{5mm}{(b)\\\rule{0ex}{19mm}}}
  \includegraphics[angle=0,origin=c,height=23mm,clip=true,trim=24mm 22mm 17mm 10mm]{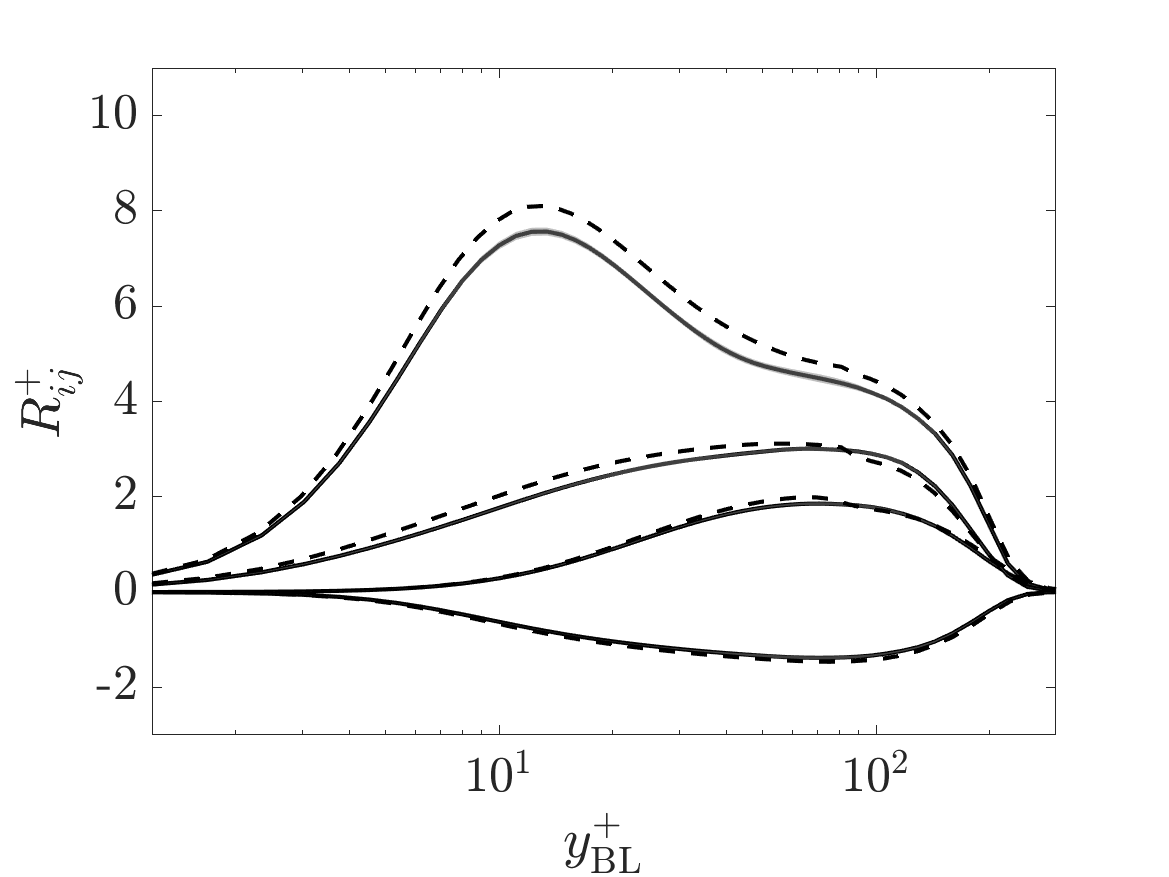}\llap{\parbox[b]{5mm}{(c)\\\rule{0ex}{19mm}}}
   \includegraphics[angle=0,origin=c,height=23mm,clip=true,trim=24mm 22mm 17mm 10mm]{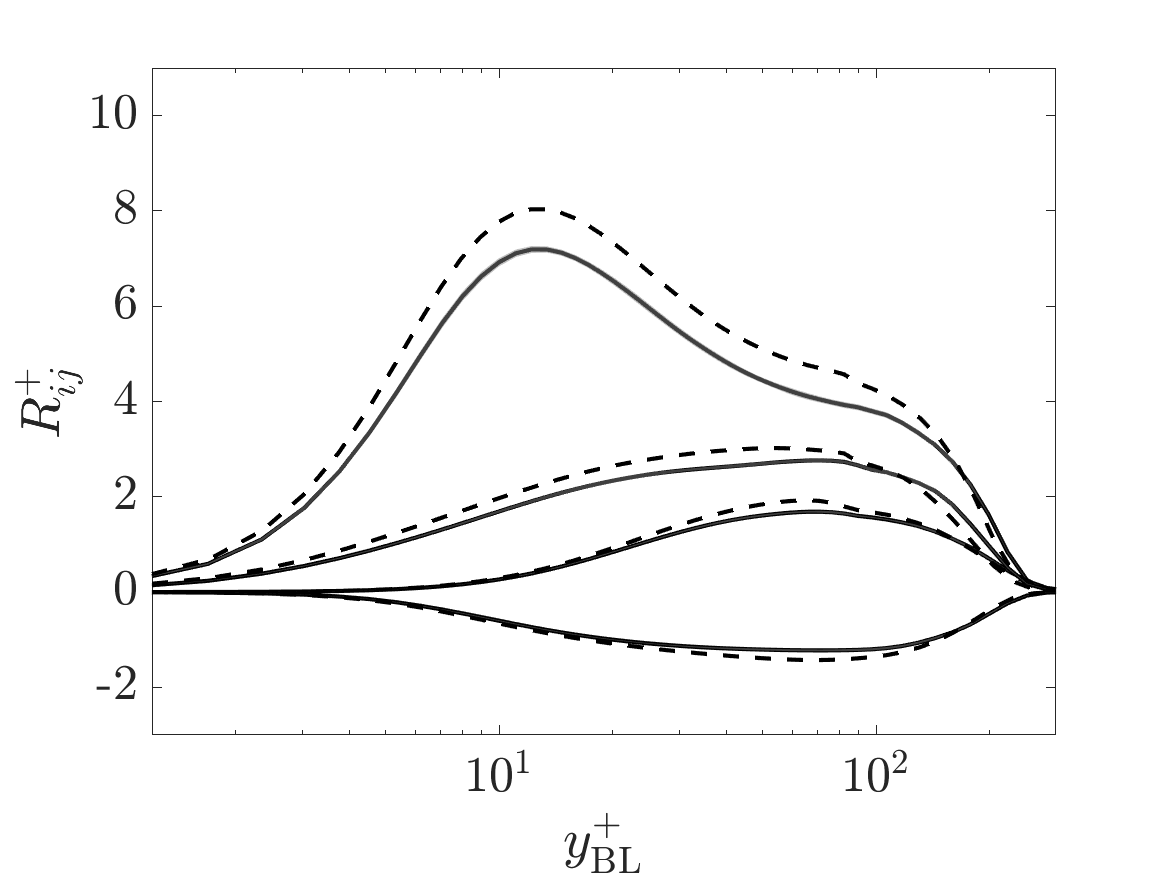}\llap{\parbox[b]{5mm}{(d)\\\rule{0ex}{19mm}}}
   \\
    \includegraphics[angle=0,origin=c,height=23mm,clip=true,trim=0mm 22mm 17mm 10mm]{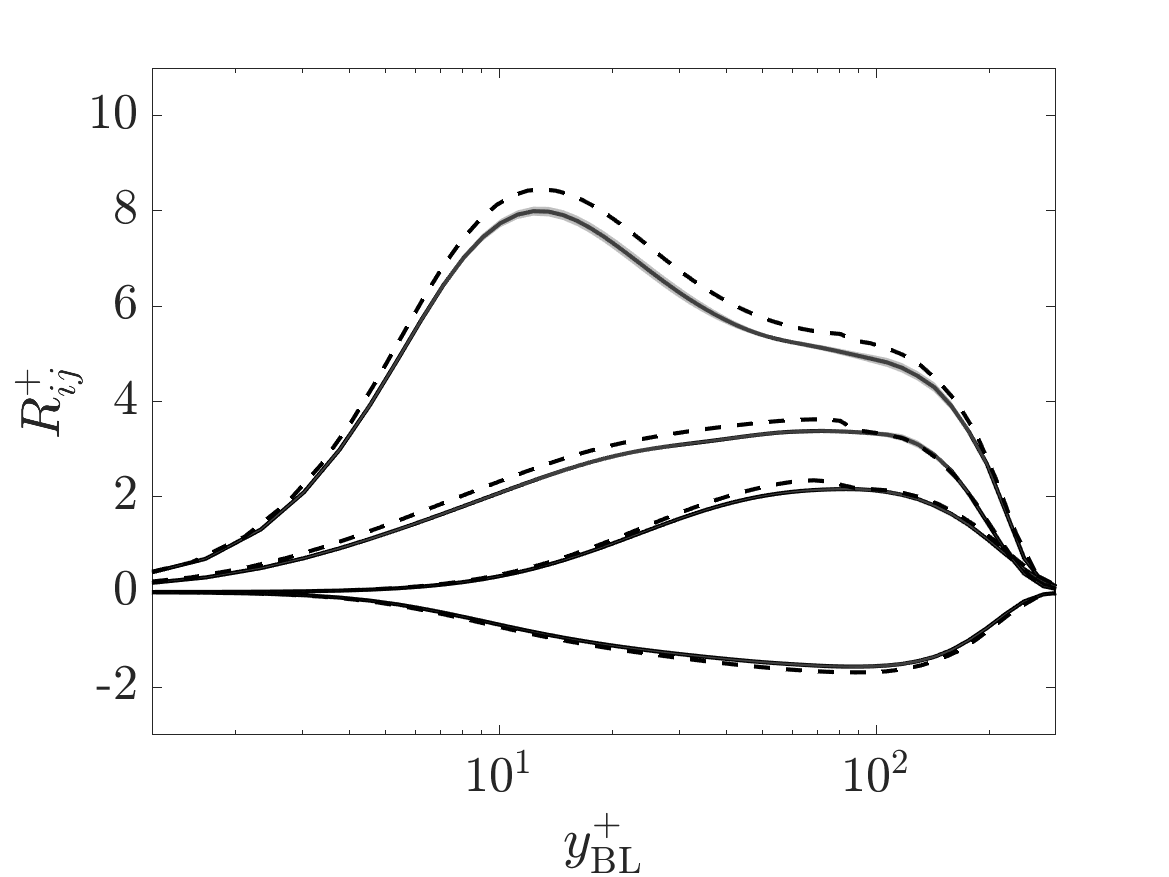}\llap{\parbox[b]{5mm}{(e)\\\rule{0ex}{19mm}}}
 \includegraphics[angle=0,origin=c,height=23mm,clip=true,trim=24mm 22mm 17mm 10mm]{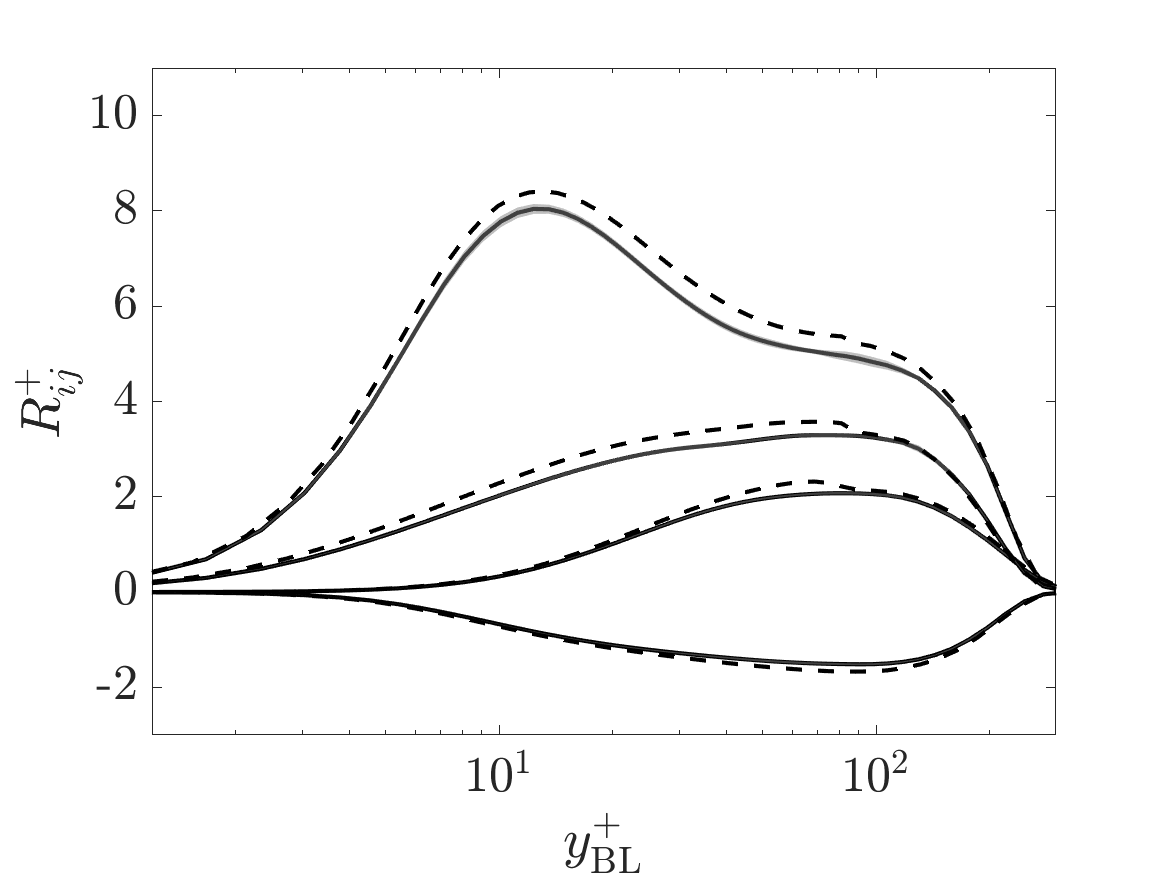}\llap{\parbox[b]{5mm}{(f)\\\rule{0ex}{19mm}}}
  \includegraphics[angle=0,origin=c,height=23mm,clip=true,trim=24mm 22mm 17mm 10mm]{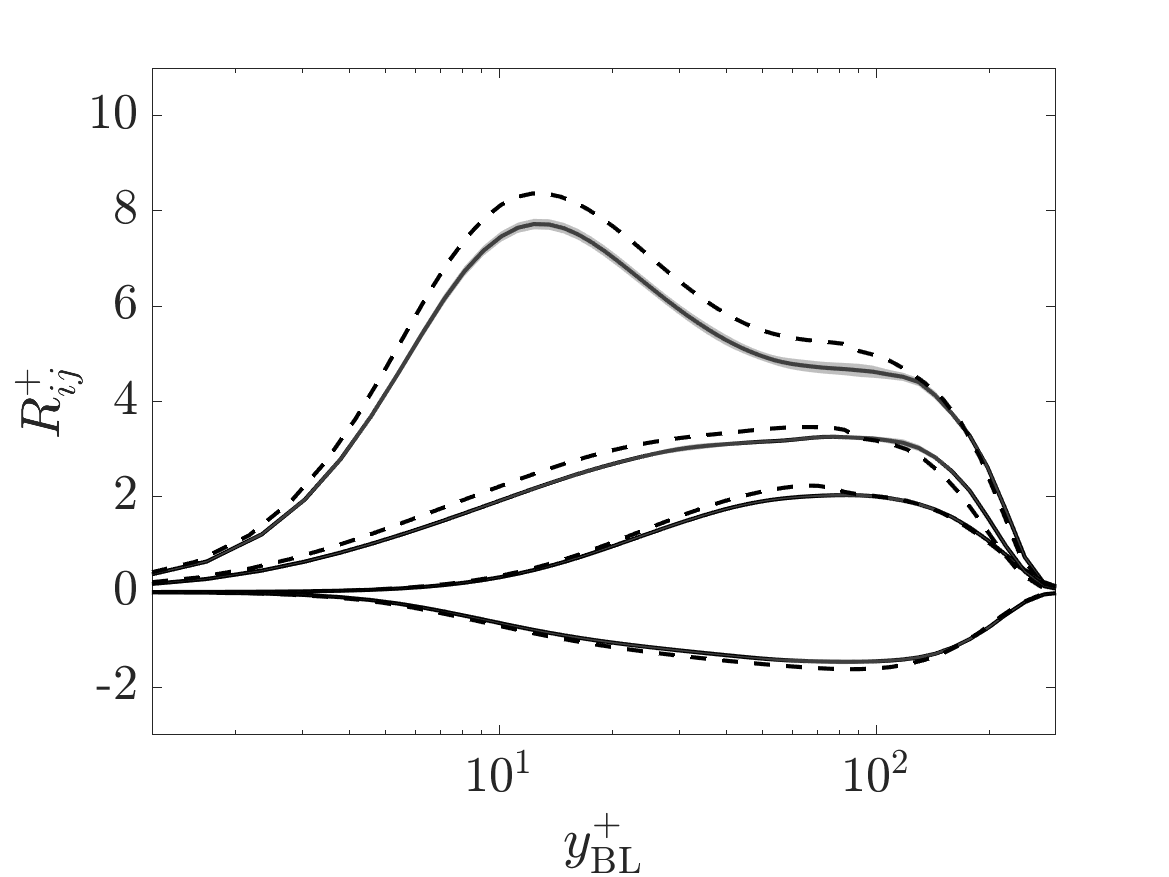}\llap{\parbox[b]{5mm}{(g)\\\rule{0ex}{19mm}}}
   \includegraphics[angle=0,origin=c,height=23mm,clip=true,trim=24mm 22mm 17mm 10mm]{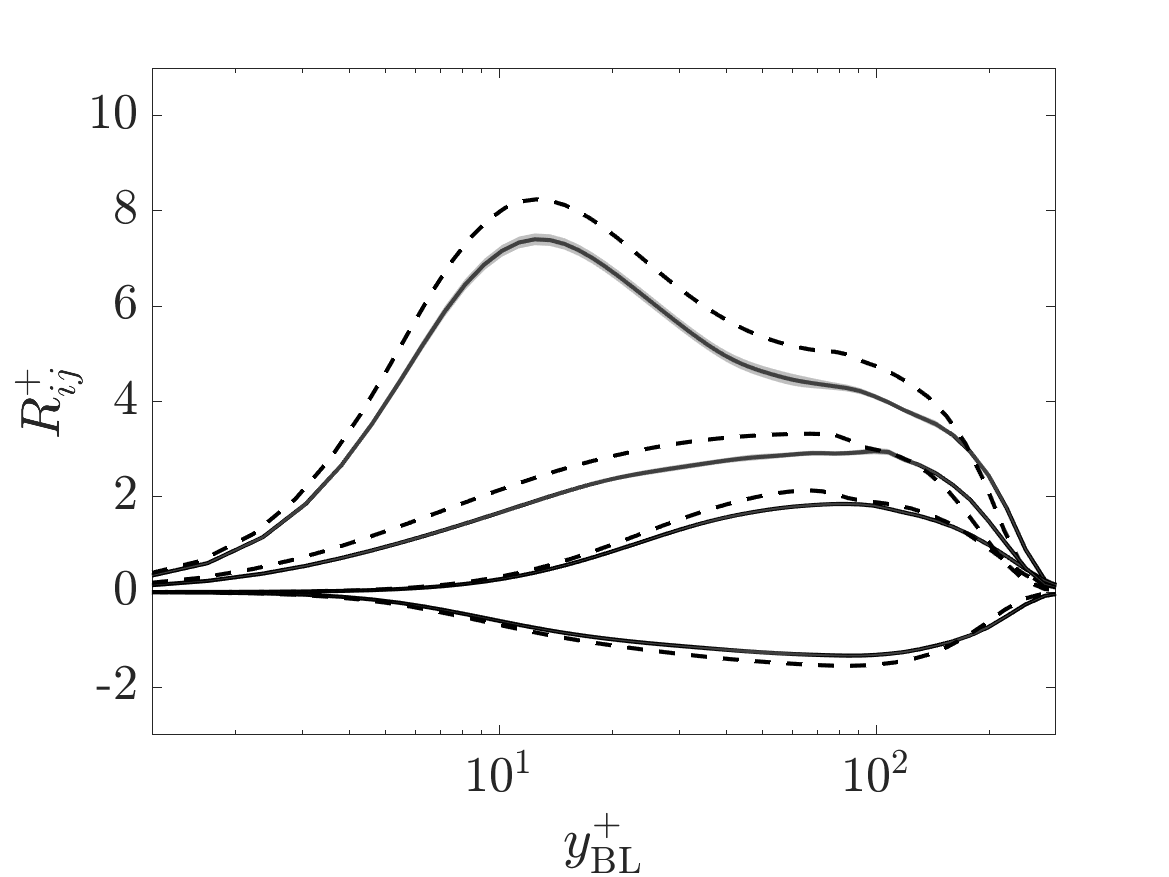}\llap{\parbox[b]{5mm}{(h)\\\rule{0ex}{19mm}}}
\\
    \includegraphics[angle=0,origin=c,height=23mm,clip=true,trim=0mm 22mm 17mm 10mm]{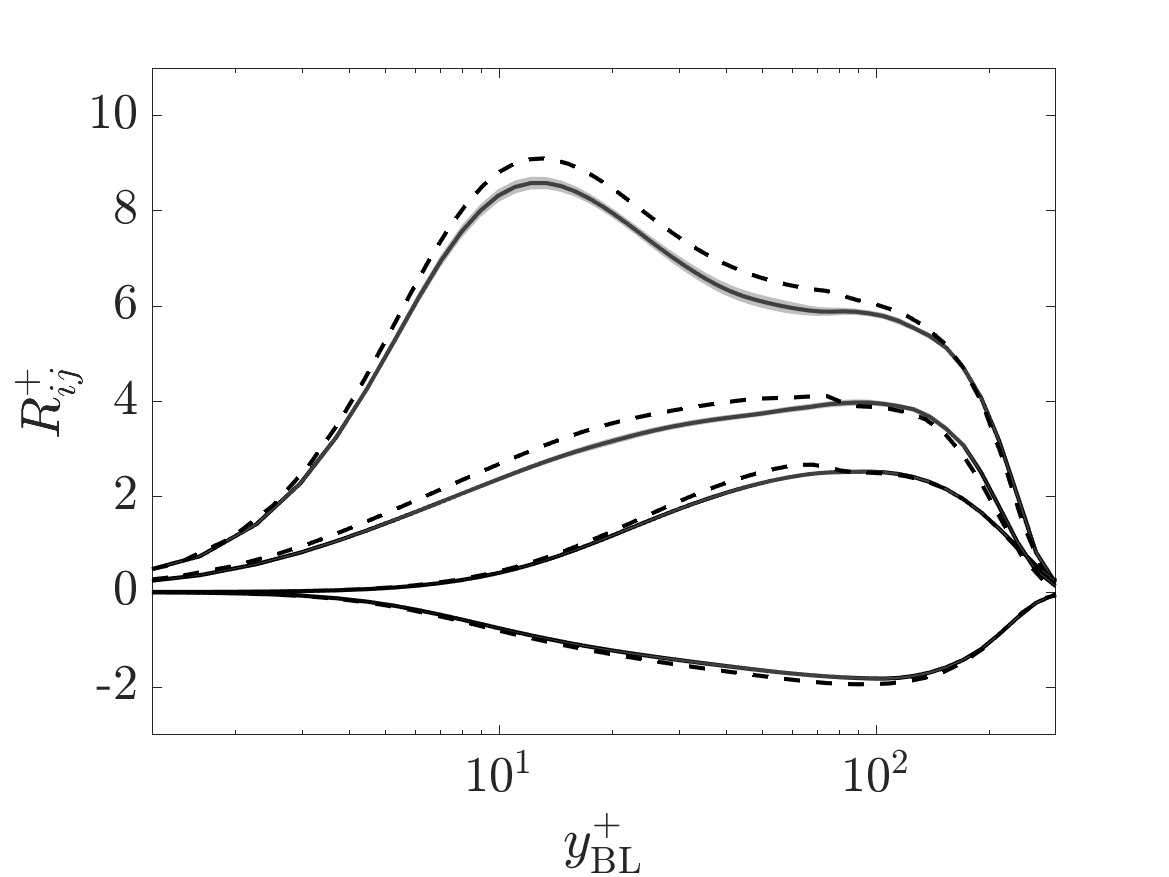}\llap{\parbox[b]{5mm}{(i)\\\rule{0ex}{19mm}}}
 \includegraphics[angle=0,origin=c,height=23mm,clip=true,trim=24mm 22mm 17mm 10mm]{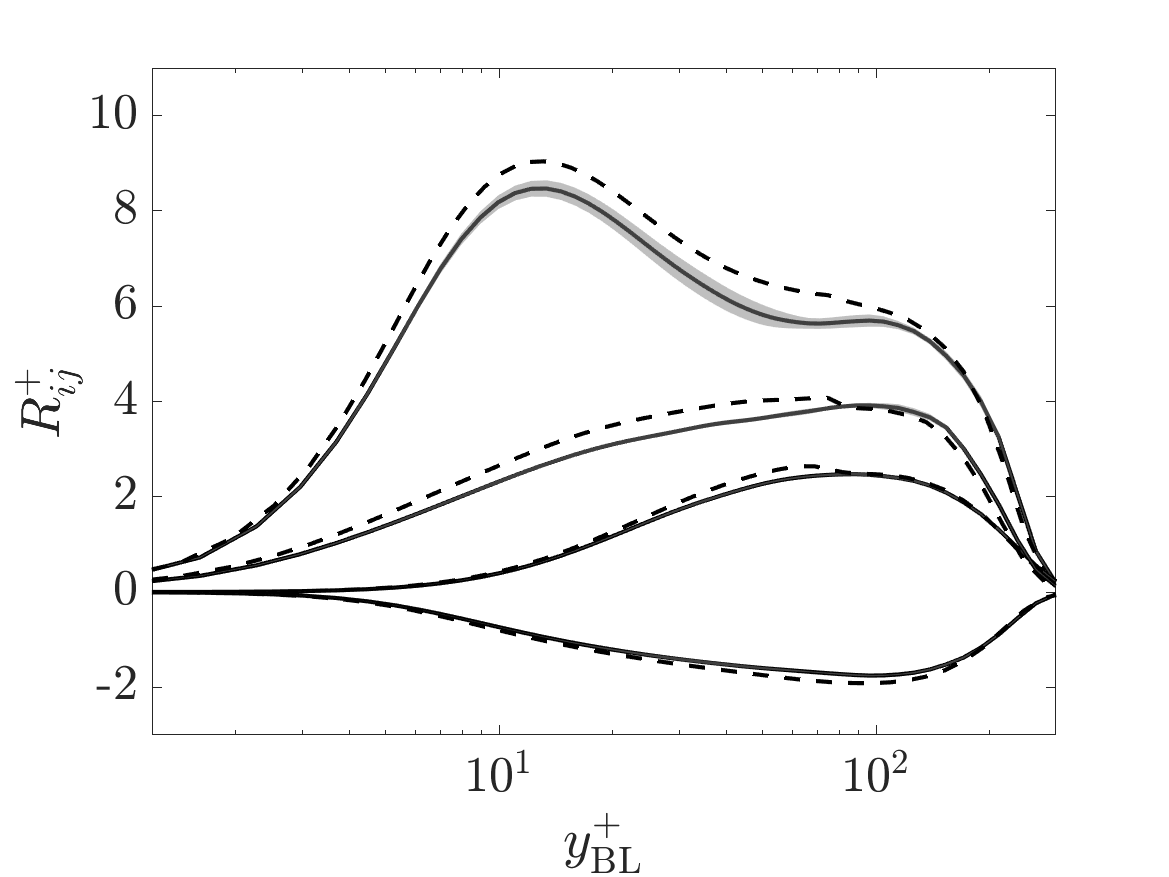}\llap{\parbox[b]{5mm}{(j)\\\rule{0ex}{19mm}}}
  \includegraphics[angle=0,origin=c,height=23mm,clip=true,trim=24mm 22mm 17mm 10mm]{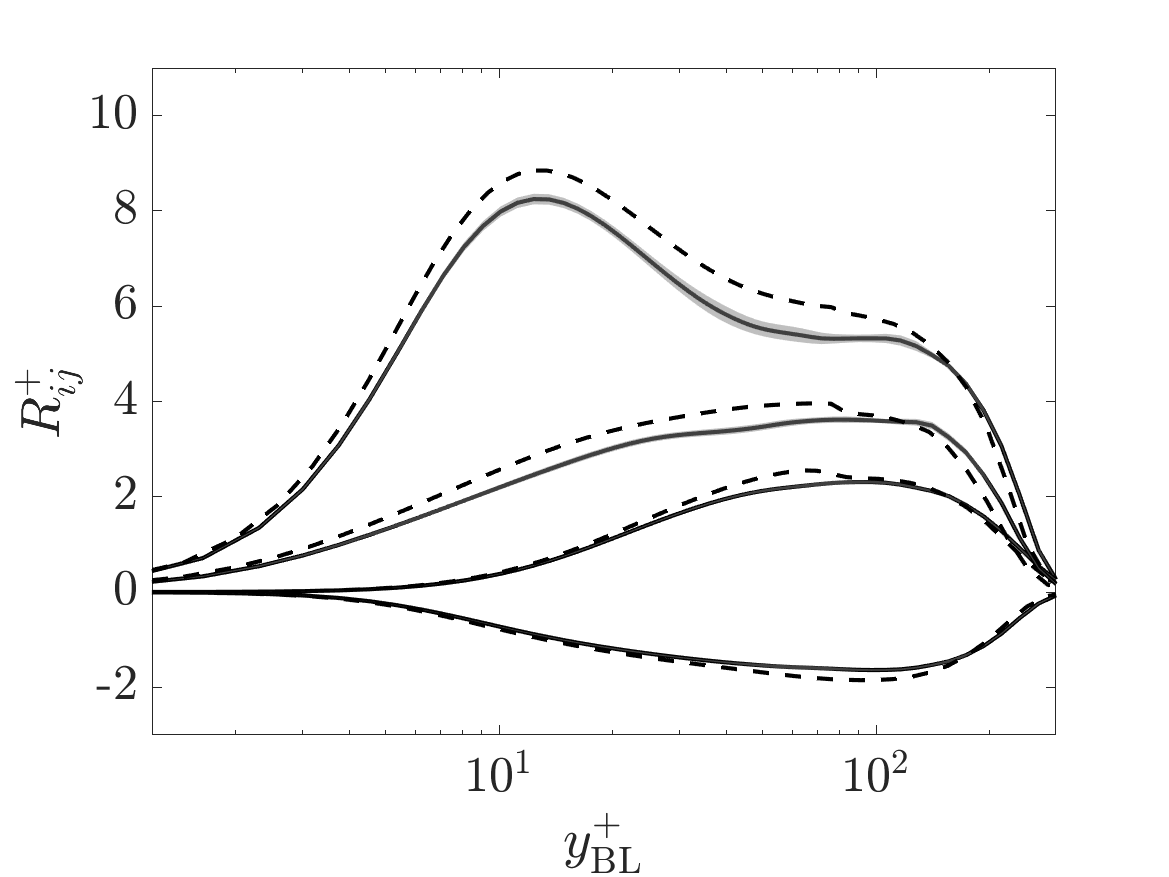}\llap{\parbox[b]{5mm}{(k)\\\rule{0ex}{19mm}}}
   \includegraphics[angle=0,origin=c,height=23mm,clip=true,trim=24mm 22mm 17mm 10mm]{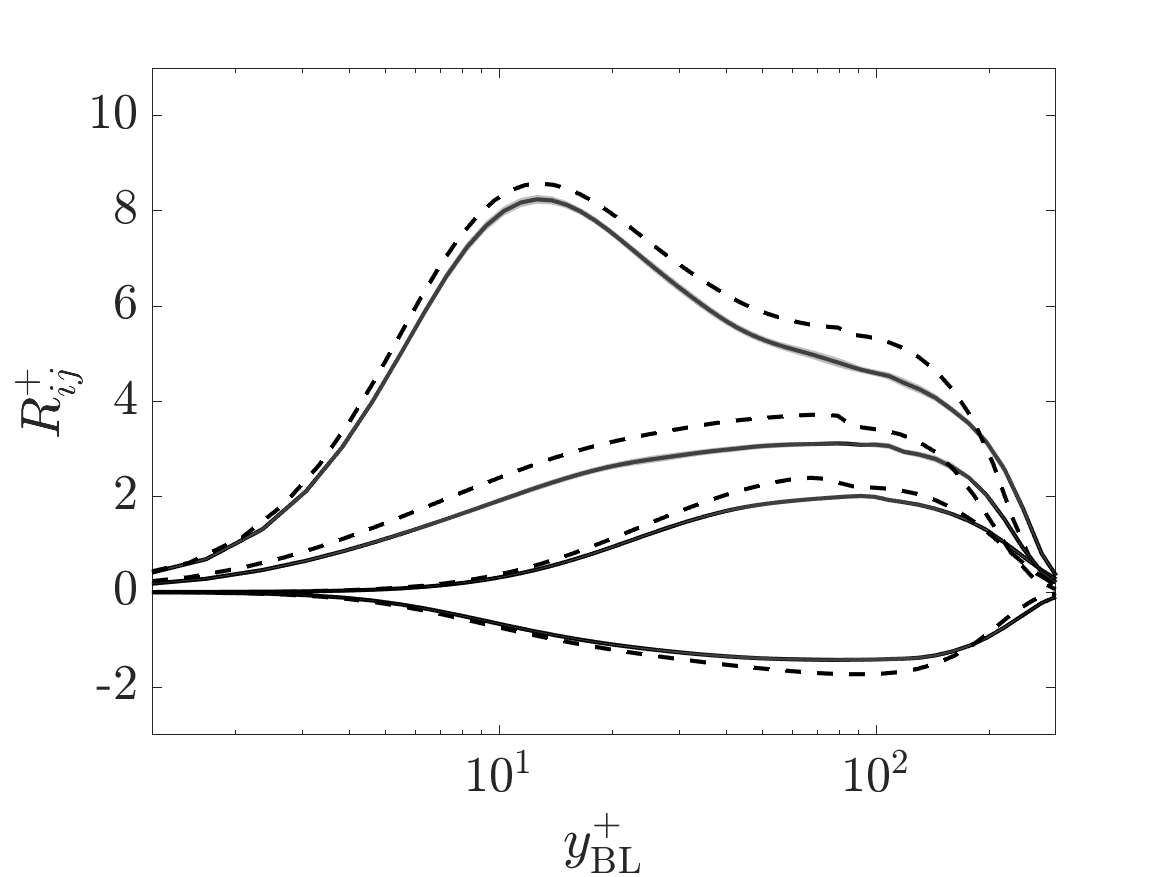}\llap{\parbox[b]{5mm}{(l)\\\rule{0ex}{19mm}}}
   \\
    \includegraphics[angle=0,origin=c,height=27.2mm,clip=true,trim=0mm 0mm 17mm 10mm]{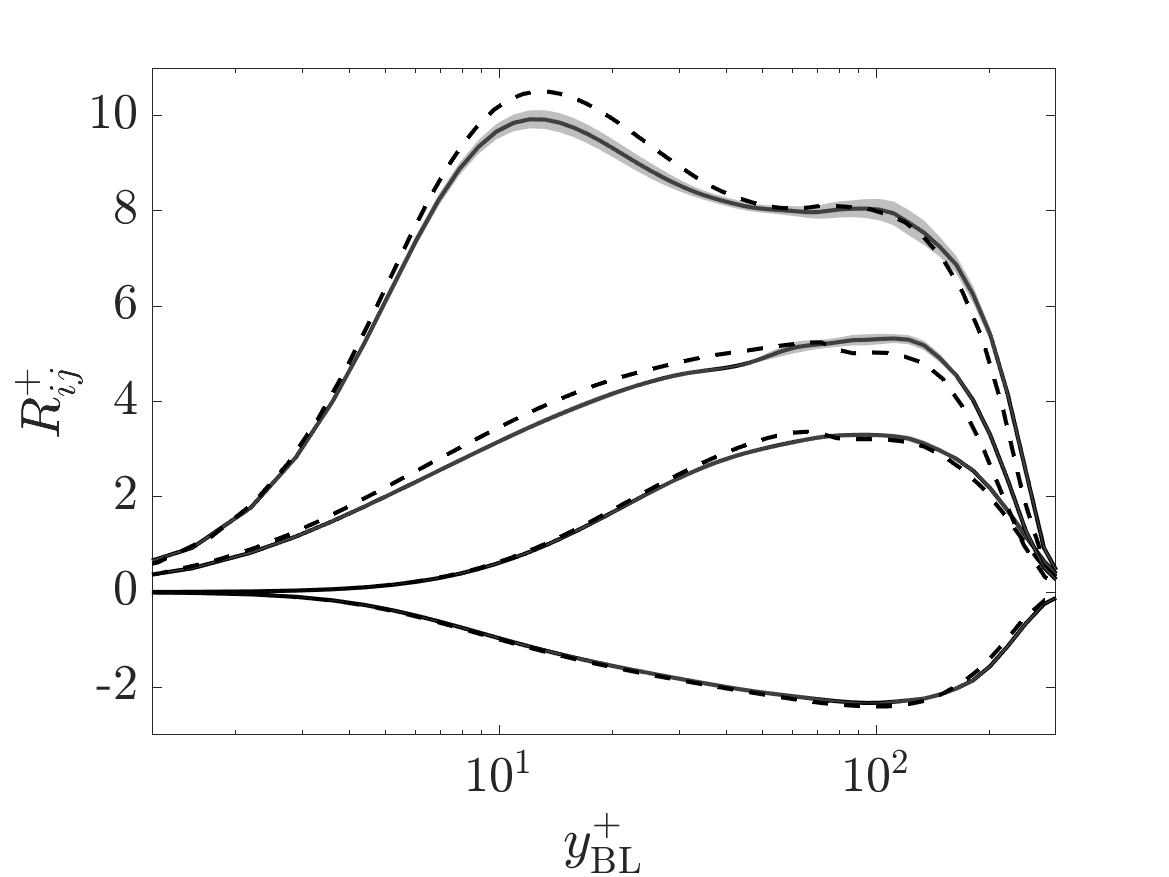}\llap{\parbox[b]{5mm}{(m)\\\rule{0ex}{23.2mm}}}
 \includegraphics[angle=0,origin=c,height=27.2mm,clip=true,trim=24mm 0mm 17mm 10mm]{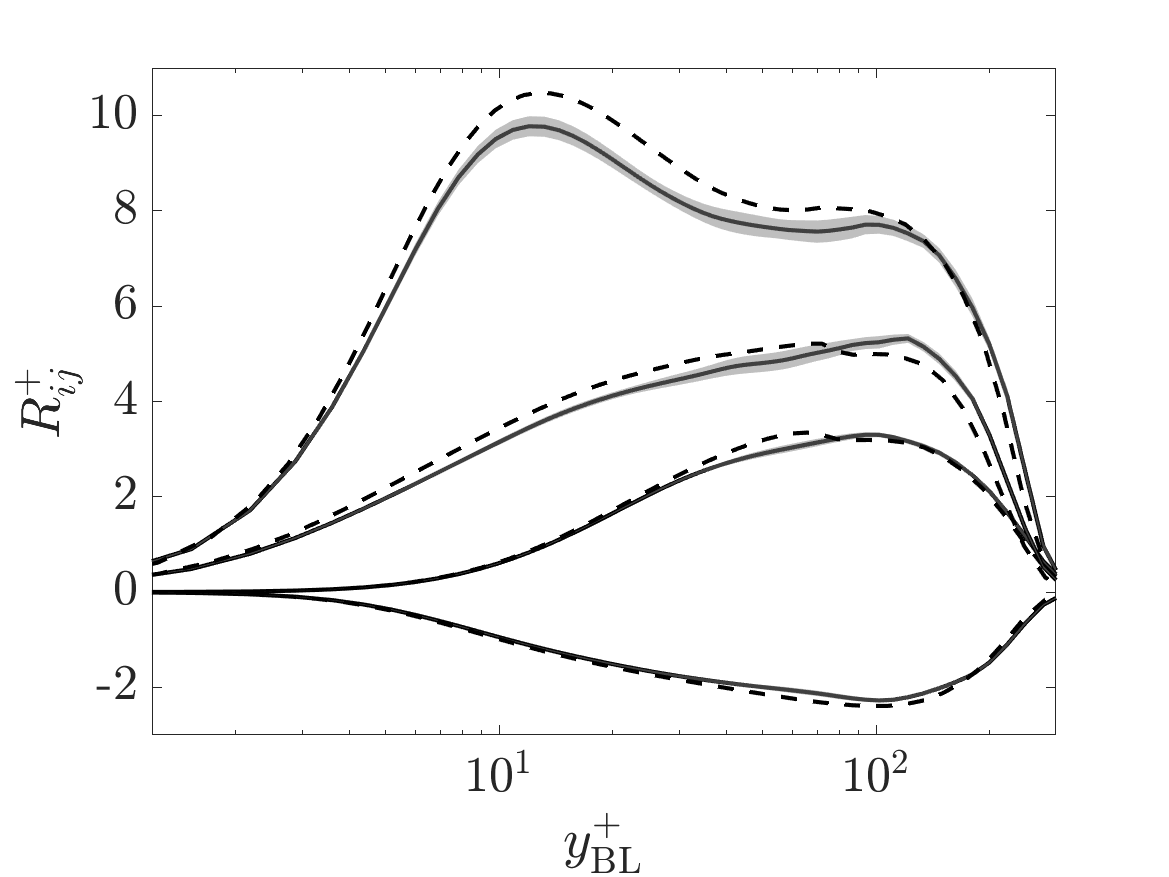}\llap{\parbox[b]{5mm}{(n)\\\rule{0ex}{23.2mm}}}
  \includegraphics[angle=0,origin=c,height=27.2mm,clip=true,trim=24mm 0mm 17mm 10mm]{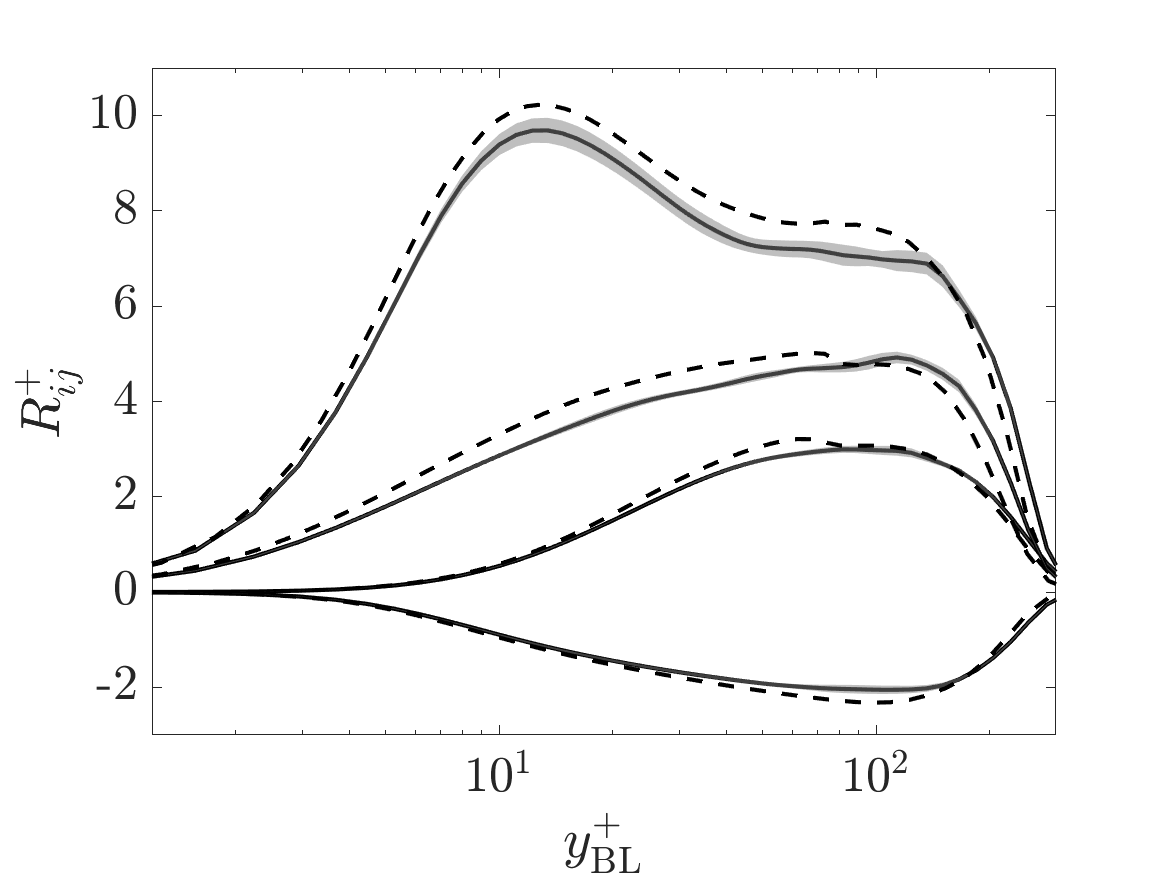}\llap{\parbox[b]{5mm}{(o)\\\rule{0ex}{23.2mm}}}
   \includegraphics[angle=0,origin=c,height=27.2mm,clip=true,trim=24mm 0mm 17mm 10mm]{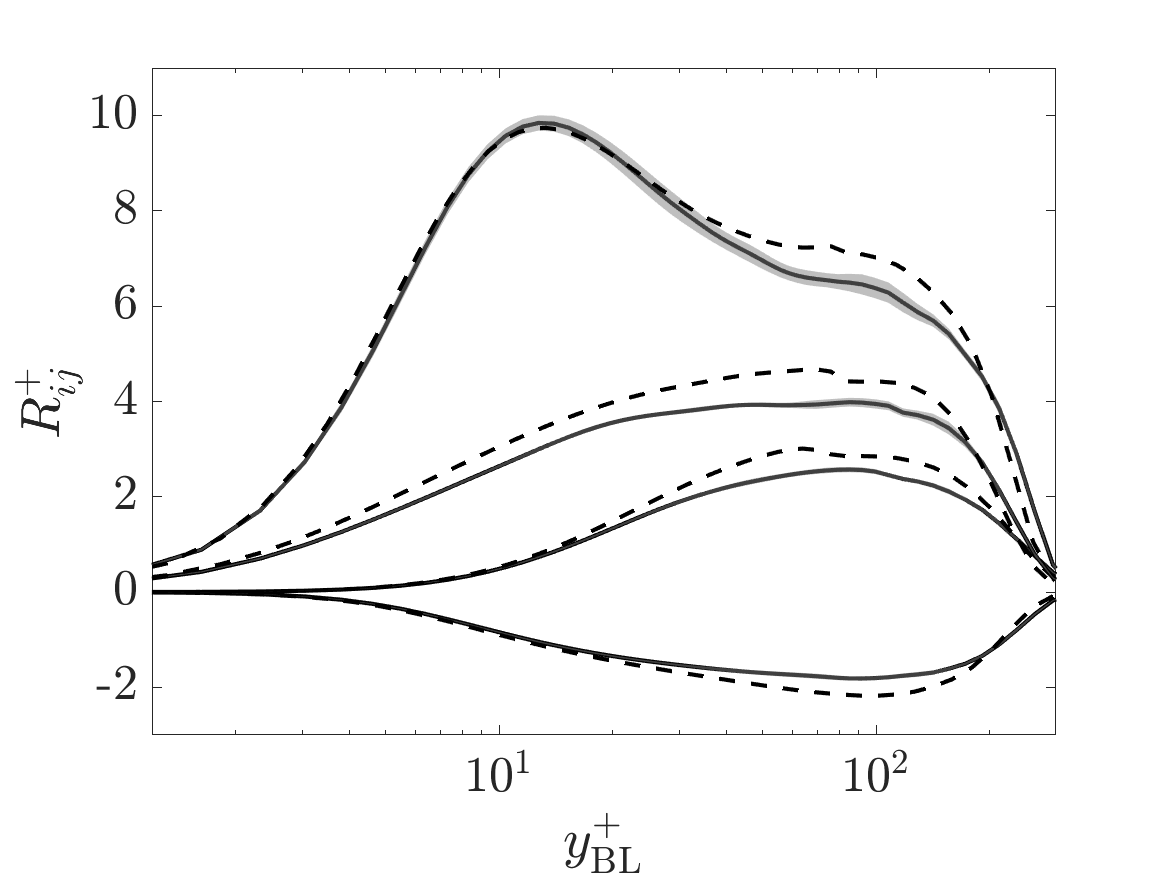}\llap{\parbox[b]{5mm}{(p)\\\rule{0ex}{23.2mm}}}
  \caption{
  \label{fig:rwt10-RS-allXZ}
  Variation of the Reynolds stresses with wall distance 
  at several streamwise and spanwise locations on the suction side of RWT-10 (solid lines)
  compared to P-5 (dashed lines) at matching $(Re_\tau,\beta_{x_\tau})$ values.
  Shaded regions correspond to 80\% confidence intervals of the inner-scaled Reynolds stress profiles due to finite time averaging.
%  Note that RWT-10 has been averaged for around half the time as RWT-5 and
%  therefore has higher uncertainties due to time averaging by a factor of around $\sqrt{2}$.   
  See the caption of figure~\ref{fig:rwt0-RS-allXZ} for more details.
  }
\end{figure}

Figures~\ref{fig:rwt0-RS-allXZ},~\ref{fig:rwt5-RS-allXZ}, and~\ref{fig:rwt10-RS-allXZ}
show the $R_{11}$, $R_{22}$, $R_{33}$, and $R_{12}$ components of the Reynolds stress
$R_{ij}=\avg{ u'_i u'_j }_\tau$
at different chord-wise and spanwise locations on the suction side
of RWT-0, RWT-5, and RWT-10,
compared with their corresponding profiles 
from P-0, P-2, and P-5, 
at matching $Re_\tau$, $\beta_{x_\tau}$, and $\alpha_{\rm eff}$.
Four streamlines are released at $x'/c=0.12$ (\ie, immediately downstream of the trip)
at four different spanwise ($z'$) locations: 
$z'/c=0.1$, $0.31$, $0.51$, and $0.72$. 
We then compute the intersection of these streamlines  
with planes located at the chord-wise locations $x'/c=0.6$, $0.7$, $0.8$, and $0.9$. 
These are the locations used for plotting the profiles,
where we note that the $z'$ locations depend on the flow field and are different 
at different $x'$ locations and for different angles of attack.
The minimum and maximum spanwise locations
are selected to avoid the effect of the symmetry plane at the root 
(although limited to a much smaller spanwise region)
and be fully inside the turbulent region of the flow 
(tripping only applied to $z'/c < 0.75$; see figure~\ref{fig:GFF-viz}).
The downstream spanwise locations are selected based on the
near-wall streamlines in the collateral region. 
%with nearly-constant deflection angles
%(which is different from the location of streamlines released farther from the wall).

% Office MacBook + office Lenovo display 

% Office MacBook + office Lenovo display 

% Office MacBook + office Lenovo display 

Tables~\ref{tab:app-rwt0-vs-p0},~\ref{tab:app-rwt5-vs-p2}, and~\ref{tab:app-rwt10-vs-p5} 
summarize the local friction Reynolds number $Re_\tau$
and the Clauser pressure-gradient parameters 
$\beta_{x_\tau}$ (streamwise) and $\beta_{z_\tau}$ (spanwise)
for all the profiles plotted in 
figures~\ref{fig:rwt0-RS-allXZ},~\ref{fig:rwt5-RS-allXZ}, and~\ref{fig:rwt10-RS-allXZ}.

\begin{table}
  \begin{center}
\def~{\hphantom{0}}
  \begin{tabular}{rcccc}
	$\frac{{\rm RWT-0}}{{\rm P-0}}$ & $z_1$ & $z_2$  & $z_3$  & $z_4$ \\
%	RWT-0 vs. P-0 & $z_1$ & $z_2$  & $z_3$  & $z_4$ \\
	$x'=0.6$ & 
	$\frac{(177,0.34,0.01)}{(176,0.34)}$ & $\frac{(176,0.33,0.03)}{(174,0.34)}$ & 
	$\frac{(178,0.33,0.06)}{(173,0.34)}$ & $\frac{(179,0.31,0.14)}{(170,0.33)}$ 
	\\
   	$x'=0.7$ & 
	$\frac{(201,0.46,0.01)}{(202,0.46)}$ & $\frac{(201,0.46,0.02)}{(201,0.46)}$ & 
	$\frac{(201,0.44,0.05)}{(199,0.44)}$ & $\frac{(211,0.40,0.11)}{(197,0.43)}$ 
	\\
	$x'=0.8$ & 
	$\frac{(223,0.73,0.01)}{(226,0.72)}$ & $\frac{(222,0.71,0.01)}{(225,0.71)}$ & 
	$\frac{(222,0.70,0.02)}{(224,0.69)}$ & $\frac{(234,0.65,0.04)}{(223,0.67)}$ 
	\\
	$x'=0.9$ & 
	$\frac{(238,1.48,0.00)}{(244,1.48)}$ & $\frac{(238,1.46,0.00)}{(244,1.46)}$ & 
	$\frac{(239,1.46,-0.03)}{(244,1.46)}$ & $\frac{(252,1.37,-0.15)}{(243,1.38)}$ 
	\\
  \end{tabular}
  \caption{
  \label{tab:app-rwt0-vs-p0}
  The set of three values in the numerator are 
  $(Re_\tau,\beta_{x_\tau},\beta_{z_\tau})$ for RWT-0 at the locations plotted in figure~\ref{fig:rwt0-RS-allXZ}.
  The pair of values in the denominator are $(Re_\tau,\beta_{x_\tau})$ for the P-0 profiles plotted for comparison.
  }
  \end{center}
\end{table}

\begin{table}
  \begin{center}
\def~{\hphantom{0}}
  \begin{tabular}{lcccc}
	$\frac{{\rm RWT-5}}{{\rm P-2}}$  & $z_1$ & $z_2$  &$z_3$  & $z_4$ \\
	$x'=0.6$ & 
	$\frac{(208,0.74,0.03)}{(222,0.70)}$ & $\frac{(207,0.72,0.07)}{(220,0.68)}$ & 
	$\frac{(208,0.68,0.12)}{(215,0.64)}$ & $\frac{(214,0.68,0.11)}{(218,0.67)}$ 
	\\
	$x'=0.7$ & 
	$\frac{(233 ,0.96 ,0.03 )}{(240 ,0.95 )}$ & $\frac{(232 ,0.94 ,0.06 )}{(239 ,0.92 )}$ & 
	$\frac{(235 ,0.88 ,0.09 )}{(237 ,0.87 )}$ & $\frac{(243 ,0.88 ,0.03 )}{(238 ,0.89 )}$ 
	\\
	$x'=0.8$ & 
	$\frac{(255 ,1.42 ,0.03 )}{(254 ,1.43 )}$ & $\frac{(254 ,1.37 ,0.04 )}{(253 ,1.37 )}$ & 
	$\frac{(258 ,1.31 ,0.03 )}{(252 ,1.32 )}$ & $\frac{(266 ,1.30 ,-0.12 )}{(252 ,1.33 )}$
	\\
	$x'=0.9$ & 
	$\frac{(267 ,2.65 ,0.02 )}{(259 ,2.67 )}$  & $\frac{(267 ,2.59 ,0.01 )}{(260 ,2.60 )}$  &
	$\frac{(273 ,2.44 ,-0.06 )}{(260 ,2.45 )}$ & $\frac{(281 ,2.51 ,-0.44 )}{(260 ,2.51 )}$
	\\
  \end{tabular}
  \caption{
  \label{tab:app-rwt5-vs-p2}
  The local friction Reynolds number and Clauser pressure-gradient parameters 
  of RWT-5 (numerator) and P-2 (denominator) 
  for the profiles plotted in figure~\ref{fig:rwt5-RS-allXZ}.
  See the caption of table~\ref{tab:app-rwt0-vs-p0} for more details.
  }
  \end{center}
\end{table}

\begin{table}
  \begin{center}
\def~{\hphantom{0}}
  \begin{tabular}{lcccc}
	$\frac{{\rm RWT-10}}{{\rm P-5}}$ & $z_1$ & $z_2$  &$z_3$  & $z_4$ \\ 
	$x'=0.6$ & 
	$\st{\frac{(240 ,1.33 ,0.05 )}{(249 ,1.26 )}}$ & $\st{\frac{(239 ,1.27 ,0.09 )}{(248 ,1.24 )}}$ & 
	$\st{\frac{(237 ,1.16 ,0.12 )}{(244 ,1.13 )}}$ & $\st{\frac{(245 ,0.99 ,0.05 )}{(237 ,1.05 )}}$ 
	\\
	$x'=0.7$ & 
	$\st{\frac{(263 ,1.72 ,0.03 )}{(266 ,1.70 )}}$ & $\st{\frac{(264 ,1.63 ,0.04 )}{(265 ,1.64 )}}$ & 
	$\st{\frac{(265 ,1.47 ,0.00 )}{(261 ,1.48 )}}$ & $\st{\frac{(273 ,1.21 ,-0.17 )}{(254 ,1.33 )}}$ 
	\\
	$x'=0.8$ & 
	$\st{\frac{(283 ,2.48 , 0.01 )}{(278 ,2.48 )}}$ & $\st{\frac{(284 ,2.40 ,-0.03 )}{(278 ,2.41 )}}$ & 
	$\st{\frac{(289 ,2.19 ,-0.19 )}{(277 ,2.19 )}}$ & $\st{\frac{(294 ,1.76 ,-0.55 )}{(272 ,1.83 )}}$
	\\
	$x'=0.9$ & 
	$\st{\frac{(292 ,4.20 ,-0.04 )}{(283 ,4.22 )}}$ & $\st{\frac{(292 ,4.17 ,-0.18 )}{(282 ,4.17 )}}$ & 
	$\st{\frac{(298 ,3.85 ,-0.46 )}{(282 ,3.87 )}}$ & $\st{\frac{(297 ,3.31 ,-1.19 )}{(284 ,3.31 )}}$
	\\ 
  \end{tabular}
  \caption{
 \label{tab:app-rwt10-vs-p5}
  The local friction Reynolds number and Clauser pressure-gradient parameters 
  of RWT-10 (numerator) and P-5 (denominator) 
  for the profiles plotted in figure~\ref{fig:rwt10-RS-allXZ}.
  See the caption of table~\ref{tab:app-rwt0-vs-p0} for more details.
  }
  \end{center}
\end{table}

\bibliographystyle{jfm}
\bibliography{references.bib}

%% End of file `jfm2esam.bib'.

\end{document}